\documentclass[conference]{IEEEtran-4}
\IEEEoverridecommandlockouts

\usepackage{bbding} 
\usepackage{graphicx} 
\usepackage{soul}
\usepackage{amsmath,amsfonts,amsthm,bm}
\usepackage{booktabs}
\usepackage{multirow}
\usepackage{bigstrut}
\usepackage{makecell}
\usepackage{algorithm}
\usepackage{algorithmicx}
\usepackage{algpseudocode}
\usepackage[table]{xcolor}
\usepackage{caption}
\usepackage[most]{tcolorbox}
\usepackage{tikz}
\usepackage{rotating}
\usepackage{wasysym}
\usepackage{cite}
\usepackage[table]{xcolor}
\usepackage[hyphens]{url}
\usepackage{xurl}
\usepackage[colorlinks,breaklinks=true]{hyperref}
\usepackage[available,functional]{ndssbadges}
\usepackage{lipsum}

\usepackage[normalem]{ulem}

\usepackage{hyperref}
\hypersetup{
	colorlinks=true,
	linkcolor=black,
	filecolor=magenta,      
	urlcolor=black,
	citecolor=black
}

\renewcommand{\paragraph}[1]{\par\noindent\textbf{#1}.}

\newcommand{\ndsstt}{\color{black}}
\newcommand{\ndssmt}{\textcolor{black}}
\newcommand{\ndssm}{\textcolor{black}}

\newcommand{\solid}{\CIRCLE}
\newcommand{\hollow}{\Circle}

\DeclareMathOperator*{\argmin}{arg\,min}

\begin{document}

\title{
Character-Level Perturbations Disrupt\\ LLM Watermarks
}

\author{
\IEEEauthorblockN{Zhaoxi Zhang$^{*\dag}$}
\thanks{$^*$ Work done during the visit at Griffith University.}
\IEEEauthorblockA{University of Technology Sydney\\
    zhaoxi.zhang-1@student.uts.edu.au}
\and
\IEEEauthorblockN{Xiaomei Zhang$^{\dag}$}
\thanks{$^{\dag}$ Equal contribution.}
\IEEEauthorblockA{Griffith University\\
    xiaomei.zhang@griffithuni.edu.au}
\and
\IEEEauthorblockN{Yanjun Zhang}
\IEEEauthorblockA{University of Technology Sydney\\
    yanjun.zhang@uts.edu.au}
\and
\IEEEauthorblockN{He Zhang}
\IEEEauthorblockA{RMIT University\\
    he.zhang@rmit.edu.au}
\and
\IEEEauthorblockN{Shirui Pan}
\IEEEauthorblockA{Griffith University\\
    s.pan@griffith.edu.au}
\and
\IEEEauthorblockN{Bo Liu}
\IEEEauthorblockA{University of Technology Sydney\\
    bo.liu@uts.edu.au}
\and
\IEEEauthorblockN{Asif Gill}
\IEEEauthorblockA{University of Technology Sydney\\
    asif.gill@uts.edu.au}
\and
\IEEEauthorblockN{Leo Yu Zhang \Envelope}
\thanks{\Envelope ~Corresponding author.}
\IEEEauthorblockA{Griffith University\\
    leo.zhang@griffith.edu.au}
}


\IEEEoverridecommandlockouts
\makeatletter\def\@IEEEpubidpullup{6.5\baselineskip}\makeatother
\IEEEpubid{\parbox{\columnwidth}{
    Network and Distributed System Security (NDSS) Symposium 2026\\
    23 - 27 February 2026, San Diego, CA, USA\\
    ISBN 979-8-9919276-8-0\\  
    https://dx.doi.org/10.14722/ndss.2026.230138\\
    www.ndss-symposium.org
}
\hspace{\columnsep}\makebox[\columnwidth]{}}

\maketitle

\begin{abstract}
Large Language Model (LLM) watermark has emerged as a promising technique for copyright protection, misuse prevention, and machine-generated content detection. 
It injects detectable signals during the LLM generation process, allowing for later identification by a corresponding detector.
To assess the robustness of watermark schemes, existing studies typically adopt watermark removal attacks, which aim to erase embedded signals by modifying the
watermarked text. 
However, we reveal that existing watermark removal attacks are suboptimal, which leads to the misconception that effective watermark removal requires either a large perturbation budget or a strong adversary’s capabilities, such as unlimited queries to the victim LLM or its watermark detector.  A systematic scrutinization of removal attack capabilities as well as the development of more sophisticated techniques remains largely underexplored. As a result, the robustness of existing watermarking schemes may be overestimated.

To bridge the gap, we first formalize the system model for LLM watermark, and characterize two realistic threat models constrained on limited access to the watermark detector. 
We then analyze how different types of perturbation vary in their attack range, i.e., the number of tokens they can affect with a single edit. We observe that character-level perturbations (e.g., typos, swaps, deletions, homoglyphs) can influence multiple tokens simultaneously by disrupting the tokenization process. We demonstrate that character-level perturbations are significantly more effective for watermark removal compared to token-level or sentence-level approaches under the most restrictive threat model. 
We further propose guided removal attacks based on the Genetic Algorithm (GA) that uses a reference detector for optimization. 
Under a practical threat model with limited black-box queries to the watermark detector, our method demonstrates strong removal performance.  
Experiments across five representative watermarking schemes and two widely-used LLMs consistently confirm the superiority of character-level perturbations and the effectiveness of the reference-detector-guided GA in removing watermarks under realistic constraints. 
\ndssm{Additionally, we argue there is an adversarial dilemma when considering potential defenses: any fixed defense can be bypassed by a suitable perturbation strategy. Motivated by this principle, we propose an adaptive compound character-level attack. Experimental results show that this approach can effectively defeat the defenses. } 
Our findings highlight significant vulnerabilities in existing LLM watermark schemes and underline the urgency for the development of new robust mechanisms.

\end{abstract}


\section{Introduction}

Large Language Models (LLMs) have become foundational components in modern AI systems \cite{zhang2022opt,touvron2023llama, openai2022gpt}, powering a wide range of consumer-facing applications and critical decision-making systems, such as chat assistants \cite{deepseek2025Terms, gemini2024Terms}, educational tools \cite{grammarlymain, chatpaper}, and content generation platforms \cite{DeepmindLyria, DeepmindImagen}, etc. 
However, as LLMs are increasingly used to generate high-quality content indistinguishable from human-written text, growing concerns have emerged about their potential misuse~\cite{zhang2022self, zhang2022evaluating, gong2025not,sun2025better, zhang2025dynamic, zhang2024unraveling}, including misinformation generation~\cite{pan2023risk}, automated phishing~\cite{hazell2023large}, and academic fraud~\cite{kasneci2023chatgpt}, etc. 
In response, there is a pressing demand for reliable mechanisms to attribute machine-generated text and distinguish it from human-written content.  
Among existing attribution techniques, the LLM watermark has emerged as one of the most promising approaches \cite{kirchenbauer2023watermark, zhao2024provable, hu2024unbiased, wu2024resilient, dathathri2024scalable, kuditipudi2024robust, christ2024undetectable, feng2025bimark}. 
LLM watermarks work by subtly adjusting the model’s generation process, embedding statistically detectable signals into the output text. 
These changes are invisible to users but can be identified by a watermark detector, which typically uses statistical tests to distinguish watermarked text from non-watermarked text.

However, the practical utility of such watermarks hinges on their robustness, i.e., the ability to remain detectable after adversarial edits. 
To assess robustness, researchers commonly conduct watermark removal attacks \cite{kirchenbauer2023watermark, kuditipudi2024robust, zhao2024provable, feng2024certified}, which attempt to erase embedded signals by modifying the watermarked text, typically via token-level (e.g., synonym replacement) or sentence-level perturbations (e.g., paraphrasing). 
However, existing approaches face several notable limitations. 
First, prior work often relies on unrealistic assumptions about the adversary's capabilities, 
such as full knowledge of the watermarking scheme, unlimited queries to the victim LLM or its watermark detector, or the availability of a similar surrogate LLM \cite{wu2024bypassing, huang2024b4, chen2024mark, chang2024watermark}. 
Second, these methods primarily rely on token-level or sentence-level perturbations \cite{krishna2024paraphrasing, kirchenbauer2024reliability, zhao2024provable}, which we \ndssm{have found to be suboptimal for efficient watermark removal (see experiments in Section~\ref{sec:compare_token_sentence} for detailed results)}. 
Third, due to the lack of guidance, these methods fail to prioritize tokens most critical to watermark removal, leading to inefficient perturbations \cite{kirchenbauer2024reliability, piet2023mark}.  
These limitations have led to a misleading perception that successful watermark removal requires either a large perturbation budget or strong adversary's capabilities. As a result, the robustness of existing watermarking schemes may be overestimated. This highlights the urgent need for a systematic investigation of watermark removal capabilities and the development of more sophisticated attack techniques.
 
To this end, we firstly conduct a systematic study on LLM watermark removal under realistic constraints.
We assume a black-box setting where the adversary can only access the output texts of the victim LLM, with no access to its architecture, parameters, or internal states (e.g., logits). Additionally, the adversary has no knowledge of the underlying watermarking scheme and is restricted to a limited number of queries per input.
Based on the adversary’s access to the original watermark detector of the victim LLM, we consider two distinct threat models:
(1) the adversary has no access to the detector; 
(2) the adversary can query the detector under a limited budget per input. 
Based on our categorization of adversary knowledge and capabilities, we classify existing watermark removal methods accordingly (refer to Table~\ref{tab:compare_exist}). Existing approaches are not applicable under the two practical threat models we consider.

To obtain the optimal perturbation operation, we begin with evaluating the removal attack performance by using different perturbation types, where the adversary has no access to the original watermark detector. 
In this challenging setting, the adversary can only apply random perturbations to the watermarked text. As a result, the effectiveness of the attack is largely determined by the attack range, which refers to the number of tokens affected by a single edit. 
Through analysis, we observe that character-level perturbations, such as typos, deletions, swaps, insertions of zero-width or whitespace characters, and homoglyph substitutions, can disrupt the tokenization process. This disruption often splits a single token into multiple subword units, allowing each edit to affect more than one token\footnote{\ndssm{
This phenomenon contrasts with the effects of character-level perturbations on adversarial robustness in traditional NLP, where small input changes are intended to cause significant changes in a model’s internal embeddings. From this perspective, token-level and character-level perturbations follow similar principles~\cite{li2018textbugger, morris2020textattack, gao2018black,zhang2025exploring,Zhang2023Masked}. In contrast, the principles and effects of token-level and character-level attacks on LLM watermarks are fundamentally different. 
}}. 
Motivated by this insight, we design a character-level watermark removal attack and evaluate its effectiveness through extensive experiments. Results show that character-level perturbations consistently outperform token-level and sentence-level methods, enabling watermark removal with a smaller perturbation budget. 

To handle the lack of direct guidance, we propose using a reference detector to improve the effectiveness of watermark removal attacks, where the adversary is allowed only a limited number of black-box queries to the original watermark detector. We first train a lightweight reference detector using data collected within the allowed query budget. This reference detector mimics the behavior of the original detector and helps guide our following Genetic Algorithm (GA) method in selecting impactful token positions. 
With the reference detector in place, the GA iteratively finds tokens most removal-relevant, 
enabling more focused and efficient perturbations while overcoming the constraint of limited queries to the original detector.

Our contributions are summarized as follows.
\begin{itemize}
    \item \textbf{Systematic formulation of LLM watermark removal.} We provide the first comprehensive analysis of watermark removal attacks against LLMs. We formally characterize threat models under different system settings, offering a structured framework for evaluating attack effectiveness.
    
    \item \textbf{Character-level perturbations and attack range analysis.} We identify the attack range, the impact scope of a single perturbation, as a key factor affecting the effectiveness of watermark removal. We show that character-level perturbations offer the widest attack range by disrupting tokenization. 
    
    \item \textbf{Reference-detector-guided removal attack and adaptive strategy.} Under the setting of limited access to the original watermark detector, we propose a GA-based removal attack guided by a reference detector, which is trained to approximate the behavior of the original detector using limited queries. 
    We also highlight a key limitation of the reference detector: the mismatch between the reference and original detectors makes gradient-based optimization unreliable. 
    \ndssm{Furthermore, we highlight an adversarial dilemma for potential defenses: for any fixed defense, there always exists an effective perturbation strategy that can bypass it. Guided by this insight, we propose an adaptive compound character-level attack.}
    Our method is evaluated across diverse watermarking schemes and consistently achieves strong performance.\footnote{We release the source code at \url{https://github.com/plll4zzx/CharacterRemoval4WM}} 
\end{itemize}

\begin{table}[t]
  \centering
  \caption{
  Comparison of our study with existing typical watermark removal methods.
  \textbf{Adversary’s capabilities \& knowledge} (CK): CK1: Limited query budget to victim LLM; CK2: No access to original detector; CK3: Limited query budget to original detector; CK4: No knowledge of victim LLM; CK5: No knowledge of watermark scheme. \textbf{Attack perturbation type \& strategies} (PS): PS1: Character-level perturbation; PS2: Token-level perturbation; PS3: Sentence-level perturbation; PS4: Gradient-free guidance. Here, \solid\ denotes ``Fully Considered'', and \hollow\ denotes ``Not Considered''. 
  }
  \resizebox{0.9\linewidth}{!}{
    \begin{tabular}{c|c|cccccccc}
    \toprule
          &       & CK1+CK2 & CK1+CK3 & CK4    & CK5    & PS1    & PS2    & PS3    & PS4 \\
    \midrule
    \multirow{4}[2]{*}{\makecell{Stealing \\ \&Removal}} & \cite{Jovanovi2024steal} & \solid & \solid & \solid & \hollow & \hollow & \solid & \hollow & \solid \\
          & \cite{zhang2024steal} & \solid & \solid & \solid & \hollow & \hollow & \solid & \hollow & \solid \\
          & \cite{wu2024bypassing} & \hollow & \hollow & \solid & \hollow & \hollow & \solid & \hollow & \solid \\
          & \cite{chen2024mark} & \solid & \solid & \solid & \hollow & \hollow & \solid & \hollow & \solid \\
    \midrule
    \multirow{9}[4]{*}{Removal} & \cite{he2024translation} & \solid & \hollow & \solid & \solid & \hollow & \hollow & \solid & \hollow \\
          & \cite{krishna2024paraphrasing} & \solid & \hollow & \solid & \solid & \hollow & \hollow & \solid & \hollow \\
          & \cite{liu2024evaluating} & \solid & \hollow & \solid & \solid & \solid & \solid & \solid & \hollow \\
          & \cite{creo2025silverspeak} & \solid & \hollow & \solid & \solid & \solid & \hollow & \solid & \hollow \\
          & \cite{piet2023mark} & \solid & \hollow & \solid & \solid & \solid & \solid & \solid & \hollow \\
          & \cite{zhang2024sand} & \solid & \hollow & \solid & \solid & \hollow & \solid & \hollow & \hollow \\
          & \cite{liang2024waterpark} & \solid & \solid & \solid & \solid & \solid & \solid & \solid & \hollow \\
          & \cite{chang2024watermark} & \hollow & \hollow & \hollow & \solid & \hollow & \solid & \hollow & \hollow \\
\cmidrule{2-10}          & \textbf{Ours} & \solid & \solid & \solid & \solid & \solid & \solid & \solid & \solid \\
    \bottomrule
    \end{tabular}%
    }
  \label{tab:compare_exist}%
\end{table}%

\section{Background}
\label{sec:preliminaries}

\subsection{LLM Generation}
\label{sec:llm_gen}

Large Language Models (LLMs) generate text autoregressively, producing one token at each step conditioned on the given prompt (and the previously generated tokens), until an end-of-sequence token is generated or a predefined maximum length is reached. The generation process at each step $t$ involves three stages: computing the logits, deriving a probability distribution, and sampling the next token \cite{zhang2022opt, touvron2023llama, openai2022gpt}. 
First, the model takes the prompt and previously generated tokens $x_{<t} = \{x_1, x_2, \dots, x_{t-1}\}$ as input and produces a logit vector $\ell_t \in \mathbb{R}^{|V|}$, where $|V|$ is the vocabulary size. Second, the logits are transformed via a softmax function into a probability distribution over the vocabulary. Third, the next token $x_t$ is sampled from this distribution and then appended to the sequence.

\subsection{Injecting Watermark}
\label{sec:add_wm} 

LLM watermark aims to inject detectable signals into LLM-generated content.  In this study, we focus on inference-time watermark, which injects watermark signals during text generation without modifying the model’s parameters \cite{kirchenbauer2023watermark, zhao2024provable, hu2024unbiased, wu2024resilient, dathathri2024scalable, kuditipudi2024robust, christ2024undetectable}. 
Compared to training-time watermark methods, such as trigger-based approaches~\cite{Cong2022ssl, Jia2021Watermarks}, inference-time methods are more flexible and cost-effective, as they do not require access to the model’s weights or retraining. 
Formally, at generation step $t$, the LLM computes a key $k_t$ using a hash function $\mathrm{hash}(\cdot)$ over the previous context $c_t = \{x_{t-h}, \cdots, x_{t-1}\}$, where $h$ is the length of context. 
Beyond the typical range, there exists a special case where $h = 0$, in which all tokens generated by the same LLM share the same key. This key is then used to subtly alter the LLM’s generation behavior. 
Existing watermarking techniques fall into three categories based on the generation stage where the watermark is introduced:
\begin{itemize}
    \item \textbf{Watermark during logits generation:} These methods aim to shift the generation process by modifying the LLM's logits. At each generation step $t$, a pseudorandom function and a key $k_t$ are used to partition the vocabulary $V$ into two subsets: a green list and a red list. 
    A fixed bias $\delta$ is added to the logits of green-list tokens, increasing their likelihood of being sampled. So, generated texts are biased toward green-list tokens and can be detected as watermarked, such as KGW \cite{kirchenbauer2023watermark}, Unigram \cite{zhao2024provable}.

    \item \textbf{Watermark during token probability distribution generation:} These methods aim to modify the token probability distribution. At each step $t$, a pseudorandom function and a key $k_t$ are used to reorder the vocabulary tokens. Tokens whose cumulative probability exceeds a threshold $\gamma$ are selected and reweighted to increase their sampling probability, resulting in watermark insertion, such as Unbias \cite{hu2024unbiased}, DIP \cite{wu2024resilient}.

    \item \textbf{Watermark during sampling:} These methods aim to embed watermarks by modifying the sampling process. Two main strategies are commonly used. 
    (1) A set of candidate tokens is first sampled from the original token probability distribution. Then, each candidate is assigned a pseudorandom score derived from a key $k_t$, and the token with the highest score is selected as $x_t$, such as SynthID \cite{dathathri2024scalable}. 
    (2) Alternatively, a pseudorandom score is derived from $k_t$ and used in inverse transform sampling to deterministically select the next token \cite{kuditipudi2024robust, christ2024undetectable}. 
\end{itemize}

\subsection{Detecting Watermark}
\label{sec:detect_wm}

Watermark detection is inherently tied to the watermark injecting strategy, as each approach modifies the token generation process differently. The goal is to evaluate whether a given text $X$ exhibits statistical traces of watermarking, typically via a global watermark score $S_w(X)$. 
A higher score indicates a stronger alignment with the watermark. 
Formally, the input text is first tokenized into a sequence $X = \{x_1, x_2, \cdots, x_t, \cdots, x_m\}$. 
For each position $t$, a key $k_t$ is derived from a hash function over the prior context $c_t = \{x_{t-h}, \cdots, x_{t-1}\}$, i.e., $k_t = \mathrm{hash}(c_t)$. Depending on the specific watermark injection method (see Section~\ref{sec:add_wm}), a watermark score $s_t$ for each token $x_t$ might be computed using the token itself and its associated key $k_t$.
If such scores are used, they are typically aggregated across the sequence to obtain the global score \( S_w(X) \); otherwise, detection relies on aggregate statistics such as token counts.  If $S_w(X)$ exceeds a predefined threshold $\tau_d$, the sequence $X$ is labelled as watermarked. Detection methods for each category of watermark scheme are detailed below:
\begin{itemize}
    \item \textbf{Watermark during logits generation:}  The detector uses $k_t$ to determine the green list at position $t$ and checks whether $x_t$ belongs to it. The global watermark score is typically computed using a one-sided z-test: $S_w(X) = \frac{|X|_G - \gamma |X|}{\sqrt{\gamma(1 - \gamma)|X|}}$, where $|X|_G$ is the number of green-list tokens in $X$, $\gamma$ is the ratio of the green list to the entire vocabulary \cite{kirchenbauer2023watermark, zhao2024provable}. 

    \item \textbf{Watermark during token probability generation:} These watermark methods modify the token probability distribution during generation, without altering the logits directly. To detect such watermarks, the detector compares how likely each token is under the watermarked LLM $M_w$ versus the original LLM $M$. Specifically, for each token $x_t$, the token-level watermark score is computed as the log-likelihood ratio: $s_t = \log \frac{M_w(x_t \mid x_{<t}, k_t)}{M(x_t \mid x_{<t})}$, where $M_w$ generates a key-conditioned distribution using $k_t$. The global watermark score is then aggregated as: $S_w(X) = \sum_{t=1}^{m} s_t$. Watermarked texts tend to follow the token probability distribution of $M_w$, resulting in a higher $S_w(X)$ than non-watermarked texts \cite{hu2024unbiased, wu2024resilient}.

    \item \textbf{Watermark during sampling:} To detect the watermark, the detector recovers the pseudorandom score assigned to $x_t$ by using the key $k_t$, and computes the token-level watermark score $s_t$ by checking whether $x_t$ aligns with the pseudorandom score. The global watermark score of $X$ is then calculated by summing all $s_t$: $S_w(X) = \sum_{t=1}^m s_t$. If $x_t$ is from watermarked text, it tends to get higher $s_t$, and watermarked text tends to get higher $S_w(X)$ \cite{kuditipudi2024robust, dathathri2024scalable, christ2024undetectable}.

\end{itemize}

\subsection{Watermark Removal Attacks}
\label{sec:wm_removal}
We study watermark removal attacks, which aim to erase the watermark embedded in LLM-generated text while preserving its original semantics. 
Formally, given a watermarked text $X = \{x_1, x_2, \cdots, x_t, \cdots, x_m\}$, the adversary applies an attack $\mathcal{A}$ to generate a perturbed text $\tilde{X} = \mathcal{A}_{\tilde{P}}(X)$, where $\tilde{P} \subset \{1, 2, \cdots, m\}$ indicates the positions selected for perturbation. 
After perturbed, the global watermark score $S_w(\tilde{X})$ is below the threshold, thereby evading watermark detection.
Existing watermark removal methods can be categorized based on the level of perturbation:
\begin{itemize} 
    \item Character-level perturbations includes typos, character deletion, character swapping, insertion zero-width character or withe space, and homoglyph substitution etc., denoted as $\tilde{X} = \mathcal{A}_{\tilde{P}}^C(X)$, where $\tilde{P}$ is randomly selected from $\{1,2, \cdots, m\}$  \cite{liang2024waterpark, piet2023mark, liu2024evaluating, creo2025silverspeak}. 
    \item Token-level perturbations include word deletion, reordering, and synonym replacement etc., denoted as $\tilde{X} = \mathcal{A}_{\tilde{P}}^T(X)$,  with $\tilde{P}$ also selected randomly \cite{zhao2024provable, kirchenbauer2023watermark, kuditipudi2024robust}. 
    \item Sentence-level perturbations apply high-level transformations such as paraphrasing or translation, denoted as $\tilde{X} = \mathcal{A}_{\tilde{P}}^S(X)$, where $\tilde{P}$ is selected by a paraphraser or translator \cite{krishna2024paraphrasing, he2024translation}. 
\end{itemize}

\paragraph{Summary}
For character-level and token-level, perturbation positions are typically chosen randomly without considering the watermark structure \cite{liang2024waterpark, piet2023mark, liu2024evaluating, creo2025silverspeak, zhao2024provable, kirchenbauer2023watermark, kuditipudi2024robust}. 
For sentence-level methods, such as paraphrasing, the position is selected according to the semantic demand \cite{krishna2024paraphrasing, he2024translation}. 
Consequently, we define these methods as \textit{random strategies}, as they lack awareness of watermark-related features while removing. 
To remove the watermark, these methods typically rely on altering as many tokens as possible. 
As a result, the systematic scrutinization of removal attack capability --- as well as the development of more sophisticated techniques --- remains largely underexplored (refer to Appendix~\ref{sec:related_work} for more discussion about related work). 

\paragraph{Evaluation Metric}
Following prior works~\cite{kirchenbauer2023watermark, zhao2024provable, dathathri2024scalable, wu2024resilient, hu2024unbiased, kuditipudi2024robust}, we adopt a set of metrics to assess the effectiveness of watermark removal attacks. These metrics can be categorized into two groups according to their purposes:
\begin{itemize}
    \item Metric for perturbation budget:
        (1) Character editing distance ($\mathrm{ED}_C(X, \tilde{X})$): Number of modified characters. 
        (2) Character editing rate ($\mathrm{ER}_C(X, \tilde{X})$): Ratio of modified characters, computed as $\frac{\mathrm{ED}_C(X, \tilde{X})}{|X|_C}$, where $|X|_C$ is the character number of $X$. 
        (3) Token editing distance ($\mathrm{ED}_T(X, \tilde{X})$): Number of modified tokens. 
        (4) Token editing rate ($\mathrm{ER}_T(X, \tilde{X})$): Ratio of modified tokens, $\frac{\mathrm{ED}_T(X, \tilde{X})}{|X|}$, where $|X|$ is the token number of $X$.
    \item Metric for removal performance: 
        (1) Watermark score dropping rate 
        \ndssm{($\frac{S_w(X) - S_w(\tilde{X})}{S^w_{max}-S_{min}}$): where, $S^w_{max}$ is defined as the maximum value of $S_w(\cdot)$ across all sentences in watermarked dataset, and $S_{min}$ is defined as the minimum value of $S_w(\cdot)$ across all sentences in non-watermarked dataset.}
        (2) Attack success rate ($\mathrm{ASR}$): The percentage of watermarked texts where the watermark is no longer detected after the attack, $\mathrm{ASR}$ is computed over a dataset consisting of watermarked samples only.
\end{itemize}
Based on our analysis of the C4 dataset~\cite{2019t5}, each token contains an average of $5.18$ characters. Therefore, when a single character is modified within a token, both the character-level and token-level editing distances are $1$, while the character-level editing rate is approximately $\frac{1}{5.18}$ of the token-level editing rate. 
For a fair comparison, all editing distance ($\mathrm{ED}$) and editing rate ($\mathrm{ER}$) in this paper refer to the token level ($\mathrm{ED}_T$ and $\mathrm{ER}_T$) unless otherwise specified.

\section{Problem Formulation}

\subsection{System Model} 
\label{sec:system_model}
We consider a LLM system where the model provider offers a black-box API of a large language model (LLM), which takes user input and returns text output embedded with a watermark \cite{DeepmindImagen, DeepmindLyria, DeepmindSynthID}. A corresponding watermark detector API is provided to determine whether a given text contains a watermark generated by the LLM. 
Based on the accessibility of the watermark detector, we define two system models:
\begin{itemize}
    \item \textbf{System Model 1:} The watermark detector is private and can only be accessed by the model provider or authorized parties. 
    \item \textbf{System Model 2:} The watermark detector is provided as a public API. Anyone can query the detector with a text input and obtain the detection result. 
\end{itemize}
Modern LLM systems typically deploy access controls to detect and block malicious behaviors, such as reverse engineering or data extraction \cite{openai2024Terms, gemini2024Terms, deepseek2025Terms}. 
Therefore, in practical scenarios, adversaries are allowed only a limited number of black-box queries to the LLM and (if applicable) the watermark detector for each input and its slightly perturbed versions.

\begin{figure}
    \centering
    \includegraphics[width=\linewidth]{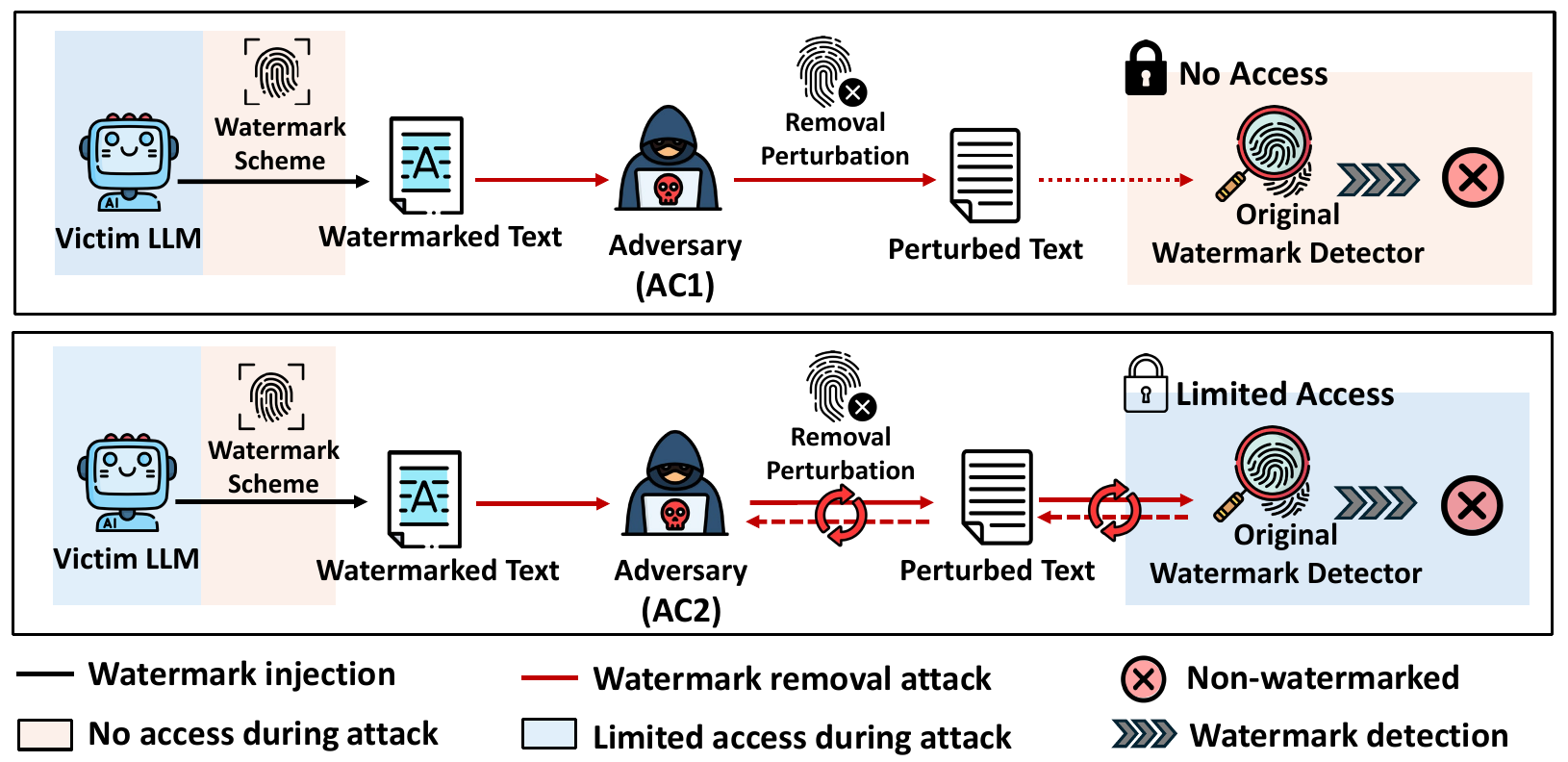}
    \caption{
    Illustration of threat models in watermark removal. 
    }
    \label{fig:threat_model}
\end{figure}

\subsection{Threat Model}
\label{sec:threat_model}

\paragraph{Adversary's Goal}
We consider adversaries who aim to remove watermarks from LLM-generated text using minimal imperceptible perturbation.
Formally, the goal is to find a perturbed text $\tilde{X}$ such that the global watermark score $S_w(\tilde{X})$ falls below a detection threshold $\tau_d$, while minimizing the editing rate: 
\begin{IEEEeqnarray}{rCl} 
\begin{aligned} 
    \argmin_{\tilde{X}} \mathrm{ER}(X, \tilde{X}), \ 
    \text{s.t. } S_w(\tilde{X}) < \tau_d.
\end{aligned} 
\end{IEEEeqnarray} 
This goal aligns with practical scenarios where adversaries seek to bypass watermark detectors while preserving the text’s utility, enabling unauthorized use of the model output. 
Moreover, analyzing the minimum perturbation required for removal offers a precise measure of watermark robustness. 

\paragraph{Adversary's Knowledge} 
According to the system model, the adversary only has access to the watermarked text generated by the victim LLM API. As a result, they have no knowledge of the underlying watermark scheme or its parameters (e.g., $\delta$, $\gamma$, $h$ described in Section~\ref{sec:add_wm}). This limitation makes watermark stealing attacks infeasible~\cite{wu2024bypassing, zhang2024steal, Jovanovi2024steal}, as such attacks typically rely on knowledge of the watermark embedding mechanism. 
Due to the black-box nature of the LLM API, the adversary cannot access the model’s architecture, weights, or internal outputs such as logits. Consequently, they are unable to perform model distillation or identify a similar open-source surrogate, which rules out attacks that approximate the token probability distribution using surrogate LLMs~\cite{chang2024watermark, diaa2024optimizing, huang2024b4}. 

\paragraph{Adversary's Capabilities}  
Based on the system model, we define two levels of adversarial capabilities (ACs), as illustrated in Figure~\ref{fig:threat_model}, each corresponding to different system models:
\begin{itemize}
    \item \textbf{AC1} (System model 1): The adversary has \textit{no access} to the original watermark detector. They can only interact with the victim LLM through a limited number of black-box queries per input.
    \item \textbf{AC2} (System model 2): The adversary has \textit{limited access} to the victim LLM and its original watermark detector. They can query them with a limited number of black-box queries per input.
\end{itemize}
These capabilities reflect practical constraints imposed by modern LLM systems, which implement access control to prevent malicious behaviors \cite{openai2024Terms, gemini2024Terms, deepseek2025Terms}.

\section{Benchmark of Removal Attack for LLM Watermark}
\label{sec:bad_char}

In this section, we systematically investigate the effectiveness of watermark removal attacks under the setting of \textbf{AC1} and highlight the clear advantages of character-level perturbations. We begin by analyzing the inherent strengths of character-level attacks, particularly their broader impact on watermarking mechanisms under constrained editing budgets. Building on this insight, we introduce a simple yet effective baseline that applies character-level perturbations in the \textbf{AC1} scenario. Finally, we conduct a comprehensive experimental evaluation comparing character-level attacks with token-level and sentence-level methods, demonstrating the superior removal effectiveness and better text quality preservation achieved by character-level approaches.

\begin{figure}
    \centering
    \includegraphics[width=0.8\linewidth]{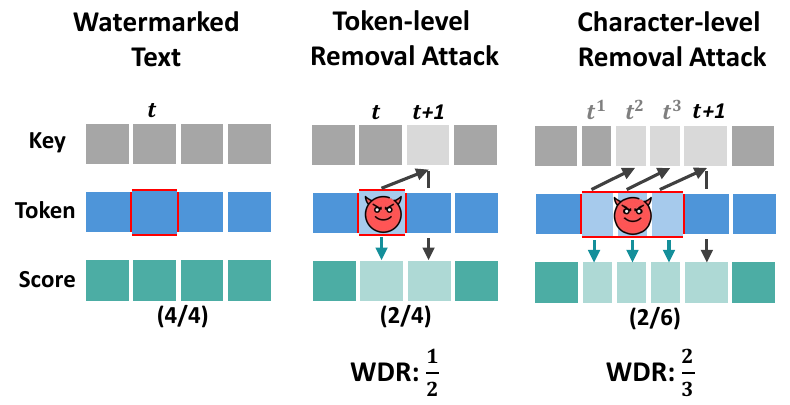}
    \caption{
    Comparison of attack range between token-level and character-level perturbations in watermark removal. 
    Under a context size of $h=1$, modifying one token affects the token itself and the key computation for $h$ subsequent tokens, resulting in an attack range of $2$ (i.e., $h+1$). 
    Modifying a single character (e.g., via homoglyph substitution) will split the affected token into at least 3 subword tokens (e.g., splitting token $t$ into sub-tokens $t^1$, $t^2$, $t^3$). This perturbation also influences the next $h$ positions, yielding an attack range of $4$ (i.e., $h+3$).
    Assuming each token watermark score $s_t \in \{0,1\}$, with dark shading indicating $s_t = 1$, the global watermark score of the watermarked text is $S_w(X) = \frac{\sum s_t}{|X|} = \frac{4}{4}$. After applying token-level perturbations, the score drops to $S_w(\tilde{X}) = \frac{2}{4}$, yielding a watermark score dropping rate ($\mathrm{WDR}$) of $\frac{1}{2}$. In contrast, character-level perturbations result in $\mathrm{WDR} = \frac{2}{3}$.
}
    \label{fig:bad_char}
\end{figure}

\subsection{Motivation of Using Character-Level Perturbation in Removal Attack}
\label{sec:bad_char_analysis}
The goal of a watermark removal attack is to reduce the global watermark score $S_w(X)$ of a text $X$ below the detection threshold. This score is computed by aggregating the individual watermark scores of all tokens in the sequence $X = \{x_1, x_2, \cdots, x_t, \cdots, x_m\}$. When a token $x_t$ is modified, its own score $s_t$ will be affected. In addition, since the watermark key $k_{t+i}$ is computed from the preceding context $c_{t+i} = \{x_{t-h+i}, \cdots, x_{t-1+i}\}$ ($x_t \in c_{t+i}$, if $1 \leq i \leq h$), a single modification can also affect the keys of the following $h$ tokens (i.e., ${k_{t+i}}, i \in [1, h]$). Consequently, this also alters the watermark scores of those subsequent tokens (i.e., ${s_{t+i}}, i \in [1, h]$).
This means that the effectiveness of a watermark removal attack depends not only on the editing rate—how many tokens are changed—but also on the attack range, i.e., how many tokens are affected by a single edit. 
Under the constraint of keeping the editing rate small, it becomes critical to choose perturbations that maximize the attack range in order to reduce the global watermark score more effectively. 

To analyze the attack range more intuitively, we consider that a token’s score is negatively affected (watermark score decreases) if either the token itself or its key is changed. In token-level attacks, modifying one token directly affects its own score and the keys of the following $h$ tokens, leading to an attack range of $h+1$. 
In contrast, character-level attacks can have a broader impact. As shown in Figure~\ref{fig:bad_char}, replacing a middle character of the $t$-th token with a homoglyph can cause the token to be split into at least three subword tokens (the subword before homoglyph, homoglyph, and the subword after homoglyph) during tokenization—specifically $x_{t^1}$, $x_{t^2}$, and $x_{t^3}$ in the perturbed text. For example, replacing ``t'' in ``letter'' with its homoglyph ``\v{t}'' (U+0165) may split the word into [``le'', ``\v{t}'', ``ter'']. This means that a single character-level edit can directly affect three tokens.
Furthermore, because the key $k_{t+1}$ depends on the previous $h$ tokens, the watermark score of $x_{t+1}$ token in the perturbed text will also be affected. As a result, a single character-level modification leads to an attack range of $h+3$.

Figure~\ref{fig:bad_char} illustrates this effect under the common setting of $h = 1$. For simplicity, we assume that each token in the watermarked text carries watermark information, with a watermark score of 1 (shown in dark color, light color means a watermark score of 0). In this case, the attack range of a token-level attack is 2, while a character-level attack reaches 4, twice as much. In terms of reducing the global watermark score, character-level attacks are significantly more effective.

\begin{tcolorbox}[colback=gray!10!white, colframe=gray!50!black, boxrule=0.8pt, arc=3pt]
\textbf{Takeaway: } 
Maximizing the attack range is crucial for effective watermark removal under a low editing rate.
\end{tcolorbox}

\subsection{A Random Strategy for Character-Level Removal Attack}
\label{sec:bad_char_rm}

Based on the above analysis, we propose a simple baseline method for watermark removal based on character-level perturbation. As described in the threat model in Section~\ref{sec:threat_model}, in \textbf{AC1}, the adversary is not allowed to interact with a local surrogate LLM and the original watermark detector of the victim LLM. As a result, the attacker adopts a random perturbation strategy: randomly selecting positions in the text and applying character-level perturbation. 

To maximize the attack range of each perturbation, we modify only a single character for each selected token, preferably one located near the center of the token. 
For example, the token ``position'' may be modified to ``po\v{s}ition'', where ``\v{s}'' (U+0161 in Unicode) is a homoglyph of ``s''. 
This method can be formulated as follows: 
\begin{IEEEeqnarray}{rCl}
\begin{aligned}
    \tilde{X} = \mathcal{A}_{\tilde{P}}^C(X), \
    \text{s.t.} \quad \tilde{P} \xleftarrow{\text{rand}} \{1, \cdots, m\},\quad 
    \frac{|\tilde{P}|}{m} \leq \epsilon,
\end{aligned}
\label{eq:bad_char_base}
\end{IEEEeqnarray}
where $\mathcal{A}^C$ is a character-level attack, $\tilde{P} \subset \{1, 2, \cdots, m\}$ is sampled uniformly at random, $m$ is the token number of $X$, and $\epsilon$ is upper bond of editing rate.

\begin{table}[t]
  \centering
\caption{Comparison of watermark removal performance under token-level, character-level, and sentence-level attacks across five watermarking schemes. For sentence-level attacks, we use DIPPER ($\dagger$) and \ndssmt{AuthorMist ($*$)}. The editing rate of DIPPER can be approximately adjusted to $\mathrm{ER} \approx 0.5$, \ndssmt{ whereas AuthorMist lacks editing rate controllability.} 
Character-level attacks consistently achieve higher $\mathrm{WDR}$ and $\mathrm{ASR}$ under comparable editing rates.
}

  \resizebox{\linewidth}{!}{
  {
    \begin{tabular}{c||c|cc|cc||ccc}
    \toprule
          &       & \multicolumn{2}{c|}{Token} & \multicolumn{2}{c||}{Char} & \multicolumn{3}{c}{Sentence} \\
          & $\mathrm{ER}$ & \ndssmt{$\mathrm{WDR}(\uparrow)$} & $\mathrm{ASR}(\uparrow)$ & \ndssmt{$\mathrm{WDR}(\uparrow)$} & $\mathrm{ASR}(\uparrow)$ & $\mathrm{ER}$ & \ndssmt{$\mathrm{WDR}(\uparrow)$ }& $\mathrm{ASR}(\uparrow)$ \\
    \midrule
    \rowcolor[rgb]{ .906,  .902,  .902} \multicolumn{9}{c}{\textbf{OPT}} \\
    \midrule
    \multirow{2}[2]{*}{KGW} & 0.1   & \ndssmt{0.0832} & 0.0224 & \ndssmt{0.1274} & 0.2228 & 0.6130$^\dagger$ & \ndssmt{0.2947$^\dagger$} & 0.7710$^\dagger$ \\
          & 0.5   & \ndssmt{0.3349} & 0.9725 & \ndssmt{0.4146} & 0.9949 & \ndssmt{2.5579$^*$} & \ndssmt{0.3710$^*$} & \ndssmt{0.9863$^*$} \\
    \midrule
    \multirow{2}[2]{*}{DIP} & 0.1   & \ndssmt{0.1847} & 0.5917 & \ndssmt{0.2000} & 0.6586 & 0.5832$^\dagger$ & \ndssmt{0.3539$^\dagger$} & 0.9784$^\dagger$ \\
          & 0.5   & \ndssmt{0.4206} & 0.9978 & \ndssmt{0.4226} & 0.9967 & \ndssmt{2.4413$^*$} & \ndssmt{0.4194$^*$} & \ndssmt{1.0000$^*$} \\
    \midrule
    \multirow{2}[2]{*}{SynthID} & 0.1   & \ndssmt{0.1698} & 0.0709 & \ndssmt{0.2191} & 0.1368 & 0.5068$^\dagger$ & \ndssmt{0.3587$^\dagger$} & 0.6229$^\dagger$ \\
          & 0.5   & \ndssmt{0.4259} & 0.9574 & \ndssmt{0.4486} & 0.9970 & \ndssmt{2.4469$^*$} & \ndssmt{0.4750$^*$} & \ndssmt{0.9933$^*$} \\
    \midrule
    \multirow{2}[2]{*}{Unigram} & 0.1   & \ndssmt{0.0437} & 0.0520 & \ndssmt{0.0788} & 0.0788 & 0.5911$^\dagger$ & \ndssmt{0.1540$^\dagger$} & 0.4377$^\dagger$ \\
          & 0.5   & \ndssmt{0.2176} & 0.8654 & \ndssmt{0.2663} & 0.9072 & \ndssmt{2.3805$^*$} & \ndssmt{0.2077$^*$} & \ndssmt{0.6815$^*$} \\
    \midrule
    \multirow{2}[2]{*}{Unbias} & 0.1   & \ndssmt{0.1760} & 0.5684 & \ndssmt{0.1964} & 0.6728 & 0.5797$^\dagger$ & \ndssmt{0.3682$^\dagger$} & 0.9823$^\dagger$ \\
          & 0.5   & \ndssmt{0.4030} & 0.9968 & \ndssmt{0.4151} & 0.9978 & \ndssmt{2.4596$^*$} & \ndssmt{0.4286$^*$} & \ndssmt{0.9965$^*$} \\
    \midrule
    \rowcolor[rgb]{ .906,  .902,  .902} \multicolumn{9}{c}{\textbf{LLaMA}} \\
    \midrule
    \multirow{2}[2]{*}{KGW} & 0.1   & \ndssmt{0.0923} & 0.1359 & \ndssmt{0.1136} & 0.2261 & 0.5240$^\dagger$ & \ndssmt{0.2461$^\dagger$} & 0.7376$^\dagger$ \\
          & 0.5   & \ndssmt{0.3472} & 0.9675 & \ndssmt{0.3747} & 0.9830 & \ndssmt{2.3925$^*$} & \ndssmt{0.3867$^*$} & \ndssmt{0.9714$^*$} \\
    \midrule
    \multirow{2}[2]{*}{DIP} & 0.1   & \ndssmt{0.1724} & 0.7188 & \ndssmt{0.1977} & 0.8150 & 0.5033$^\dagger$ & \ndssmt{0.3355$^\dagger$} & 0.9292$^\dagger$ \\
          & 0.5   & \ndssmt{0.4069} & 0.9988 & \ndssmt{0.3983} & 0.9988 & \ndssmt{2.3680$^*$} & \ndssmt{0.4234$^*$} & \ndssmt{0.9875$^*$} \\
    \midrule
    \multirow{2}[2]{*}{SynthID} & 0.1   & \ndssmt{0.1499} & 0.1064 & \ndssmt{0.1849} & 0.1824 & 0.5112$^\dagger$ & \ndssmt{0.3089$^\dagger$} & 0.6801$^\dagger$ \\
          & 0.5   & \ndssmt{0.3739} & 0.9686 & \ndssmt{0.3842} & 0.9970 & \ndssmt{2.4235$^*$} & \ndssmt{0.4200$^*$} & \ndssmt{0.9933$^*$} \\
    \midrule
    \multirow{2}[2]{*}{Unigram} & 0.1   & \ndssmt{0.0243} & 0.0562 & \ndssmt{0.0529} & 0.1564 & 0.5320$^\dagger$ & \ndssmt{0.1775$^\dagger$} & 0.8625$^\dagger$ \\
          & 0.5   & \ndssmt{0.1233} & 0.5366 & \ndssmt{0.1855} & 0.8290 & \ndssmt{2.3342$^*$} & \ndssmt{0.2328$^*$} & \ndssmt{0.8812$^*$} \\
    \midrule
    \multirow{2}[2]{*}{Unbias} & 0.1   & \ndssmt{0.1812} & 0.7463 & \ndssmt{0.1983} & 0.7750 & 0.4804$^\dagger$ & \ndssmt{0.3264$^\dagger$} & 0.9625$^\dagger$ \\
          & 0.5   & \ndssmt{0.4105} & 0.9938 & \ndssmt{0.3963} & 0.9963 & \ndssmt{2.3913$^*$} & \ndssmt{0.4217$^*$} & \ndssmt{0.9958$^*$} \\
    \bottomrule
    \end{tabular}%
    }}
  \label{tab:compare_token_sentence}%
\end{table}%

\begin{table}[htbp]
  \centering
  \caption{
  \ndssmt{Comparison of AUC scores before (BA) and after applying watermark removal attacks. }
  }
  \resizebox{0.9\linewidth}{!}{
  {\ndsstt
    \begin{tabular}{c|c|ccc|ccc}
    \toprule
          &       & \multicolumn{3}{c|}{Token} & \multicolumn{3}{c}{Char} \\
          & BA & $\mathrm{ER}$=0.1 & $\mathrm{ER}$=0.3 & $\mathrm{ER}$=0.5 & $\mathrm{ER}$=0.1 & $\mathrm{ER}$=0.3 & $\mathrm{ER}$=0.5 \\
    \midrule
    \rowcolor[rgb]{ .906,  .902,  .902} \multicolumn{8}{c}{\textbf{OPT}} \\
    \midrule
    KGW   & 1.0000 & 0.9997 & 0.9838 & 0.8898 & 0.9990 & 0.9305 & 0.7043 \\
    DIP   & 1.0000 & 0.9786 & 0.7453 & 0.5700 & 0.9714 & 0.7091 & 0.5670 \\
    SynthID & 0.9999 & 0.9947 & 0.8544 & 0.6337 & 0.9904 & 0.7639 & 0.5528 \\
    Unigram & 1.0000 & 1.0000 & 0.9975 & 0.9625 & 0.9999 & 0.9945 & 0.9608 \\
    Unbias & 1.0000 & 0.9795 & 0.7393 & 0.5896 & 0.9721 & 0.7142 & 0.5506 \\
    \midrule
    \rowcolor[rgb]{ .906,  .902,  .902} \multicolumn{8}{c}{\textbf{LLaMA}} \\
    \midrule
    KGW   & 1.0000 & 0.9997 & 0.9836 & 0.8758 & 0.9996 & 0.9710 & 0.8447 \\
    DIP   & 0.9998 & 0.9722 & 0.7061 & 0.5783 & 0.9561 & 0.6854 & 0.5702 \\
    SynthID & 1.0000 & 0.9933 & 0.8358 & 0.6342 & 0.9869 & 0.7874 & 0.6077 \\
    Unigram & 1.0000 & 1.0000 & 0.9982 & 0.9859 & 0.9993 & 0.9879 & 0.9374 \\
    Unbias & 0.9999 & 0.9692 & 0.7124 & 0.5748 & 0.9584 & 0.7005 & 0.5651 \\
    \bottomrule
    \end{tabular}%
    }}
  \label{tab:auc}%
\end{table}%

\subsection{Experimental Setup}

\subsubsection{Watermarked Data}
We follow the experimental setup used in prior work~\cite{zhao2024provable, kirchenbauer2023watermark, liang2024waterpark, kirchenbauer2024reliability}. Specifically, the prompts used to generate watermarked text are sampled from the RealNewsLike subset of the C4 dataset~\cite{2019t5}. Watermarked text is generated by two victim LLMs: LLaMA-3-8B~\cite{touvron2023llama} and OPT-1.3B~\cite{zhang2022opt}. All experiments are conducted on NVIDIA A100 GPUs. 
We evaluate five representative watermark methods: KGW~\cite{kirchenbauer2023watermark}, DIP~\cite{wu2024resilient}, SynthID~\cite{dathathri2024scalable}, Unigram~\cite{zhao2024provable}, and Unbias~\cite{hu2024unbiased}, using the official implementations provided by MarkLLM~\cite{pan2024markllm}. 
For KGW, Unigram, DIP, and Unbias, we set $\gamma = 0.5$. In KGW and Unigram, $\gamma$ denotes the proportion of the green list in the vocabulary. In DIP and Unbias, $\gamma$ represents the cumulative probability of the selected tokens. 
For KGW and Unigram, we set $\delta = 2$, which indicates the logit bias added to green list tokens. 
For KGW, we set the context length $h = 1$. Other hyperparameters follow the default configurations in MarkLLM.

\subsubsection{Implementation Details} 
We compare our baseline character-level attack with token-level and sentence-level approaches for watermark removal. 
In the token-level attack, we randomly select tokens and replace them with synonyms generated using Gensim~\cite{gensim}. 
For sentence-level attacks, we use DIPPER and \ndssm{AuthorMist} to rewrite watermarked text. DIPPER is a paraphrasing model specifically designed for watermark removal \cite{krishna2024paraphrasing}, \ndssm{while AuthorMist \cite{authormist2025} is trained via reinforcement learning to paraphrase AI-generated text into a more human-like style, aiming to evade detection while preserving semantic. As an aggressive paraphraser, AuthorMist extensively alters the writing style and vocabulary of the input, resulting in significantly longer outputs with most words modified, and consequently a very high $\mathrm{ER}$.}
For the character-level attack, we consider five types of visually imperceptible perturbations:
(1) \textbf{Typo}: replace a character with a nearby keyboard key (e.g., ``t''~$\rightarrow$~``r'').
(2) \textbf{Deletion}: remove the selected character from the token.
(3) \textbf{Swap}: swap a character with its adjacent one (e.g., ``their''~$\rightarrow$~``thier'').
(4) \textbf{Insertion}: insert a zero-width character (e.g., U+200B) or a whitespace at a chosen position.
(5) \textbf{Homoglyph substitution}: replace a character with a visually similar Unicode homoglyph (e.g., ``g''~$\rightarrow$~``\v{g}'', U+011F). 
\ndssm{
Typo, Deletion, and Insertion usually create at least $2$ sub-tokens. Homoglyph substitution changes often lead to $3$ or more sub-tokens. Swap typically produces at least $2\sim4$ sub-tokens.
}

\begin{table}[htbp]
  \centering
  \caption{
Comparison of text quality after watermark removal using three types of perturbations across five watermarking schemes. 
Evaluation is based on BLEU$(\uparrow)$, ROUGE-F1$(\text{RF}, \uparrow)$, and PPL rate$(\text{PR}, \downarrow)$. 
Sentence-level methods include DIPPER ($\dagger$), with editing rates roughly aligned to $\mathrm{ER} \approx 0.5$, \ndssmt{and AuthorMist ($*$), which yields high and less controllable editing rates. 
}}

  \resizebox{\linewidth}{!}{
  {
    \begin{tabular}{c||c|ccc|ccc||cccc}
    \toprule
          &       & \multicolumn{3}{c|}{Token} & \multicolumn{3}{c||}{Char} & \multicolumn{4}{c}{Sentence} \\
          & $\mathrm{ER}$ & BLEU  & RF & PR & BLEU  & RF & PR & $\mathrm{ER}$ & BLEU  & RF & PR\\
    \midrule
    \rowcolor[rgb]{ .906,  .902,  .902} \multicolumn{12}{c}{\textbf{OPT}} \\
    \midrule
    \multirow{2}[2]{*}{KGW} & 0.1   & 0.7669 & 0.9078 & 1.5886 & 0.7714 & 0.8678 & 1.1198 & 0.6130$^\dagger$ & 0.3830$^\dagger$ & 0.7685$^\dagger$ & -0.0425$^\dagger$ \\
          & 0.5   & 0.1664 & 0.5586 & 15.9911 & 0.1739 & 0.4402 & 1.7028 & \ndssmt{2.5579$^*$} & \ndssmt{0.0875$^*$} & \ndssmt{0.4544$^*$} &\ndssmt{ -0.1708$^*$} \\
    \midrule
    \multirow{2}[2]{*}{DIP} & 0.1   & 0.7683 & 0.9079 & 1.6342 & 0.7713 & 0.8675 & 1.1080 & 0.5832$^\dagger$ & 0.3944$^\dagger$ & 0.7756$^\dagger$ & 0.0391$^\dagger$ \\
          & 0.5   & 0.1615 & 0.5584 & 16.7235 & 0.1737 & 0.4395 & 1.8324 & \ndssmt{2.4413$^*$} & \ndssmt{0.0890$^*$} & \ndssmt{0.4596$^*$} &\ndssmt{ -0.0947$^*$ }\\
    \midrule
    \multirow{2}[2]{*}{SynthID} & 0.1   & 0.7680 & 0.9085 & 1.8014 & 0.7732 & 0.8688 & 1.3514 & 0.5068$^\dagger$ & 0.4551$^\dagger$ & 0.8072$^\dagger$ & 0.1647$^\dagger$ \\
          & 0.5   & 0.1621 & 0.5572 & 20.8919 & 0.1751 & 0.4425 & 2.7043 & \ndssmt{2.4469$^*$} &\ndssmt{ 0.1079$^*$} & \ndssmt{0.4778$^*$} & \ndssmt{0.2673$^*$} \\
    \midrule
    \multirow{2}[2]{*}{Unigram} & 0.1   & 0.7658 & 0.9075 & 1.7040 & 0.7711 & 0.8669 & 1.0857 & 0.5911$^\dagger$ & 0.3904$^\dagger$ & 0.7676$^\dagger$ & -0.0263$^\dagger$ \\
          & 0.5   & 0.1643 & 0.5517 & 16.9014 & 0.1737 & 0.4351 & 1.4794 & \ndssmt{2.3805$^*$} & \ndssmt{0.0966$^*$} & \ndssmt{0.4596$^*$} &\ndssmt{ -0.1705$^*$} \\
    \midrule
    \multirow{2}[2]{*}{Unbias} & 0.1   & 0.7683 & 0.9080 & 1.5725 & 0.7707 & 0.8676 & 1.1270 & 0.5797$^\dagger$ & 0.3801$^\dagger$ & 0.7733$^\dagger$ & 0.0292$^\dagger$ \\
          & 0.5   & 0.1626 & 0.5580 & 16.5869 & 0.1740 & 0.4391 & 1.8644 & \ndssmt{2.4596$^*$} & \ndssmt{0.0871$^*$} & \ndssmt{0.4578$^*$} & \ndssmt{-0.1012$^*$} \\
    \midrule
    \rowcolor[rgb]{ .906,  .902,  .902} \multicolumn{12}{c}{\textbf{LLaMA}} \\
    \midrule
    \multirow{2}[2]{*}{KGW} & 0.1   & 0.7688 & 0.9088 & 1.7966 & 0.7718 & 0.8685 & 0.5560 & 0.5240$^\dagger$ & 0.4574$^\dagger$ & 0.8013$^\dagger$ & 0.3457$^\dagger$ \\
          & 0.5   & 0.1616 & 0.5570 & 23.0072 & 0.1763 & 0.4412 & 0.4952 & \ndssmt{2.3925$^*$} & \ndssmt{0.1030$^*$} & \ndssmt{0.4767$^*$} & \ndssmt{0.1460$^*$} \\
    \midrule
    \multirow{2}[2]{*}{DIP} & 0.1   & 0.7680 & 0.9088 & 1.8353 & 0.7711 & 0.8676 & 0.5454 & 0.5033$^\dagger$ & 0.4800$^\dagger$ & 0.8111$^\dagger$ & 0.3818$^\dagger$ \\
          & 0.5   & 0.1619 & 0.5581 & 24.4797 & 0.1729 & 0.4399 & 0.4720 & \ndssmt{2.3680$^*$} & \ndssmt{0.1075$^*$} & \ndssmt{0.4834$^*$} & \ndssmt{0.1549$^*$} \\
    \midrule
    \multirow{2}[2]{*}{SynthID} & 0.1   & 0.7683 & 0.9086 & 1.8186 & 0.7713 & 0.8678 & 0.5212 & 0.5112$^\dagger$ & 0.4643$^\dagger$ & 0.8074$^\dagger$ & 0.3171$^\dagger$ \\
          & 0.5   & 0.1608 & 0.5570 & 23.6556 & 0.1759 & 0.4413 & 0.4372 & \ndssmt{2.4235$^*$} & \ndssmt{0.1060$^*$} & \ndssmt{0.4737$^*$} & \ndssmt{0.0791$^*$} \\
    \midrule
    \multirow{2}[2]{*}{Unigram} & 0.1   & 0.7691 & 0.9098 & 1.7713 & 0.7731 & 0.8699 & 0.5425 & 0.5320$^\dagger$ & 0.4385$^\dagger$ & 0.7908$^\dagger$ & 0.2600$^\dagger$ \\
          & 0.5   & 0.1602 & 0.5571 & 20.2958 & 0.1786 & 0.4473 & 0.4701 & \ndssmt{2.3342$^*$} & \ndssmt{0.1006$^*$} & \ndssmt{0.4790$^*$} & \ndssmt{0.1947$^*$} \\
    \midrule
    \multirow{2}[2]{*}{Unbias} & 0.1   & 0.7683 & 0.9084 & 1.8765 & 0.7724 & 0.8678 & 0.5438 & 0.4804$^\dagger$ & 0.4799$^\dagger$ & 0.8112$^\dagger$ & 0.3939$^\dagger$ \\
          & 0.5   & 0.1619 & 0.5585 & 24.3046 & 0.1755 & 0.4407 & 0.4522 & \ndssmt{2.3913$^*$} & \ndssmt{0.1024$^*$} & \ndssmt{0.4789$^*$} & \ndssmt{0.1555$^*$} \\
    \bottomrule
    \end{tabular}%
    }}
  \label{tab:text_quality_baseline}%
\end{table}%

\subsection{Evaluation}
\label{sec:bad_char_exp} 

In this section, we first compare the effectiveness of character-, token-, and sentence-level perturbations for watermark removal, showing the superiority of character-level attacks (Section~\ref{sec:compare_token_sentence}). We then evaluate the impact of each perturbation type on text quality (Section~\ref{sec:text_quality_baseline}) and analyze how editing rate affects removal performance (Section~\ref{sec:editing_dist_asr}). We further compare different character-level perturbation operations and find homoglyph insertion to be the most effective (Section~\ref{sec:char-perturb-comparison}). In addition, we discuss the frequency-based zero-feedback attack (Section~\ref{sec:freq-zero-feedback-ac1}). Appendix~\ref{sec:text_len_asr} and \ref{sec:random-char-other-language} present additional results on text length and cross-lingual generalization, respectively.

\subsubsection{Comparison with Token-Level and Sentence-Level Removal Attacks}
\label{sec:compare_token_sentence}
Table~\ref{tab:compare_token_sentence} compares the effectiveness of watermark removal using token-level, character-level, and sentence-level perturbations across five watermark schemes. 
Each method is evaluated on texts generated by two LLMs: OPT and LLaMA. For sentence-level attacks, we use the DIPPER and \ndssm{AuthorMist} as paraphrasers. DIPPER is configured with a lexical diversity of $20$ and an order diversity of $0$. These settings yield an $\mathrm{ER}$ of approximately $0.5$, ensuring a fair comparison with other methods. For token-level and character-level attacks, we report results under both low ($\mathrm{ER}=0.1$) and high ($\mathrm{ER}=0.5$) perturbation budgets.

Our results show that character-level attacks consistently outperform both token-level and sentence-level methods across all settings. 
The advantage of character-level attacks is especially notable under low editing budgets. For instance, on KGW, SynthID, and Unigram watermarks with $\mathrm{ER} = 0.1$, character-level attacks achieve an average $\mathrm{ASR}$ of $0.1672$, nearly double that of token-level ($0.0740$), aligning with our analysis in Section~\ref{sec:bad_char_analysis} on the benefits of broader attack ranges. 
For weaker watermark schemes like DIP and Unbias, token- and character-level attacks perform similarly. Both achieve high $\mathrm{ASR}$ and $\mathrm{WDR}$ across $\mathrm{ER}$ settings, indicating token-level perturbations are already effective, with limited gains from character-level attacks.

\ndssm{
Following prior work \cite{liang2024waterpark, piet2023mark,kirchenbauer2024reliability,kirchenbauer2023watermark, krishna2024paraphrasing,zhang2024sand}, our previous experiments primarily used the optimal detection thresholds provided by watermark designers to ensure the best detection performance. 
Further, to evaluate whether our attacks remain effective under varying detection thresholds $\tau_d$, we compare the Area Under the ROC Curve (AUC) of the original watermark detector before and after the attack. Specifically, we treat original non-watermarked texts as negative samples, and watermarked texts and attacked watermarked texts as positive samples before and after the attack, respectively. By sliding $\tau_d$, we compute the true positive rate and false positive rate at each point and the resulting AUC.
As shown in Table~\ref{tab:auc},  the original detector achieves an average AUC of 1.0 before attack (BA), indicating a strong ability to distinguish watermarked from non-watermarked text. After character-level attacks, the average AUC drops to $0.9833$, $0.8244$, and $0.6861$ at $\mathrm{ER}= 0.1, 0.3, 0.5$, respectively; for token-level attacks, the corresponding AUCs are $0.9887$, $0.8556$, and $0.7294$. 
These results indicate that, at the same $\mathrm{ER}$, character-level attacks consistently achieve lower AUC values, suggesting better removal performance.
}
 
\subsubsection{Impact on Text Quality}
\label{sec:text_quality_baseline}
Table~\ref{tab:text_quality_baseline} evaluates the impact of watermark removal on text quality across three types of perturbations using OPT-generated watermarked text. 
We evaluate text quality using three widely adopted metrics: BLEU \cite{papineni2002bleu}, ROUGE-F1 \cite{lin2004rouge}, and perplexity rate ($\text{PPL rate} = \frac{\text{PPL}(\tilde{X}) - \text{PPL}(X)}{\text{PPL}(X)}$), where PPL is measured by the victim LLM. 
Higher BLEU and ROUGE-F1 indicate better preservation of semantic quality, while lower PPL rate reflects better fluency.
Results are reported at editing rates $\mathrm{ER}=0.1$ and $0.5$, using 100 token inputs. 
Character-level attacks achieve comparable or superior semantic fidelity to token-level attacks, especially under higher perturbation.
For OPT, at $\mathrm{ER} = 0.1$, their average BLEU/ROUGE-F1 scores ($0.7715$/$0.8677$) are close to token-level ($0.7675$/$0.9080$); at $\mathrm{ER} = 0.5$, the scores ($0.1741$/$0.4393$) remain similar to token-level ($0.1634$/$0.5568$), though ROUGE-F1 is slightly lower. 
This is because ROUGE is based on n-gram overlap, and character-level perturbations increase token fragmentation, thereby decreasing the proportion of overlapping tokens. 
Character-level attacks also yield much lower perplexity, indicating better fluency. At $\mathrm{ER}=0.1$, the average PPL rate is $1.1584$ compared to $1.6601$ for token-level. 
The same pattern can also be observed on watermarked texts generated by LLaMA.
Although sentence-level attacks produce more fluent text, they require a very high editing rate to be effective.

\begin{figure}
    \centering
    \includegraphics[width=0.5\textwidth]{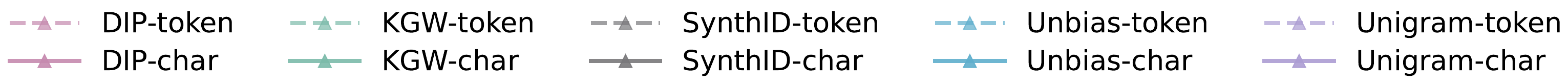} \par
    \centering
    \includegraphics[width=0.2\textwidth]{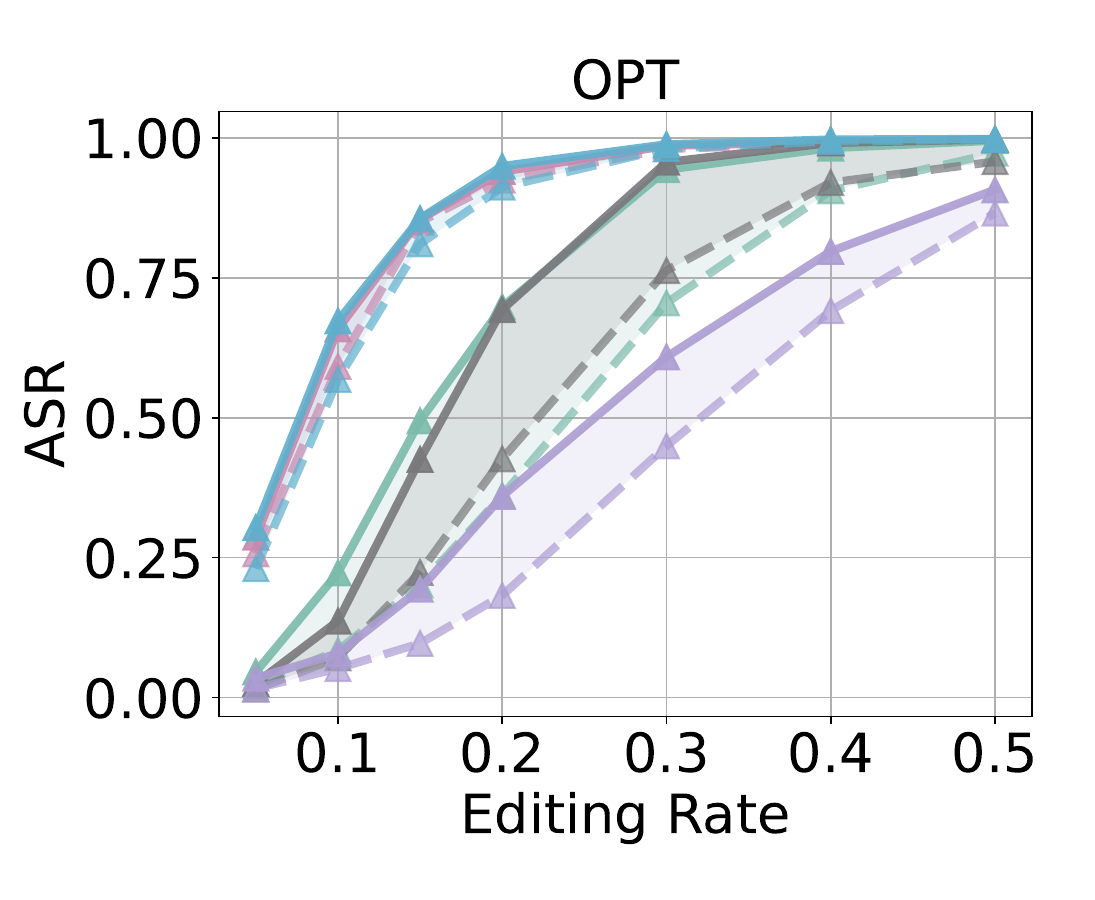}
    \includegraphics[width=0.2\textwidth]{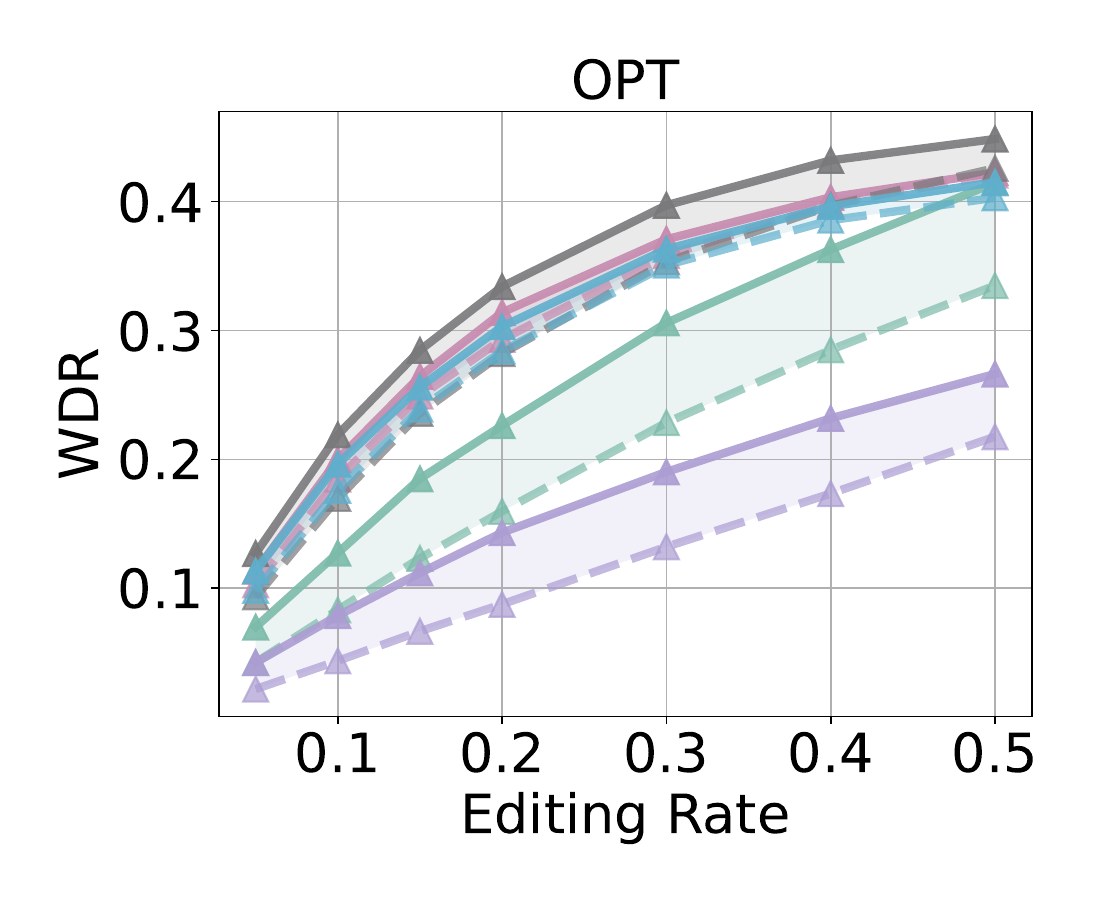}
    \includegraphics[width=0.2\textwidth]{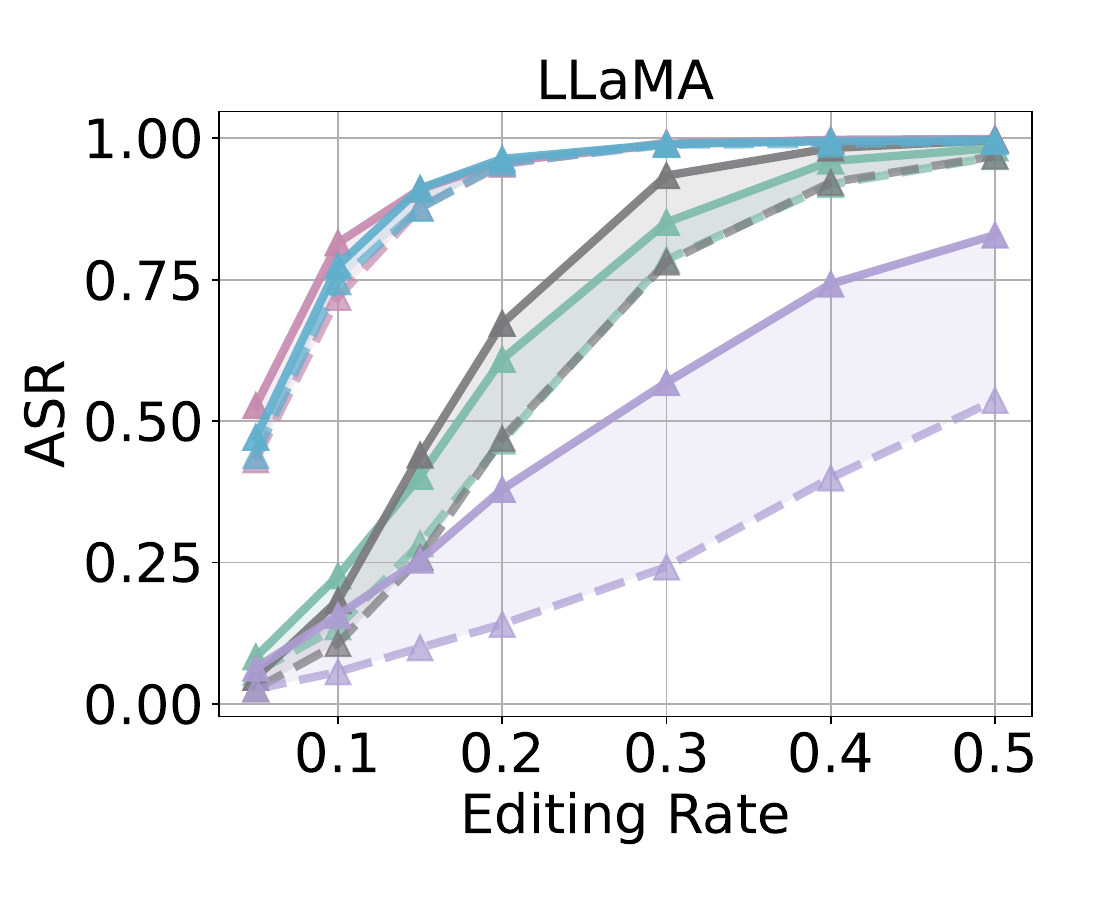}
    \includegraphics[width=0.2\textwidth]{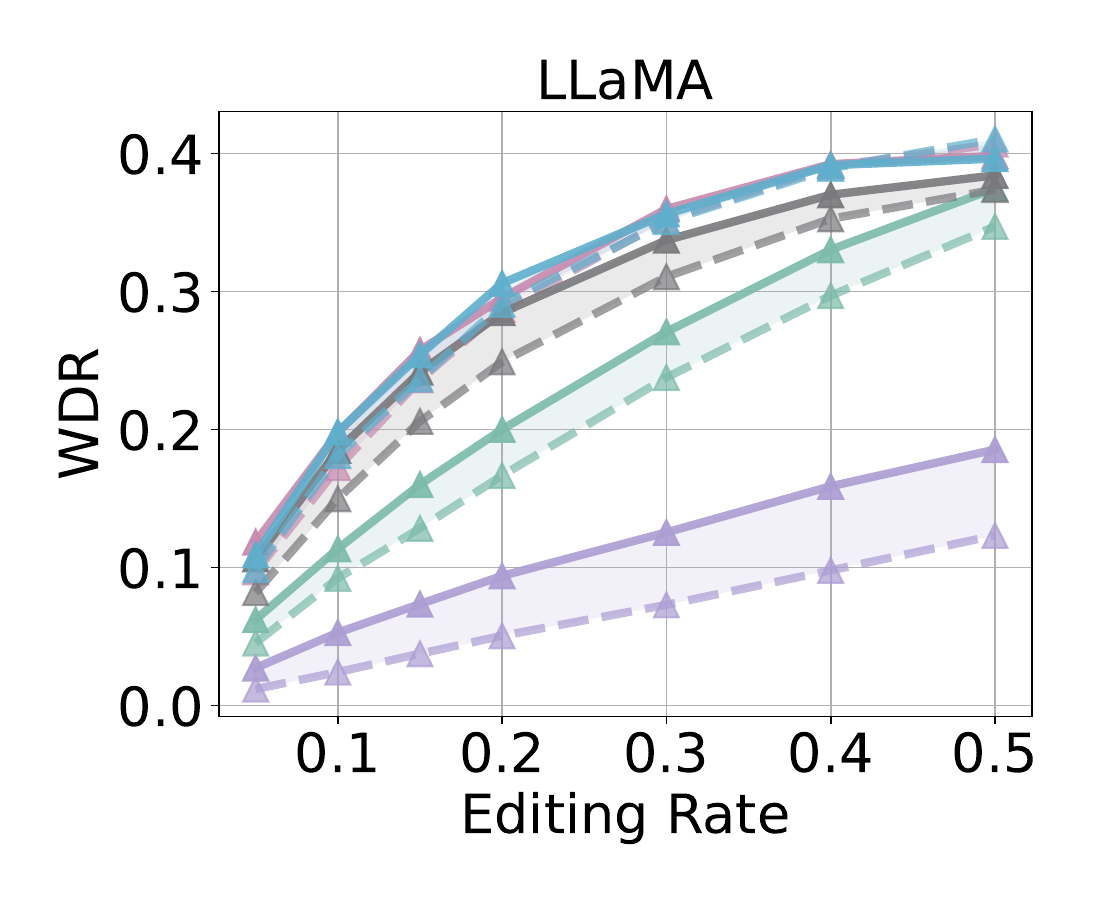}
    \caption{
    $\mathrm{ASR}$ and $\mathrm{WDR}$ of token- and character-level attacks at varying $\mathrm{ER} \in [0.05, 0.5]$, evaluated on 100-token texts.
    }
    \label{fig:editing_dist_asr}
\end{figure}

\begin{table}[t]
  \centering
  \caption{
  \ndssmt{Performance of frequency-based zero-feedback watermark removal attacks in the \textbf{AC1} setting.}
  }
  \resizebox{\linewidth}{!}{
  {\ndsstt
    \begin{tabular}{c|c|cc|cc|cc|cc}
    \toprule
          &       & \multicolumn{2}{c|}{Token (OPT)} & \multicolumn{2}{c|}{Char (OPT)} & \multicolumn{2}{c|}{Token (LLaMA)} & \multicolumn{2}{c}{Char (LLaMA)} \\
          & $\mathrm{ER}$ & $\mathrm{WDR}(\uparrow)$ & $\mathrm{ASR}(\uparrow)$ & $\mathrm{WDR}(\uparrow)$ & $\mathrm{ASR}(\uparrow)$ & $\mathrm{WDR}(\uparrow)$ & $\mathrm{ASR}(\uparrow)$ & $\mathrm{WDR}(\uparrow)$ & \multicolumn{1}{c}{$\mathrm{ASR}(\uparrow)$} \\
    \midrule
    \multirow{2}[2]{*}{KGW} & 0.1   & 0.0880 & 0.0788 & 0.1542 & 0.3322 & 0.1003 & 0.1893 & 0.1340 & 0.2786 \\
          & 0.5   & 0.3457 & 0.9623 & 0.4045 & 0.9932 & 0.3677 & 0.9643 & 0.3995 & 0.9821 \\
    \midrule
    \multirow{2}[2]{*}{DIP} & 0.1   & 0.1794 & 0.5827 & 0.2049 & 0.7050 & 0.1928 & 0.7167 & 0.2085 & 0.7542 \\
          & 0.5   & 0.4239 & 0.9964 & 0.4135 & 0.9964 & 0.4287 & 1.0000 & 0.4195 & 0.9917 \\
    \midrule
    \multirow{2}[2]{*}{SynthID} & 0.1   & 0.1818 & 0.0673 & 0.2367 & 0.1481 & 0.1668 & 0.1077 & 0.2065 & 0.1919 \\
          & 0.5   & 0.4611 & 0.9596 & 0.4824 & 1.0000 & 0.3956 & 0.9562 & 0.4120 & 0.9966 \\
    \midrule
    \multirow{2}[2]{*}{Unigram} & 0.1   & 0.0982 & 0.1336 & 0.1249 & 0.2705 & 0.0934 & 0.3716 & 0.0493 & 0.1609 \\
          & 0.5   & 0.4747 & 0.9966 & 0.4530 & 0.9932 & 0.4548 & 0.9962 & 0.2589 & 0.9693 \\
    \midrule
    \multirow{2}[2]{*}{Unbias} & 0.1   & 0.1820 & 0.5603 & 0.2090 & 0.6773 & 0.1911 & 0.7208 & 0.2019 & 0.7792 \\
          & 0.5   & 0.4267 & 0.9965 & 0.4113 & 0.9965 & 0.4123 & 0.9958 & 0.4117 & 1.0000 \\
    \bottomrule
    \end{tabular}%
    }}
  \label{tab:zero_feed}%
\end{table}%

\subsubsection{Impact of Editing Rate}
\label{sec:editing_dist_asr}
Figure~\ref{fig:editing_dist_asr} compares the effectiveness of token-level and character-level watermark removal attacks under different $\mathrm{ER}$. 
Overall, character-level attacks consistently outperform token-level attacks in both attack success rate ($\mathrm{ASR}$) and watermark score dropping rate ($\mathrm{WDR}$), across all watermark schemes and editing rates. 
This performance gap is especially evident for KGW, Unigram, and SynthID when $\mathrm{ER} \leq 0.2$, where character-level attacks achieve a significantly higher average ASR than token-level ones ($0.2840$ vs. $0.1461$ for OPT; $0.2197$ vs. $0.1325$ for LLaMA). 
These findings align with our earlier analysis (Section~\ref{sec:bad_char_analysis}), which highlights the broader impact range of character-level attacks. 
In summary, character-level attacks are more effective than token-level attacks at removing watermarks, making them a more reliable tool for evaluating watermark robustness.

\begin{figure*}
    \centering
    \includegraphics[width=0.195\linewidth]{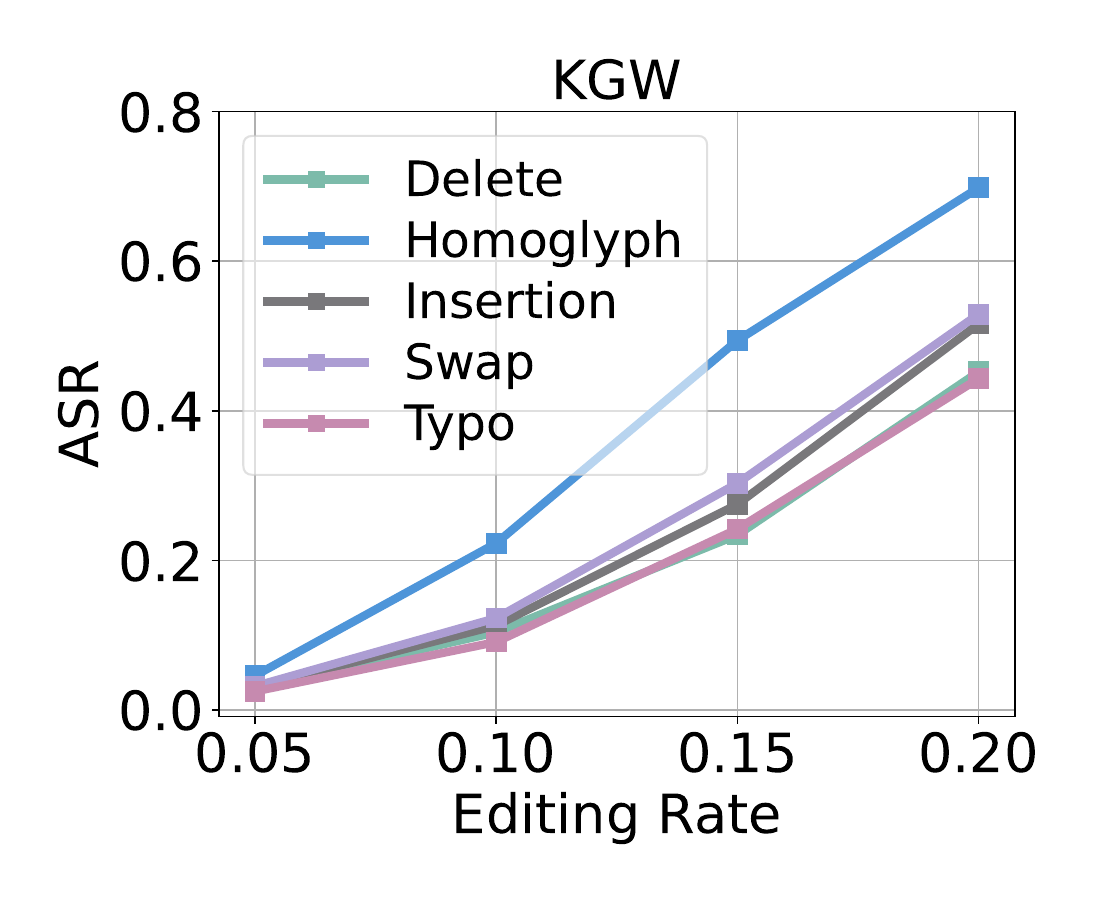}
    \includegraphics[width=0.195\linewidth]{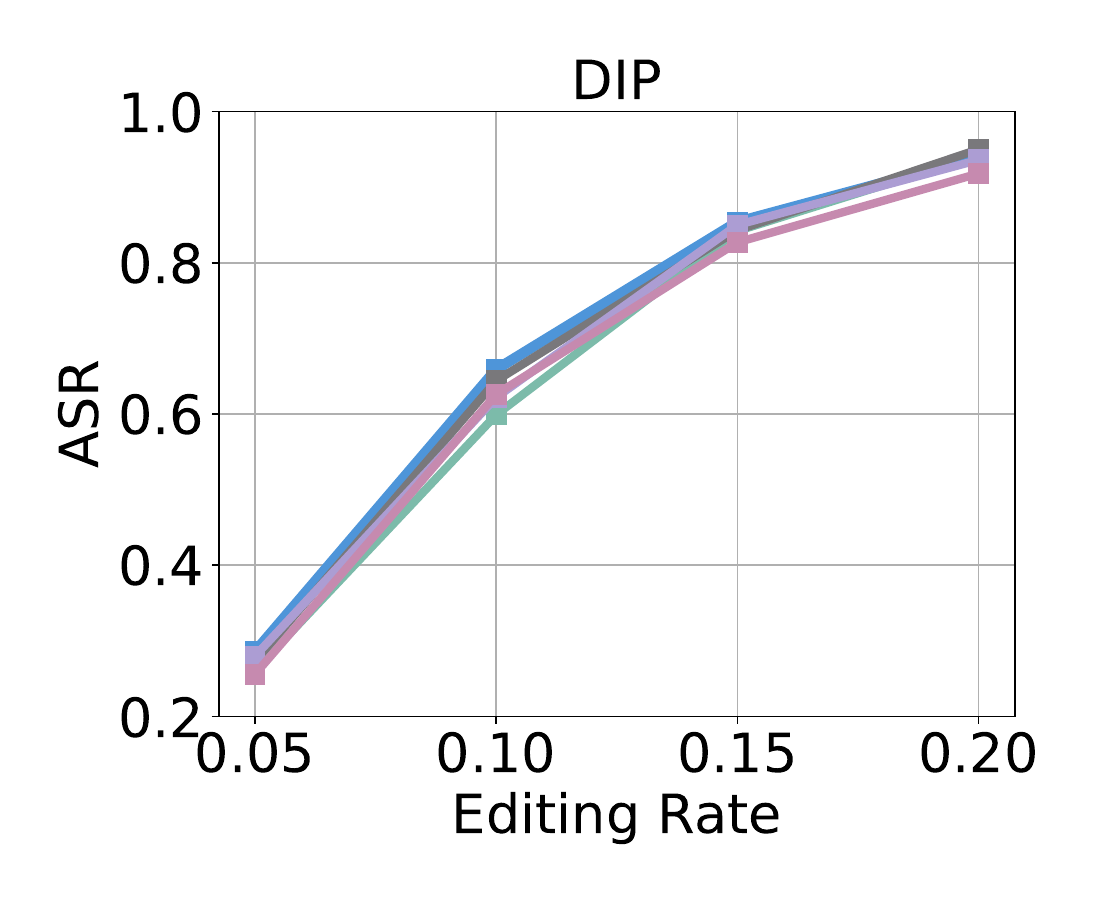}
    \includegraphics[width=0.195\linewidth]{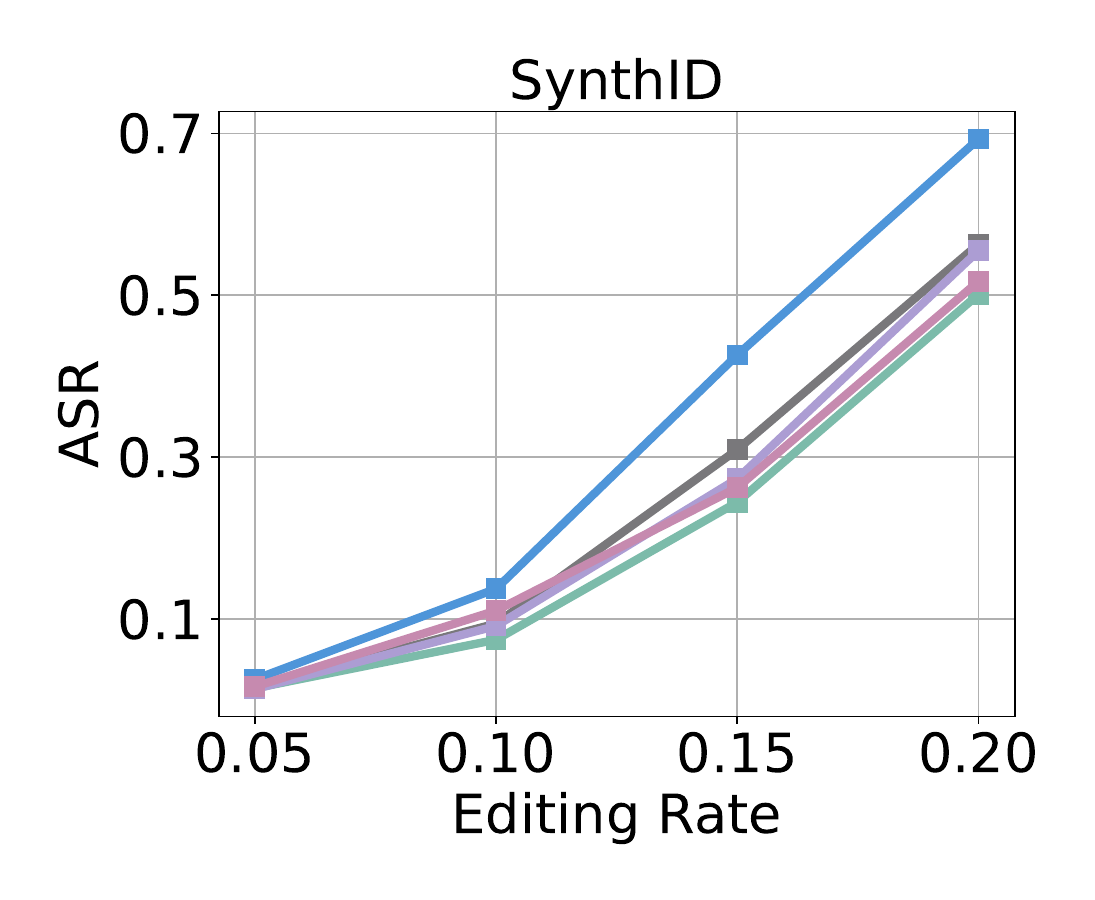}
    \includegraphics[width=0.195\linewidth]{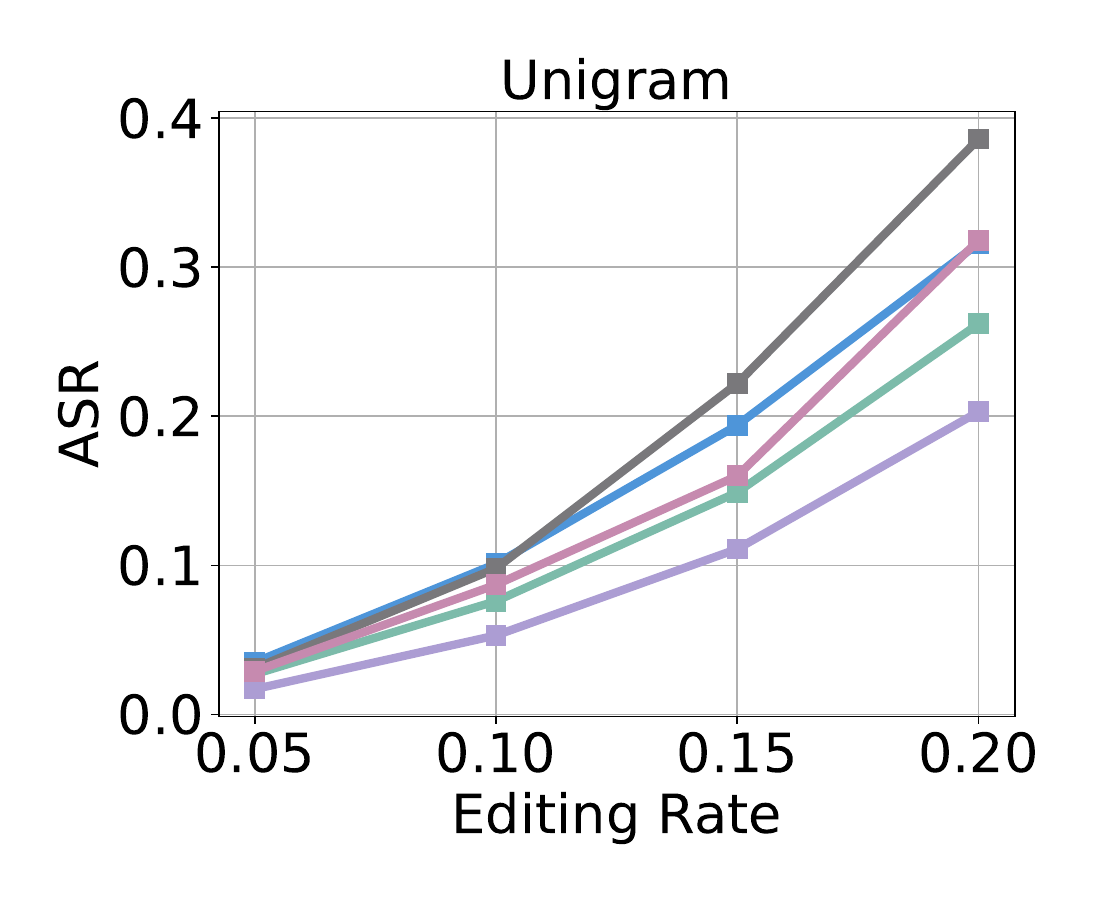}
    \includegraphics[width=0.195\linewidth]{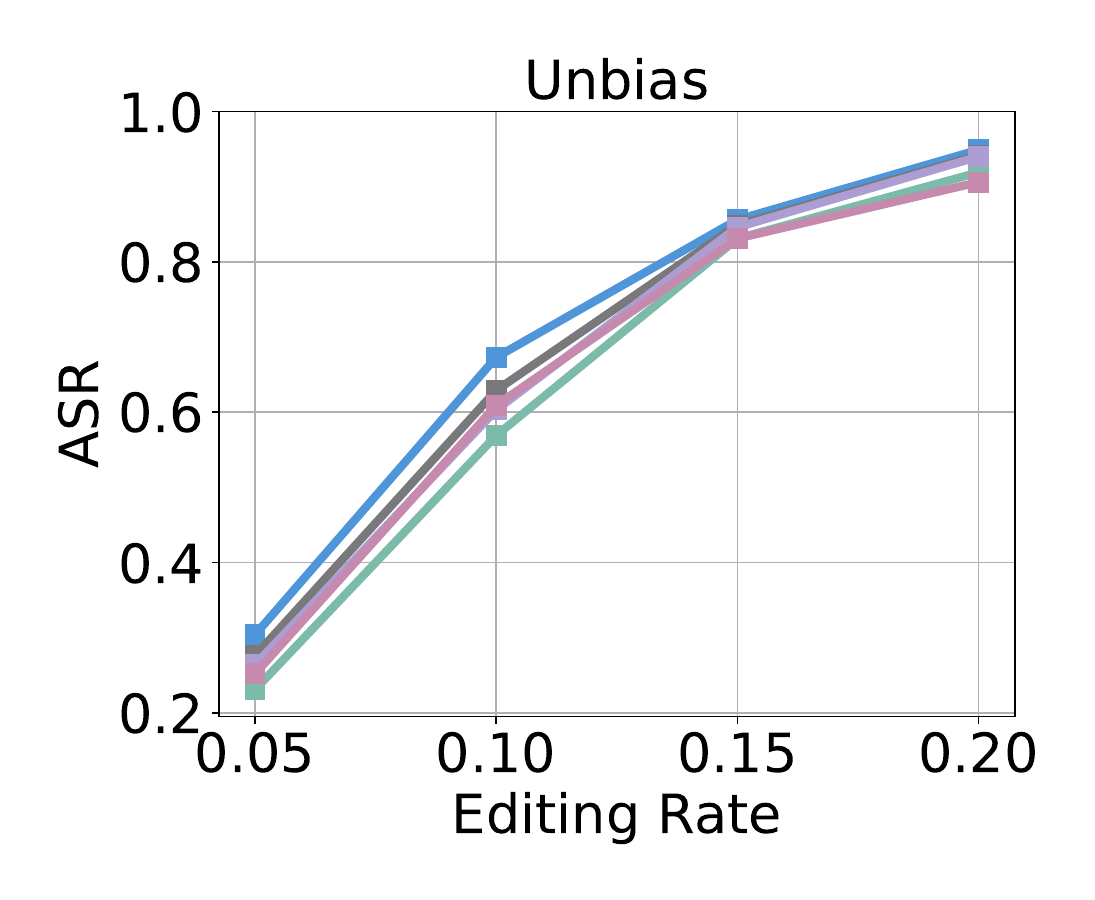}

    \caption{$\mathrm{ASR}$ of five character-level perturbation types, including typos, deletions, swaps, insertions, and homoglyph substitutions, across five watermark schemes. The length of watermarked text is 100 tokens, and they are generated by OPT.}
    \label{fig:char-perturb-comparison}
\end{figure*}

\begin{table*}[htbp]
  \centering
  \caption{\ndssmt{
  Human evaluation results for watermarked text and perturbed text from the three types of attacks. 
  The table reports the average scores across all raters for each dimension, along with the $95\%$ confidence interval of the overall score based on the t-distribution. 
  We use CEFR levels\cite{cefr} to indicate English proficiency (ranging from A1 for beginners to C2 for proficient users), and mark native English speakers with ``$^\#$''.
}}
  \resizebox{\linewidth}{!}{
  {\ndsstt
    \begin{tabular}{clllll|cccc|cccc|cccc|cccc}
    \toprule
    \multirow{2}[4]{*}{\textbf{ID}} & \multicolumn{1}{c}{\multirow{2}[4]{*}{\textbf{Age}}} & \multicolumn{1}{c}{\multirow{2}[4]{*}{\textbf{Gen}}} & \multicolumn{1}{c}{\multirow{2}[4]{*}{\textbf{CEFR}}} & \multicolumn{1}{c}{\multirow{2}[4]{*}{\textbf{Edu}}} & \multicolumn{1}{c|}{\multirow{2}[4]{*}{\textbf{Field}}} & \multicolumn{4}{c|}{\textbf{WM}} & \multicolumn{4}{c|}{\textbf{Sentence}} & \multicolumn{4}{c|}{\textbf{Token}} & \multicolumn{4}{c}{\textbf{Char}} \\
\cmidrule{7-22}          &       &       &       &       &       & \textbf{Gram} & \textbf{Corr} & \textbf{Flue} & \textbf{Overall} & \textbf{Gram} & \textbf{Corr} & \textbf{Flue} & \textbf{Overall} & \textbf{Gram} & \textbf{Corr} & \textbf{Flue} & \textbf{Overall} & \textbf{Gram} & \textbf{Corr} & \textbf{Flue} & \textbf{Overall} \\
    \midrule
    1     & \multicolumn{1}{c}{25} & \multicolumn{1}{c}{F} & \multicolumn{1}{l}{B2} & \multicolumn{1}{c}{PhD} & \multicolumn{1}{c|}{IT} & 2.758  & 2.636  & 2.788  & 2.73±0.16 & 2.322  & 2.610  & 2.593  & 2.51±0.14 & 1.298  & 1.404  & 1.404  & 1.37±0.16 & 2.255  & 2.294  & 2.333  & 2.29±0.19 \\
    2     & \multicolumn{1}{c}{24} & \multicolumn{1}{c}{M} & \multicolumn{1}{l}{C2$^\#$} & \multicolumn{1}{c}{Master} & \multicolumn{1}{c|}{Nurs} & 2.455  & 2.636  & 2.636  & 2.58±0.20 & 2.424  & 2.508  & 2.492  & 2.47±0.14 & 1.298  & 1.263  & 1.175  & 1.25±0.14 & 2.118  & 2.255  & 2.333  & 2.24±0.19 \\
    3     & \multicolumn{1}{c}{35} & \multicolumn{1}{c}{F} & \multicolumn{1}{l}{C2$^\#$} & \multicolumn{1}{c}{PhD} & \multicolumn{1}{c|}{Chem} & 2.576  & 2.727  & 2.788  & 2.70±0.17 & 2.525  & 2.441  & 2.661  & 2.54±0.14 & 1.316  & 1.333  & 1.298  & 1.32±0.16 & 2.118  & 2.216  & 2.490  & 2.27±0.19 \\
    4     & \multicolumn{1}{c}{24} & \multicolumn{1}{c}{M} & \multicolumn{1}{l}{C1} & \multicolumn{1}{c}{PhD} & \multicolumn{1}{c|}{IT} & 2.727  & 2.788  & 2.848  & 2.79±0.15 & 2.542  & 2.508  & 2.678  & 2.58±0.14 & 1.404  & 1.333  & 1.316  & 1.35±0.16 & 2.157  & 2.353  & 2.431  & 2.31±0.19 \\
    5     & \multicolumn{1}{c}{27} & \multicolumn{1}{c}{F} & \multicolumn{1}{l}{C1} & \multicolumn{1}{c}{Master} & \multicolumn{1}{c|}{Chem} & 2.424  & 2.545  & 2.667  & 2.55±0.18 & 2.407  & 2.390  & 2.525  & 2.44±0.15 & 1.158  & 1.298  & 1.439  & 1.30±0.16 & 2.157  & 2.157  & 2.275  & 2.20±0.19 \\
    6     & \multicolumn{1}{c}{27} & \multicolumn{1}{c}{M} & \multicolumn{1}{l}{C2$^\#$} & \multicolumn{1}{c}{PhD} & \multicolumn{1}{c|}{IT} & 2.697  & 2.606  & 2.515  & 2.61±0.18 & 2.254  & 2.627  & 2.593  & 2.49±0.15 & 1.404  & 1.263  & 1.123  & 1.26±0.16 & 2.196  & 2.196  & 2.314  & 2.24±0.18 \\
    7     & \multicolumn{1}{c}{26} & \multicolumn{1}{c}{F} & \multicolumn{1}{l}{C1} & \multicolumn{1}{c}{PhD} & \multicolumn{1}{c|}{IT} & 2.394  & 2.576  & 2.758  & 2.58±0.18 & 2.475  & 2.542  & 2.559  & 2.53±0.14 & 1.368  & 1.175  & 1.456  & 1.33±0.16 & 2.196  & 2.157  & 2.235  & 2.20±0.18 \\
    8     & \multicolumn{1}{c}{33} & \multicolumn{1}{c}{M} & \multicolumn{1}{l}{C1} & \multicolumn{1}{c}{Master} & \multicolumn{1}{c|}{Nurs} & 2.515  & 2.424  & 2.606  & 2.52±0.25 & 2.356  & 2.373  & 2.542  & 2.42±0.15 & 1.351  & 1.211  & 1.281  & 1.28±0.16 & 1.765  & 2.275  & 2.490  & 2.18±0.18 \\
    9     & \multicolumn{1}{c}{27} & \multicolumn{1}{c}{F} & \multicolumn{1}{l}{C2$^\#$} & \multicolumn{1}{c}{PhD} & \multicolumn{1}{c|}{Chem} & 2.182  & 2.818  & 2.818  & 2.61±0.18 & 2.441  & 2.441  & 2.593  & 2.49±0.14 & 1.211  & 1.193  & 1.228  & 1.21±0.15 & 1.980  & 2.196  & 2.412  & 2.20±0.19 \\
    10    & \multicolumn{1}{c}{30} & \multicolumn{1}{c}{M} & \multicolumn{1}{l}{C1} & \multicolumn{1}{c}{PhD} & \multicolumn{1}{c|}{IT} & 2.545  & 2.576  & 2.788  & 2.64±0.17 & 2.610  & 2.492  & 2.576  & 2.56±0.14 & 1.316  & 1.211  & 1.211  & 1.25±0.14 & 2.216  & 2.176  & 2.373  & 2.25±0.19 \\
    11    & \multicolumn{1}{c}{21} & \multicolumn{1}{c}{F} & \multicolumn{1}{l}{C2$^\#$} & \multicolumn{1}{c}{Master} & \multicolumn{1}{c|}{Nurs} & 2.545  & 2.727  & 2.727  & 2.67±0.17 & 2.169  & 2.475  & 2.525  & 2.39±0.15 & 1.070  & 1.351  & 1.158  & 1.19±0.15 & 2.314  & 2.098  & 2.235  & 2.22±0.18 \\
    12    & \multicolumn{1}{c}{28} & \multicolumn{1}{c}{M} & \multicolumn{1}{l}{B2} & \multicolumn{1}{c}{PhD} & \multicolumn{1}{c|}{IT} & 2.788  & 2.697  & 2.697  & 2.73±0.16 & 2.339  & 2.593  & 2.542  & 2.49±0.14 & 1.105  & 1.281  & 1.351  & 1.25±0.16 & 2.039  & 2.314  & 2.529  & 2.29±0.19 \\
    13    & \multicolumn{1}{c}{23} & \multicolumn{1}{c}{F} & \multicolumn{1}{l}{C1} & \multicolumn{1}{c}{Master} & \multicolumn{1}{c|}{IT} & 2.545  & 2.667  & 2.879  & 2.70±0.17 & 2.424  & 2.492  & 2.610  & 2.51±0.14 & 1.386  & 1.158  & 1.035  & 1.19±0.16 & 2.235  & 2.275  & 2.314  & 2.27±0.18 \\
    14    & \multicolumn{1}{c}{30} & \multicolumn{1}{c}{M} & \multicolumn{1}{l}{C1} & \multicolumn{1}{c}{PhD} & \multicolumn{1}{c|}{Chem} & 2.727  & 2.606  & 2.758  & 2.70±0.17 & 2.153  & 2.576  & 2.593  & 2.44±0.14 & 1.316  & 1.439  & 1.088  & 1.28±0.14 & 2.176  & 2.294  & 2.294  & 2.25±0.19 \\
    15    & \multicolumn{1}{c}{34} & \multicolumn{1}{c}{F} & \multicolumn{1}{l}{C1} & \multicolumn{1}{c}{PhD} & \multicolumn{1}{c|}{IT} & 2.212  & 2.333  & 2.455  & 2.33±0.25 & 2.220  & 2.136  & 2.458  & 2.27±0.17 & 1.246  & 1.123  & 1.000  & 1.12±0.13 & 2.275  & 2.039  & 2.039  & 2.12±0.19 \\
        \midrule
    \textbf{Avg} &       & \multicolumn{2}{l}{} &       &       & \textbf{2.54±0.10} & \textbf{2.62±0.07} & \textbf{2.72±0.07} & \textbf{2.63±0.06} & \textbf{2.38±0.08} & \textbf{2.48±0.07} & \textbf{2.57±0.03} & \textbf{2.48±0.04} & \textbf{1.28±0.06} & \textbf{1.27±0.05} & \textbf{1.24±0.08} & \textbf{1.26±0.04} & \textbf{2.15±0.08} & \textbf{2.22±0.05} & \textbf{2.34±0.07} & \textbf{2.24±0.03} \\
    \bottomrule
    \end{tabular}%
    } }
  \label{tab:human}%
\end{table*}%

\subsubsection{Comparison Among Different Character Perturbations}
\label{sec:char-perturb-comparison}
Figure~\ref{fig:char-perturb-comparison} presents the $\mathrm{ASR}$ of five character-level perturbation types: Deletion, Homoglyph substitution, Insertion, Swap, and Typo, across five watermark schemes as the editing rate increases. Overall, all perturbations show improved $\mathrm{ASR}$ with increasing editing rates. Notably, Homoglyph substitution consistently achieves higher $\mathrm{ASR}$ than other methods, especially at lower editing rates (e.g., 0.05 and 0.1). 
This effectiveness can be explained by the attack range analysis in Section~\ref{sec:bad_char_analysis}. Homoglyph substitution tends to disrupt tokenization more severely by splitting a single token into at least three subword tokens. In contrast, other perturbations (e.g., deletion, insertion of whitespace or zero-width characters, swap, and typo) typically cause only a two-subword split. 
Therefore, we adopt homoglyph substitution as the primary character-level type in the following study.

\subsubsection{\ndssm{Frequency-Based Zero-Feedback Attack}}
\label{sec:freq-zero-feedback-ac1}
\ndssm{In this experiment, we aim to evaluate whether general knowledge about watermarks can enhance removal effectiveness under the \textbf{AC1}. As discussed in Section~\ref{sec:add_wm}, watermark schemes affect the frequency of generated tokens, we therefore use token frequency as general knowledge for removal. Specifically, we compute each token's frequency in watermarked (\( \mathrm{fq}_w \)) and non-watermarked (\( \mathrm{fq}_n \)) texts, and define the frequency metric as \( \mathrm{fq}_w / \mathrm{fq}_n \).
Tokens in each text are ranked by this metric in descending order, and perturbations are applied until the target editing rate (ER) is reached. For character-level attacks, perturbations target high-ranking tokens directly. 
For token-level attacks, we substitute high-ranking tokens with their lower-frequency synonyms. 
}

\ndssm{Table~\ref{tab:zero_feed} shows the effectiveness of this strategy. 
The results show that character-level attacks still outperform token-level attacks across all settings. 
Compared to the random strategy in Equation~(\ref{eq:bad_char_base}) and Table~\ref{tab:compare_token_sentence}, this frequency-based approach yields similar performance in terms of ASR and WDR for all watermarks except Unigram. 
When $\mathrm{ER} = 0.1$, the average ASR improvement over the random strategy is only $0.009$ (OPT) and $0.0685$ (LLaMA) for token-level; $0.0429$ (OPT) and $0.0020$ (LLaMA) for character-level attacks. When $\mathrm{ER} = 0.5$, the average improvement turns to $-0.0024$ (OPT) and $0.0895$ (LLaMA) for token-level; $-0.0001$ (OPT) and $0.0272$ (LLaMA) for character-level attacks, suggesting no consistent benefit from frequency.  
An exception is the Unigram watermark. Unlike other watermarking methods that use token-specific green lists, it applies a shared green list to all tokens to enhance robustness \cite{zhao2024provable}. This causes the frequency of green tokens in Unigram watermarked text to be significantly higher than others, making the frequency-based attack more effective.
}

\subsubsection{\ndssm{Human Evaluation of Text Quality under Random Character-Level Watermark Removal}}
\label{sec:text_quality_baseline_he}

\ndssm{We conducted a human evaluation to assess visual imperceptibility. Following the setup in~\cite{dathathri2024scalable, zhao2024provable}, we recruited 15 participants, who were selected based on factors including age (20-35), gender (8 female, 7 male), education (5 Master's, 10 PhD), language (all participants are proficient in English, including 5 native English speakers), and professional background (including IT, nursing, chemistry). Each participant rated 200 anonymized texts—including original watermarked samples and outputs from three watermark removal attacks—based on grammaticality/coherence, correctness, fluency, and overall quality, using a 0–3 scale (with higher scores indicating better quality). Table~\ref{tab:human} reports the average scores and confidence interval from each rater across the four evaluation dimensions. The results show that watermarked texts received the highest average overall score ($2.626$), followed by sentence-level ($2.476$) and character-level attacks ($2.235$), both of which scored significantly higher than token-level attacks ($1.263$). These results indicate that character-level attacks better preserve text quality and provide stronger visual imperceptibility than token-level attacks. }

\section{Guided Character-level Attack for LLM Watermark}
\label{sec:guided_bad_char}
As discussed in Section~\ref{sec:threat_model}, the objective of watermark removal is to reduce the global score $S_w(\tilde{X})$ of a modified text $\tilde{X}$ below the detection threshold. However, a closer look at the watermark injection and detection process in Section~\ref{sec:add_wm} and ~\ref{sec:detect_wm} shows that not all token modifications effectively lower $S_w(\tilde{X})$. 
For example, in KGW~\cite{kirchenbauer2023watermark}, only modifications that convert green tokens into red tokens reduce $S_w(\tilde{X})$; conversely, converting red tokens into green ones strengthens the watermark. 
Therefore, to achieve effective watermark removal with minimal editing rate, it is essential to identify and perturb removal-relevant tokens, i.e.,  whose modification is most likely to decrease the global watermark score. This requires guidance beyond the random strategy. 

In this section, we introduce Genetic Algorithm (GA) to guide watermark removal under the \textbf{AC2} setting, where the adversary has limited black-box access to the original watermark detector. 
Since GA involves iterative evaluation of numerous perturbed candidates, directly querying the original detector throughout the optimization would be impractical under limited query budgets.  
To address this constraint, we train a lightweight reference detector to approximate the original detector’s behavior, allowing the GA to locate removal-relevant tokens even under strict query budgets. 
We also evaluate the GA under an excessive setting in which the original detector is fully accessible, verifying its ability to identify tokens relevant to watermark removal (refer to Appendix~\ref{sec:ga_ori}).

\begin{algorithm}
\small
    \caption{GA-based Removal with Reference Detector}
    \begin{algorithmic}[1]
        \Require Watermarked text $X$, the number of token in text $m$, reference detector $D_{\text{ref}}$, iteration rounds $n$, population size $p$, parent size $p_s$, 
        loss threshold $\delta_l$, score threshold $\tau_l$, weight for editing rate $\lambda$, maximum editing rate $\epsilon$. 
        \State \textbf{// Filter high-gradient tokens from $\{1,\cdots,m\}$}
        \State Initial position set $\mathcal{P}$ 
        \State Initial population $\textbf{P}_1=\{\tilde{\mathcal{P}}^{(q)}\}_{q=1}^{p}$,  $\tilde{\mathcal{P}}^{(q)} \subset \mathcal{P}, \frac{|\tilde{\mathcal{P}}^{(q)}|}{m}\leq \epsilon$
        \For {iteration $j = 1$ to $n$}
            \For {$q = 1$ to $p$}
                \State $\tilde{X}^{(q)} = \mathcal{A}_{\tilde{\mathcal{P}}^{(q)}}^C(X), \tilde{\mathcal{P}}^{(q)} \in \textbf{P}_j$
                \State Compute reference score $w^{(q)} = D_{\text{ref}}(\tilde{X}^{(q)})$
                \State Compute editing rate $e^{(q)}=\mathrm{ER}(\tilde{X}^{(q)}, X)$
                \If{$w^{(q)} > \tau_l$} 
                    \State \textbf{// Stage 1: Minimize reference score only}
                    \State $\mathcal{L}^{(q)} = w^{(q)}$
                \Else 
                    \State \textbf{// Stage 2: Joint optimization}
                    \State $\mathcal{L}^{(q)} = w^{(q)} - \lambda \cdot (\epsilon-e^{(q)})$
                \EndIf
            \EndFor
            \State Select parents $Q_j = \text{top-}{p_s}(\textbf{P}_j)$ in ascending $\mathcal{L}^{(q)}$
            \State Best perturbed text $\tilde{X}=\argmin_{\tilde{X}^{(q)}} \mathcal{L}^{(q)}, q \in [1,p]$
            \State $\mathcal{L}_j=\min_{q \in [1,p]}(\mathcal{L}^{(q)})$
            \If{$|\mathcal{L}_j - \mathcal{L}_{j-1}| < \delta_l$  and $j>1$}
                \State $Q_j = Q_{j-1}$ \textbf{// Keep previous parent}
            \EndIf
            \State Next population: $\textbf{P}_{j+1} \xleftarrow[\text{crossover}]{\text{mutation}} Q_j$
        \EndFor
        \State \Return Final perturbed text $\tilde{X}$
    \end{algorithmic}
    \label{alg:genetic}
\end{algorithm}

\subsection{Genetic Algorithm-Based Attack with Limited Access to the Original Watermark Detector}
\label{sec:ga_ref}

\subsubsection{Reference Detector}
As discussed in Section~\ref{sec:add_wm}, watermarking alters the token selection behavior of LLMs, resulting in measurable statistical deviations between watermarked and non-watermarked outputs. These differences make it feasible to predict the watermark through learned models. 
We follow the setting of previous work \cite{pang2024no}, which assumes that the watermark detectors return a confidence score (e.g., the global watermark score). This setting is aligned with real-world scenarios, where practical AI detection APIs often return soft predictions (e.g., confidence scores or watermark probabilities) rather than just binary labels \cite{openai2023new, grammarly, GPTZero}. Motivated by these, we design the reference detector as a regression model that predicts the global watermark score $S_w(X)$ for an input $X$.

However, reference detectors inherently differ from original detectors.  While the original detector is rule-based and tightly coupled with a specific watermarking scheme (Section~\ref{sec:detect_wm}), the reference detector is data-driven and typically relies on a neural network to approximate its behavior. As a result, they may be overly sensitive to high-gradient tokens, reacting strongly to local changes that have limited effect under the original detector. 
This mismatch poses challenges for removal strategies guided by the reference detector, particularly those relying on gradient-based optimization \cite{li2018textbugger, morris2020textattack, gao2018black} (more analysis about the mismatch refer Appendix~\ref{sec:mismatch}).

\begin{tcolorbox}[colback=gray!10!white, colframe=gray!50!black, boxrule=0.8pt, arc=3pt]
\textbf{Takeaway: } 
Gradient-based optimization guided by the reference detector is unreliable for watermark removal.
\end{tcolorbox}

\subsubsection{Genetic Algorithm}
Since GA is gradient-free, it is less affected by the mismatch between the reference detector and the original detector. 
Under the \textbf{AC2} setting, where access to the original detector is limited, the GA leverages the prediction from the reference detector to identify the removal-relevant positions. Specifically, it seeks a minimal subset of token positions whose perturbation substantially reduces the watermark score predicted by the reference detector $D_{\text{ref}}$. The objective function is defined as follows: 
\begin{IEEEeqnarray}{lll}
\begin{aligned}
    \argmin_{\tilde{P} \subset \{1, \cdots, m\}} & D_{\text{ref}}(\mathcal{A}_{\tilde{P}}^C(X))+\lambda \cdot \frac{|\tilde{P}|}{m},
    \label{eq:ga_object}
\end{aligned}
\end{IEEEeqnarray}
where $\lambda$ is the weight for editing rate,  $\tilde{P}\subset \{1, \cdots, m\}$ is an individual in the GA population that represents a set of token positions of the input text, and $m$ is the number of tokens in $X$. 
This objective is consistent with the adversary's goal in Section~\ref{sec:threat_model} after applying Lagrange multipliers.
In each iteration of GA, it generates perturbed texts, evaluates them with reference detector, and updates the best solution if a significant improvement is observed. The best individuals are selected as parents $Q$, which are then used to generate the next population through crossover and mutation. The process continues until a maximum number of iterations is reached. 
To further improve the stability of the optimization process and mitigate the impact of the mismatch between the reference detector and the original detector, we incorporate three key components into the GA framework: filtering high-gradient tokens, a two-stage optimization objective, and a convergence threshold. 
We summarize the full procedure in Algorithm~\ref{alg:genetic}.

\paragraph{Filtering High-Gradient Tokens}
As discussed above, the reference detector tends to assign disproportionately large gradients to a few influential tokens, which may not align with the original detector. To mitigate this mismatch, we exclude such high-gradient tokens from the initial population, preventing them from misleading the search process. 
Specifically, we first compute the gradient magnitude $\|\frac{\partial D_{\mathrm{ref}}(X)}{\partial x_t}\|_2$ for each token $x_t$ with respect to the output of the reference detector $D_{\mathrm{ref}}$. We then calculate the mean $\mu$ and standard deviation $\sigma$ of all gradient values in the text $X$. Tokens whose gradient magnitudes exceed $\mu + \alpha \cdot \sigma$ are considered high-gradient tokens and are filtered out, where $\alpha$ is a scaling factor controlling the sensitivity of filtering. 

\paragraph{Two-Stage Optimization Objective}
The GA is designed to jointly optimize two objectives: (i) minimizing the predicted watermark score $D_{\text{ref}}(\tilde{X})$ and (ii) minimizing the editing rate. However, these objectives typically converge at different paces. In practice, the editing rate often decreases more quickly, as reducing the number of perturbations is generally easier than effectively suppressing $D_{\text{ref}}(\tilde{X})$.
This early convergence of the editing rate can hinder further reduction in $D_{\text{ref}}(\tilde{X})$, as a smaller editing rate restricts the search space. To address this issue, we adopt a two-stage optimization strategy. In the first stage, the GA focuses solely on minimizing the watermark score until it falls below a predefined threshold $\tau_l$. Once this threshold is reached, the algorithm enters the second stage, where both $D_{\text{ref}}(\tilde{X})$ and editing rate are jointly optimized.  
The overall objective is formally defined as follows:
\begin{equation}
    \mathcal{L}=
    \begin{cases}  
        D_{\text{ref}}(\tilde{X}), & \text{if} \ D_{\text{ref}}(\tilde{X})>\tau_l, \\
        D_{\text{ref}}(\tilde{X})-\lambda \cdot (\epsilon-\mathrm{ER}(X, \tilde{X})), & \text{otherwise}, 
    \end{cases}
    \label{eq:genetic_loss_base}
\end{equation}
where $\lambda$ is the weight for editing rate, $\epsilon$ is the upper bond for editing rate.

\paragraph{Convergence Threshold $\delta_l$}
Due to the mismatch between the decision boundaries of the reference detector $D_{\text{ref}}$ and the original detector $D_{\text{ori}}$, small decreases in $D_{\text{ref}}(\tilde{X})$ may not correspond to meaningful reductions in $D_{\text{ori}}(\tilde{X})$. Relying solely on marginal improvements in $D_{\text{ref}}$ may mislead the optimization process, causing the GA to converge to suboptimal or ineffective perturbed text. To mitigate this issue, we introduce a convergence threshold $\delta_l$ (Line 20 in Algorithm~\ref{alg:genetic}). If the improvement in loss between two consecutive iterations is less than $\delta_l$, the algorithm retains the previous best solution. This prevents the GA from overreacting to insignificant changes in $D_{\text{ref}}(\tilde{X})$, thereby improving the stability of the optimization.

\begin{figure}
    \centering
    \includegraphics[height=0.28\linewidth]{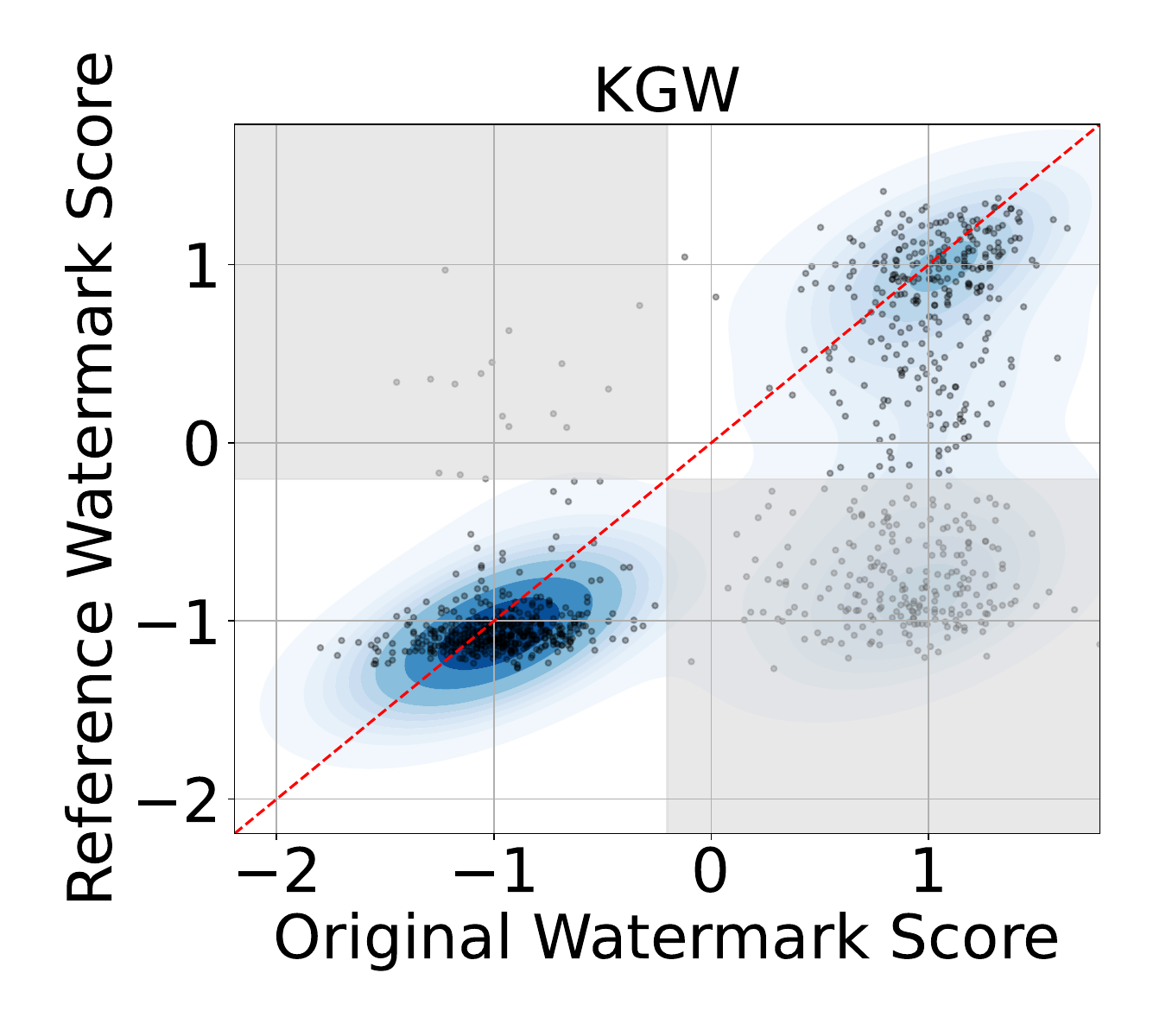}
    \includegraphics[height=0.28\linewidth]{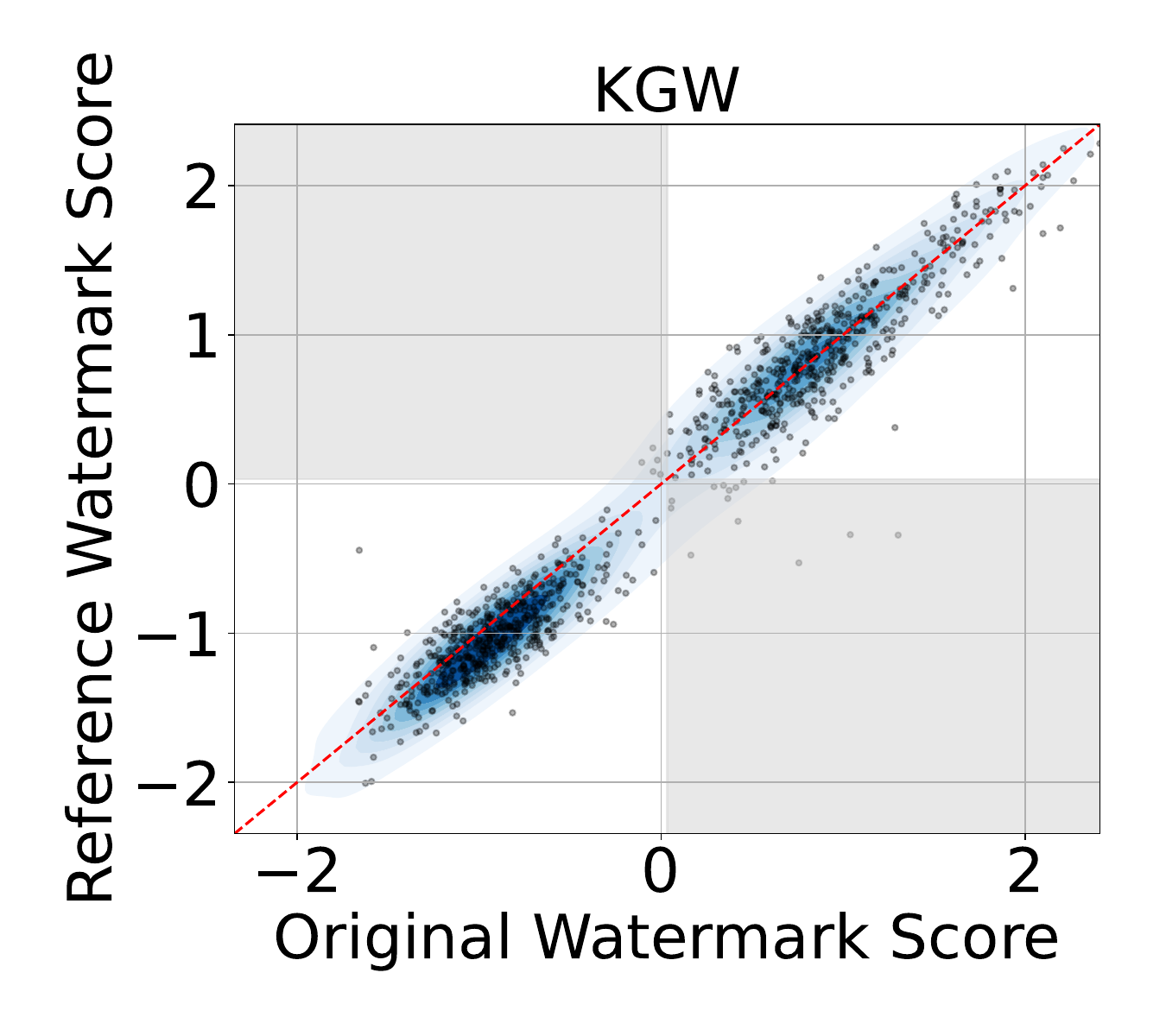}
    \includegraphics[height=0.28\linewidth]{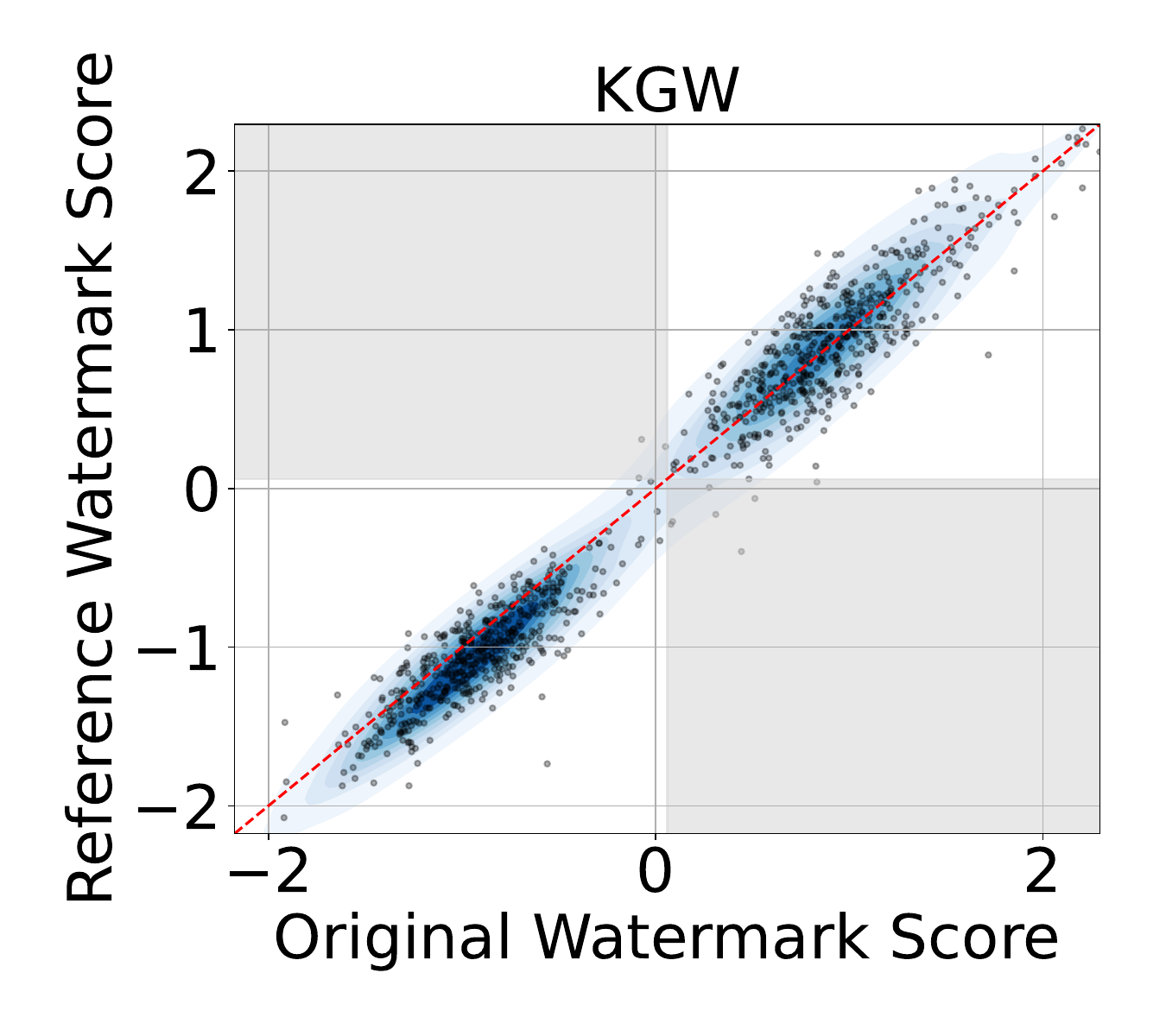}
    \caption{
    Scatter plots comparing reference detector predictions (y-axis) with original detector scores (x-axis) for KGW, with Ref-0, Ref-5, and Ref-9 shown from left to right.
    }
    \label{fig:ref_distri_part}
\end{figure}

\paragraph{Best-of-$N$ with Reference Detector} 
To highlight the advantages of the GA, we introduce a simplified version as a baseline, referred to as the Best-of-$N$ Attack. This method sets iteration rounds in Algorithm~\ref{alg:genetic} as $1$. 
In this case, the adversary generates $N$ perturbed candidates $\{\tilde{X}^{(q)}\}_{q=1}^{N}$, and evaluates them using the reference detector $D_{\mathrm{ref}}$. The candidate with the lowest watermark score $D_{\mathrm{ref}}(\tilde{X}^{(q)})$ is selected and submitted to the original detector as a transfer attack.

\begin{table*}[htbp]
  \centering
  \caption{
  Comparison of watermark removal performance across three guided strategies (Best-of-$N$, Genetic Algorithm, and Sand) using both token-level and character-level perturbations. 
  }
  \resizebox{\linewidth}{!}{
  {
    \begin{tabular}{cl|cc|cc|cc|cc|cc|cc|cc|cc}
    \toprule
          &       & \multicolumn{2}{c|}{Best-of-$N$ (10) Token} & \multicolumn{2}{c|}{Best-of-$N$ (10) Char} & \multicolumn{2}{c|}{\ndssmt{Best-of-$N$ (1500) Token}} & \multicolumn{2}{c|}{\ndssmt{Best-of-$N$ (1500) Char}} & \multicolumn{2}{c|}{GA Token} & \multicolumn{2}{c|}{GA Char} & \multicolumn{2}{c|}{Sand Token} & \multicolumn{2}{c}{Sand Char} \\
          &       & \ndssmt{$\mathrm{WDR}(\uparrow)$} & $\mathrm{ASR}(\uparrow)$ & \ndssmt{$\mathrm{WDR}(\uparrow)$} & $\mathrm{ASR}(\uparrow)$ & \ndssmt{$\mathrm{WDR}(\uparrow)$} & \ndssmt{$\mathrm{ASR}(\uparrow)$} & \ndssmt{$\mathrm{WDR}(\uparrow)$} & \ndssmt{$\mathrm{ASR}(\uparrow)$} & \ndssmt{$\mathrm{WDR}(\uparrow)$} & $\mathrm{ASR}(\uparrow)$ & \ndssmt{$\mathrm{WDR}(\uparrow)$} & $\mathrm{ASR}(\uparrow)$ & \ndssmt{$\mathrm{WDR}(\uparrow)$} & $\mathrm{ASR}(\uparrow)$ & \ndssmt{$\mathrm{WDR}(\uparrow)$} & $\mathrm{ASR}(\uparrow)$ \\
    \midrule
    \multirow{5}[1]{*}{OPT} & KGW   & \ndssmt{0.0985} & 0.0842 & \ndssmt{0.1591} & 0.2626 & \ndssmt{0.0956} & \ndssmt{0.1164} & \ndssmt{0.1490} & \ndssmt{0.3082} & \ndssmt{0.1128} & 0.1233 & \ndssmt{0.2110} & 0.4966 & \ndssmt{0.0305} & 0.0103 & \ndssmt{0.0763} & 0.0685 \\
          & DIP   & \ndssmt{0.1833} & 0.5791 & \ndssmt{0.2033} & 0.6871 & \ndssmt{0.1797} & \ndssmt{0.6151} & \ndssmt{0.2057} & \ndssmt{0.6727} & \ndssmt{0.1864} & 0.6187 & \ndssmt{0.2433} & 0.8453 & \ndssmt{0.0686} & 0.0935 & \ndssmt{0.1624} & 0.5000 \\
          & SynthID & \ndssmt{0.1896} & 0.0606 & \ndssmt{0.2536} & 0.1650 & \ndssmt{0.1886} & \ndssmt{0.0774} & \ndssmt{0.2611} & \ndssmt{0.2121} & \ndssmt{0.2262} & 0.1145 & \ndssmt{0.3093} & 0.4209 & \ndssmt{0.1091} & 0.0168 & \ndssmt{0.2160} & 0.0976 \\
          & Unigram & \ndssmt{0.0639} & 0.0741 & \ndssmt{0.1271} & 0.2907 & \ndssmt{0.0916} & \ndssmt{0.1370} & \ndssmt{0.1508} & \ndssmt{0.4110} & \ndssmt{0.1629} & 0.4829 & \ndssmt{0.1950} & 0.6575 & \ndssmt{0.0382} & 0.0274 & \ndssmt{0.0915} & 0.1164 \\
          & Unbias & \ndssmt{0.1787} & 0.5532 & \ndssmt{0.2067} & 0.6667 & \ndssmt{0.1781} & \ndssmt{0.5496} & \ndssmt{0.2006} & \ndssmt{0.6738} & \ndssmt{0.1864} & 0.6187 & \ndssmt{0.2530} & 0.8369 & \ndssmt{0.0609} & 0.1489 & \ndssmt{0.1717} & 0.5284 \\
    \midrule
    \multirow{5}[1]{*}{LLaMA} & KGW   & \ndssmt{0.0980} & 0.1667 & \ndssmt{0.1216} & 0.2558 & \ndssmt{0.0950} & \ndssmt{0.1750} & \ndssmt{0.1295} & \ndssmt{0.2643} & \ndssmt{0.1176} & 0.2321 & \ndssmt{0.1693} & 0.4214 & \ndssmt{0.0384} & 0.0357 & \ndssmt{0.0849} & 0.1179 \\
          & DIP   & \ndssmt{0.1827} & 0.6833 & \ndssmt{0.2056} & 0.8175 & \ndssmt{0.1869} & \ndssmt{0.7042} & \ndssmt{0.2116} & \ndssmt{0.8000} & \ndssmt{0.1805} & 0.6917 & \ndssmt{0.2580} & 0.9042 & \ndssmt{0.0889} & 0.3458 & \ndssmt{0.1720} & 0.6708 \\
          & SynthID & \ndssmt{0.1604} & 0.1010 & \ndssmt{0.1915} & 0.1996 & \ndssmt{0.1586} & \ndssmt{0.1010} & \ndssmt{0.2054} & \ndssmt{0.2054} & \ndssmt{0.1779} & 0.1448 & \ndssmt{0.2447} & 0.3603 & \ndssmt{0.0369} & 0.0034 & \ndssmt{0.1601} & 0.1010 \\
          & Unigram & \ndssmt{0.0296} & 0.0651 & \ndssmt{0.0596} & 0.1724 & \ndssmt{0.0292} & \ndssmt{0.0575} & \ndssmt{0.0557} & \ndssmt{0.1686} & \ndssmt{0.0694} & 0.1686 & \ndssmt{0.1888} & 0.7356 & \ndssmt{0.0127} & 0.0345 & \ndssmt{0.0148} & 0.0307 \\
          & Unbias & \ndssmt{0.1843} & 0.7167 & \ndssmt{0.2074} & 0.7875 & \ndssmt{0.1804} & \ndssmt{0.6750} & \ndssmt{0.2055} & \ndssmt{0.7958} & \ndssmt{0.1841} & 0.7208 & \ndssmt{0.2379} & 0.8667 & \ndssmt{0.0749} & 0.2833 & \ndssmt{0.1614} & 0.6292 \\
    \bottomrule
    \end{tabular}%
    }}
  \label{tab:guided_results}%
\end{table*}%

\begin{table}[htbp]
\centering
  \centering
  \caption{
  Effect of reference detector quality on watermark removal. We compare $\mathrm{ASR}(\uparrow)$ of GA, Best-of-$N$, and Sand attacks using three $D_{\text{ref}}$: Ref-0, Ref-5, and Ref-9. 
}
  \resizebox{\linewidth}{!}{
  {
    \begin{tabular}{cc|cc|cc|cc|cc|cc}
    \toprule
          &       & \multicolumn{2}{c|}{KGW} & \multicolumn{2}{c|}{DIP} & \multicolumn{2}{c|}{SynthID} & \multicolumn{2}{c|}{Unigram} & \multicolumn{2}{c}{Unbias} \\
          & Ref & Token & Char  & Token & Char  & Token & Char  & Token & Char  & Token & Char \\
    \midrule
    \multirow{3}[2]{*}{GA} & 0 & 0.3112 & 0.3596 & 0.3714 & 0.5504 & 0.0456 & 0.1145 & 0.4169 & 0.6007 & 0.4432 & 0.6418 \\
          & 5 & 0.3270 & 0.3801 & 0.5276 & 0.7842 & 0.0633 & 0.4209 & 0.4268 & 0.6096 & 0.5431 & 0.7730 \\
          & 9 & 0.3744 & 0.4966 & 0.5622 & 0.8453 & 0.0535 & 0.3603 & 0.4343 & 0.6575 & 0.5795 & 0.8369 \\
    \midrule
    \multirow{3}[2]{*}{\makecell{Best-\\of-$N$}} & 0 & 0.0924 & 0.2464 & 0.5931 & 0.6623 & 0.0487 & 0.1379 & 0.0915 & 0.2927 & 0.5622 & 0.6395 \\
          & 5 & 0.1170 & 0.2567 & 0.5714 & 0.6883 & 0.0751 & 0.1866 & 0.0894 & 0.2886 & 0.5451 & 0.6395 \\
          & 9 & 0.0842 & 0.2626 & 0.5791 & 0.6871 & 0.0606 & 0.1650 & 0.0741 & 0.2907 & 0.5922 & 0.7021 \\
    \midrule
    \multirow{3}[2]{*}{Sand} & 0 & 0.0068 & 0.0240 & 0.0540 & 0.0288 & 0.0404 & 0.0000 & 0.0274 & 0.0822 & 0.1277 & 0.0461 \\
          & 5 & 0.0137 & 0.0822 & 0.0683 & 0.4568 & 0.0101 & 0.1178 & 0.0308 & 0.1336 & 0.0745 & 0.4078 \\
          & 9 & 0.0103 & 0.0685 & 0.0935 & 0.5000 & 0.0168 & 0.0976 & 0.0274 & 0.1164 & 0.1489 & 0.5284 \\
    \bottomrule
    \end{tabular}%
    }}
  \label{tab:ref_asr}%

\end{table}%
\begin{table}[t]
  \centering
  \caption{
   $\mathrm{ASR}$ and $\mathrm{WDR}$ of gradient-based transfer attacks using TextBugger and DeepWordBug.
   }

  \resizebox{0.7\linewidth}{!}{
    \begin{tabular}{l|cc|cc}
    \toprule
          & \multicolumn{2}{c|}{Textbugger} & \multicolumn{2}{c}{DeepWordBug} \\
          & \ndssmt{$\mathrm{WDR}(\uparrow)$} & $\mathrm{ASR}(\uparrow)$ & \ndssmt{$\mathrm{WDR}(\uparrow)$} & $\mathrm{ASR}(\uparrow)$ \\
    \midrule
    KGW   & \ndssmt{0.0450} & 0.0095 & \ndssmt{0.0422} & 0.0095 \\
    DIP   & \ndssmt{0.0239} & 0.0288 & \ndssmt{0.0293} & 0.0360 \\
    SynthID & \ndssmt{0.0748} & 0.0067 & \ndssmt{0.0729} & 0.0067 \\
    Unigram & \ndssmt{0.0273} & 0.0189 & \ndssmt{0.0092} & 0.0034 \\
    Unbias & \ndssmt{0.0560} & 0.0490 & \ndssmt{0.0293} & 0.0390 \\
    \bottomrule
    \end{tabular}%
    }
  \label{tab:gradient}%
\end{table}%

\subsection{Experimental Setup}
\label{sec:exp_set_up_ref}

\subsubsection{Reference Detector}
To support guided attacks, we train a separate reference detector for each watermark scheme. For each watermark scheme, we construct a dataset containing $5000$ watermarked and non-watermarked text samples. The reference detector is implemented by fine-tuning a BERT \cite{devlin2019bert} regression model to predict the normalized global watermark score. 
To improve the reliability and generalization of the reference detector, we apply light data augmentation by using token-level and character-level perturbations. Given the limited access under the \textbf{AC2} setting, we restrict the augmentation to a small number of variants per sample. Specifically, we consider three configurations for each watermark scheme: Ref-0 (no augmentation), Ref-5 (five augmented variants per sample), and Ref-9 (nine augmented variants per sample).

\subsubsection{Baseline Method}
We adapt the watermark removal method introduced in \cite{zhang2024sand} as a baseline for our study, referring to it as ``Sand''. In their original design, perturbations are added to the text incrementally over $N$ rounds. At each round, the victim model is used to verify whether the perturbation degrades text quality. If not, the perturbation is accepted. 
In contrast, our work focuses on evaluating the robustness of watermark schemes under the \textbf{AC2} setting, rather than preserving text quality. To accommodate these constraints, we modify their strategy: we still apply perturbations incrementally over $N$ rounds, but at each round, we evaluate whether $D_{\text{ref}}(\tilde{X})$ decreases, if so, the perturbation is accepted.

\subsection{Evaluation}
In this section, we present a comprehensive evaluation of our GA-based watermark removal method guided by a reference detector. We begin by assessing its effectiveness across multiple watermarking schemes (Section~\ref{sec:guided_removal}), followed by an analysis of the reference detector’s quality and its impact on removal performance (Sections~\ref{sec:ref_preformance}, ~\ref{sec:ref_to_removal}, and Appendix~\ref{sec:app_ref}). We also evaluate gradient-based attacks using the reference detector (Section~\ref{sec:gradient_ref}) and discuss the computational complexity and resource overhead of our methods (Section~\ref{sec:complexity}). Additional ablations are provided in Appendix, including the effect of GA iterations (Appendix~\ref{sec:ga_n_asr}), the choice of $N$ in Best-of-$N$ (Appendix~\ref{sec:bon_ref}), and high-gradient token filtering (Appendix~\ref{sec:high_grad}).

\subsubsection{Performance Comparison of Guided Attack}
\label{sec:guided_removal}
Table~\ref{tab:guided_results} presents the effectiveness of the GA-based watermark removal attack, Best-of-$N$ attack and Sand attack \cite{zhang2024sand}. 
For GA, we set the iteration rounds $n = 15$ and population size $p = 100$, resulting in $1500$ queries to the reference detector. 
For the Best-of-$N$ strategy, we consider two settings: (1) $N=10$, and 
\ndssmt{(2) $N=1500$, which uses the same query budget as GA’s query budget. }
All experiments are conducted with text length set to $100$, Ref-9 is chosen as the reference detector. 
The editing rate of all methods is constrained by an upper bound
$\epsilon=0.1$, meaning that each attack can modify at most 10 tokens per text. 
The results show that GA consistently outperforms both Best-of-$N$ ($N=10$) and Sand in terms of $\mathrm{ASR}$ and WDR. For example, under character-level perturbations, GA achieves $\mathrm{ASR}$ scores of $0.6514$ on OPT and $0.6615$ on LLaMA, significantly higher than Best-of-$N$ ($0.4414$ / $0.4466$) and Sand ($0.2622$ / $0.3099$).
\ndssmt{Under the same query budget ($N=1500$), Best-of-$N$ achieves $\mathrm{ASR}$ scores of $0.4555$ (OPT) and $0.4468$ (LLaMA) with character-level perturbations, similar to its performance at $N=10$. This is due to the lack of optimization and potential mismatch with the reference detector, which limits the benefit of additional queries. In contrast, GA still outperforms Best-of-$N$ under the same query budget, suggesting that GA can more effectively utilize the guidance from reference detectors and mitigate mismatch issues between the reference and original detectors.} 
Additionally, Sand performs worse than Best-of-$N$ under both perturbation types. This is likely due to its incremental design: at each step, Sand applies a small perturbation and accepts it only if the reference detector score decreases. However, due to the mismatch between the reference and original detectors, small perturbations often fail to provide reliable feedback. 
Moreover, character-level attacks consistently outperform token-level attacks, reaffirming their superior effectiveness in watermark removal.

\subsubsection{Performance of Reference Detector}
\label{sec:ref_preformance}

Figure~\ref{fig:ref_distri_part} visualizes the predicted watermark scores from the reference detectors versus the original detectors for KGW (refer to Appendix~\ref{sec:ref_preformance_app} for additional results on other watermark schemes). 
The x-axis shows the global watermark scores computed by the original detector, while the y-axis shows the predictions of the reference model. The red dashed line ($y = x$) indicates perfect prediction. Shaded regions correspond to misclassified samples, while white regions in the top-right and bottom-left represent true positives and true negatives, respectively. 
As data augmentation increases, the reference detector's predictions become more aligned with the original scores, and the number of misclassified samples decreases.

\subsubsection{Impact of Reference Detector in Removal Attack}
\label{sec:ref_to_removal}
Table~\Ref{tab:ref_asr} evaluates the impact of reference detector quality, improved through data augmentation, on the effectiveness of watermark removal. We report $\mathrm{ASR}$ results on five watermarking schemes (generated by OPT) using three removal methods: GA, Best-of-$N$, and Sand, each tested with three reference detectors: Ref-0, Ref-5, and Ref-9. 
As data augmentation increases, the quality of the reference detector improves, resulting in more effective watermark removal. 
This trend indicates that enhanced detector quality enables more reliable guidance during removal. 
GA consistently achieves higher $\mathrm{ASR}$ than Best-of-$N$ and Sand under the same reference model. Sand is the most sensitive to detector quality, likely due to its incremental strategy: it adds one perturbation at a time and decides whether to retain it based on small changes in $D_{\text{ref}}(\tilde{X})$. As shown in Figure~\ref{fig:ref_distri_part}, minor variations in $D_{\text{ref}}$ may not accurately reflect the original detector’s behavior, which severely limits Sand’s effectiveness.

\subsubsection{Gradient-Based Adversarial Textual Attack is Not Effective}
\label{sec:gradient_ref}
Table~\ref{tab:gradient} reports the performance of gradient-based adversarial attacks that attempt to transfer from a reference detector to the original detector. Following our analysis in Section~\ref{sec:ga_ref}, the inherent mismatch between the reference and original detectors prevents gradient-based methods from reliably identifying watermark-relevant tokens. 
We evaluate two representative methods: TextBugger \cite{li2018textbugger} and DeepWordBug \cite{gao2018black}, both of which rank tokens by gradient magnitude and iteratively apply hybrid perturbations (token-level and character-level) to reduce the reference detector’s predicted watermark score below a decision threshold using minimal edits. 
As shown in the table, both methods achieve an $\mathrm{ASR} < 0.1$ across all watermark schemes. These results are significantly lower than those of Best-of-$N$ and GA-based strategies, demonstrating that gradient-based transfer attacks are largely ineffective for watermark removal in this setting.

\begin{table}[t]
  \centering
  \caption{
   ASR comparison under different adaptive settings against 4 defenses. 
   ``GA'' denotes the normal GA-based attack in Algorithm~\ref{alg:genetic}.  ``Adaptive GA ($\cdot$)'' represents the adaptive GA-based attacks in Equation~(\ref{eq:genetic_ocr_loss}). Detector type indicates whether the victim watermark detector is the original ($D_{\text{ori}}$) or enhanced with defense modules (such as $F_{\text{SC}}$).
  }
  \resizebox{0.9\linewidth}{!}{

    \begin{tabular}{cl|ccccc}
    \toprule
          & Detector type & KGW   & DIP   & SynthID & Unigram & Unbias \\
    \midrule
    GA    & $D_{\text{ori}}$ & 0.4214 & 0.9042 & 0.3603 & 0.7548 & 0.8667 \\
    \midrule
    \multirow{2}[2]{*}{\makecell{Adaptive\\GA (SC)}} & $D_{\text{ori}}$ & 0.5429 & 0.9125 & 0.3644 & 0.9502 & 0.9125 \\
          & $D_{\text{ori}}\oplus F_{\text{SC}}$ & 0.4250 & 0.8917 & 0.4049 & 0.8697 & 0.8917 \\
    \midrule
    \multirow{2}[2]{*}{\makecell{Adaptive\\GA (OCR)}} & $D_{\text{ori}}$ & 0.5214 & 0.8333 & 0.4122 & 0.5896 & 0.8333 \\
          & $D_{\text{ori}}\oplus F_{\text{OCR}}$ & 0.5036 & 0.9500 & 0.5405 & 0.4776 & 0.9333 \\
    \midrule  
    \multirow{2}[2]{*}{\ndssmt{\makecell{Adaptive\\GA (DE)}}} & \ndssmt{$D_{\text{ori}}$} & \ndssmt{0.4507} & \ndssmt{0.8583} & \ndssmt{0.4595} & \ndssmt{0.9776} & \ndssmt{0.9000} \\
          & \ndssmt{$D_{\text{ori}}\oplus F_{\text{DE}}$} & \ndssmt{0.3028} & \ndssmt{0.9083} & \ndssmt{0.3311} & \ndssmt{0.7612} & \ndssmt{0.8583} \\
    \midrule
    \multirow{2}[2]{*}{\ndssmt{\makecell{Adaptive\\GA (UN)}}} & \ndssmt{$D_{\text{ori}}$} & \ndssmt{0.4718} & \ndssmt{0.9333} & \ndssmt{0.4459} & \ndssmt{0.9776} & \ndssmt{0.8750} \\
          & \ndssmt{$D_{\text{ori}}\oplus F_{\text{UN}}$} & \ndssmt{0.4507} & \ndssmt{0.9167} & \ndssmt{0.4324} & \ndssmt{0.7016} & \ndssmt{0.8667} \\
    \bottomrule
    \end{tabular}%
    }
  \label{tab:adaptive}%
\end{table}%

\subsubsection{\ndssm{Query Cost Overhead}}
\label{sec:complexity}
\ndssm{Our attack is designed for the offline setting, where a reference detector is trained once and reused for multiple attacks. Its training data can be collected incrementally, making the cost flexible and amortizable. To train the reference detector, we collect $5000$ watermarked and $5000$ non-watermarked samples. We consider three detector variants (Ref-0/5/9), which require 1, 6, and 10 queries per watermarked sample, respectively—resulting in a total of $5000$, $30000$, and $50000$ queries. In comparison, the online Best-of-$N$ attack uses N queries per sample; at $N = 10$, attacking $5000$ samples requires $50000$ queries, which is equal to or greater than the number of queries required to train the reference detectors. 
}

\ndssm{In terms of runtime on an NVIDIA A100 GPU, Best-of-$N$ attacks complete within 1s$/$sample. GA converges in $<20$ iterations (each $1\sim3$s), with total time $<30$s$/$sample.}

\section{Discussion}
\label{sec:defense}

In this section, we explore potential defenses against character-level watermark removal attacks. A natural defense strategy is to apply preprocessing techniques to reverse character-level perturbations. We consider four representative defense mechanisms: 
(1) Spell-checking and correction (SC), which automatically detects and corrects spelling errors in the text; 
(2) Optical character recognition (OCR), which renders the text as an image and re-extracts its content; 
\ndssm{(3) Unicode normalization (UN), which converts semantically equivalent characters into a standard form to ensure consistent text representation; 
(4) Deletion (DE) of anomalous characters, which directly removes invalid or suspicious symbols.}

\ndssm{In these adaptive scenarios, where the attacker designs perturbations that proactively account for defense mechanisms, watermark robustness presents an adversarial dilemma. For any fixed defense, there always exists at least one perturbation strategy capable of bypassing it.} To systematically study this problem, we propose an adaptive compound character-level attack based on a two-level optimization framework. The inner optimization selects, for each token position, the perturbation combination ($C^{\ddag}$) that minimizes the edit distance $\mathrm{ED}_C$ after applying the defense function, indicating that it best bypasses the defense. The outer optimization searches over token subsets to find those whose modification most reduces the watermark score $D_{\text{ref}}$. This can be formalized as:
\begin{IEEEeqnarray}{lll}
\begin{aligned}
\argmin_{\tilde{P} \subset \{1, \cdots, m\}}D_{\text{ref}}  \Bigl( 
\argmin_{C^{\ddag}} \mathrm{ED}_C\bigl(\tilde{X}, F_{\text{def}}(\tilde{X})\bigr) 
\Bigr)+\lambda \cdot \frac{|\tilde{P}|}{m}, 
\label{eq:genetic_ocr_loss}
\end{aligned}
\end{IEEEeqnarray}
where $\tilde{X} = \mathcal{A}_{\tilde{P}}^{C^{\ddag}}(X)$, $C^{\ddag}$ denotes compound character-level perturbations, $F_{\text{def}}$ denotes one of OCR, SC, UN, or DE.  
Compound character-level perturbations are defined as the application of one or more character modifications at the same position. Such as:
Swap + homoglyph substitution (e.g., ``compound'' $\rightarrow$ ``compu\v{o}nd''); 
Typo + homoglyph substitution (e.g., ``compound'' $\rightarrow$ ``comp\v{i}und''); 
Zero-width character insertion + homoglyph substitution (e.g., ``compound'' $\rightarrow$ ``compo\{U+200B\}\v{u}nd'').

\ndssm{To the best of our knowledge, existing defenses lack adversarial robustness. However, even when enhanced with robustness mechanisms, they remain vulnerable, because compound character-level perturbations introduce complex distortions that hinder accurate recovery of the original token. The more complex the perturbation, the greater the ambiguity in recovery. Consequently, even if the defense removes suspicious artifacts (e.g., special characters or misspellings), it still cannot reliably recover the original token, thereby disrupting the watermark key and signal.}

\paragraph{Evaluation}
We evaluate the effectiveness of our adaptive GA-based attack under common character-level defenses. Specifically, we consider widely-used tools as modules: LanguageTool \cite{LanguageTool} for SC, Python Tesseract \cite{ocr} for OCR, Python unicodedata \cite{unicodedata} for UN.

\ndssm{Table~\ref{tab:adaptive} compares the ASR of two types of GA-based attacks: normal GA and adaptive GA, across five watermark schemes. Each adaptive GA is evaluated under two detector settings: the original detector $D_{\text{ori}}$ and the defended detector $D_{\text{ori}} \oplus F_{\text{def}}$, where $F_{\text{def}}$ represents specific defense modules (SC, OCR, UN, DE). The maximum editing rate is set to $\epsilon=0.1$. 
For SC, DE, and UN, adaptive GA raises the average ASR from $0.6631$ to $0.7365, 0.7293, 0.7407$, and after applying these defenses, ASR slightly drops to $0.6966, 0.6323, 0.6736$, respectively, but remains close to the normal GA. 
This shows that compound character-level perturbations are effective and resistant to these defenses.}
For OCR, adaptive GA achieves an ASR of $0.6380$, slightly lower than normal GA, but applying OCR increases ASR to $0.6810$, indicating that the perturbations induce OCR recognition errors that disrupt watermark detection.
An exception is the Unigram watermark, where adaptive GA (with OCR) shows a notable ASR drop, as Unigram computes the key independently of context, making it harder to find effective perturbations. 
Overall, these results demonstrate that our adaptive GA approach can effectively enhance watermark removal while exhibiting resilience to potential character-level defenses.

\section{Conclusion}
In this work, we demonstrate that character-level perturbations provide a significantly stronger watermark removal capability than token-level and sentence-level methods, due to their larger attack range. We further show that,  under limited access to the original watermark detector, training a reference detector and optimizing perturbation positions using a Genetic Algorithm can offer effective guidance for watermark removal. This enables adversaries to successfully remove watermarks with a low perturbation budget. To address potential defense mechanisms such as spell checking and OCR, we propose an adaptive strategy by using compound perturbations. Overall, our findings reveal that the vulnerability of current LLM watermarking schemes has been substantially underestimated. 

Our results highlight the urgent need for dedicated defenses against watermark removal attacks. One promising research direction is to enhance robustness against such attacks by improving tokenization strategies and watermark schemes. 
Moreover, while this work primarily focuses on watermark removal, we also see research opportunities in watermark spoofing. With improved guidance techniques, adversaries may exploit character-level perturbations to spoof watermark and falsely attribute texts to LLMs. 

\section*{Acknowledgments}
The first two authors are funded by the China Scholarship Council (CSC) from the Ministry of Education, China. 
Leo’s work is partially supported by the Australian Research Council (ARC) under grant numbers LP240100315 and DP250102634.

\bibliographystyle{ieeetr}
\bibliography{ref}

\appendices
\section{}
\subsection{Related Work}
\label{sec:related_work}
Some works have evaluated the robustness of various watermarks against character-level perturbations --- such as typos, misspellings, and homoglyph substitutions --- demonstrating that these attacks can be effective at degrading detection performance~\cite{kirchenbauer2023watermark, liang2024waterpark, creo2025silverspeak, piet2023mark, liu2024evaluating}. However, these works do not investigate the underlying reasons why character-level attacks achieve higher removal success, nor do they explore more sophisticated attacks that systematically exploit the advantages of such perturbations. 
Liang et al. proposed an adaptive black-box transfer attack by using a gradient-based textual adversarial attack \cite{liang2024waterpark}. 
Due to the mismatch between the decision boundary of the reference detector and the original watermark detector, their method relies on injecting substantial noise to ensure effectiveness.
Stealing detailed information of watermark schemes is another option of guided removal attack \cite{Jovanovi2024steal, zhang2024steal, wu2024bypassing, chen2024mark}. 
While effective in some scenarios, these attacks are computationally expensive and require strong assumptions (i.e., adversaries need to know detailed information about watermark schemes). Specifically, each stealing-based method is tailored to a narrow class of watermark schemes, limiting their general applicability in practice.

\begin{figure}
    \centering
    \includegraphics[width=\linewidth]{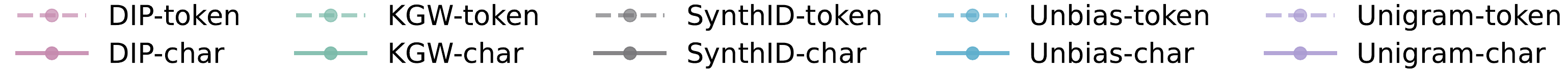}
    \includegraphics[height=0.4\linewidth]{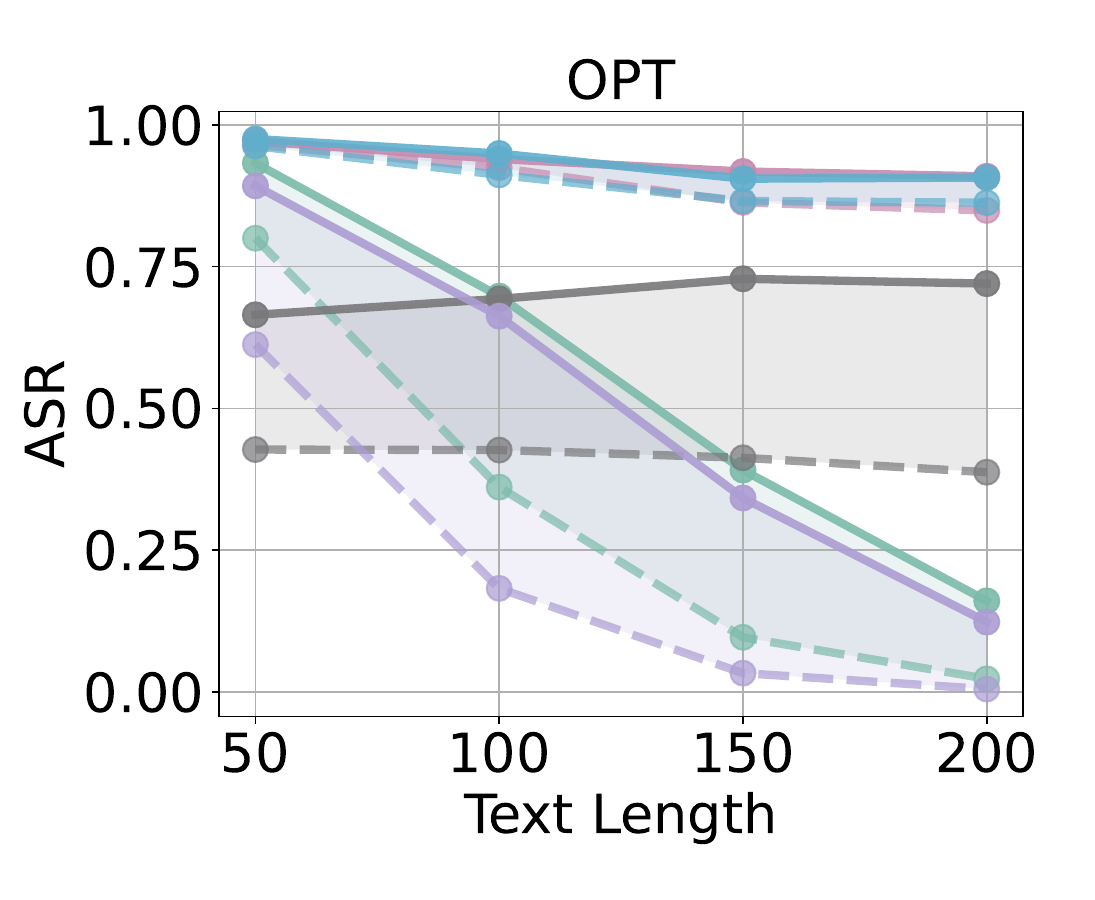}
    \includegraphics[height=0.4\linewidth]{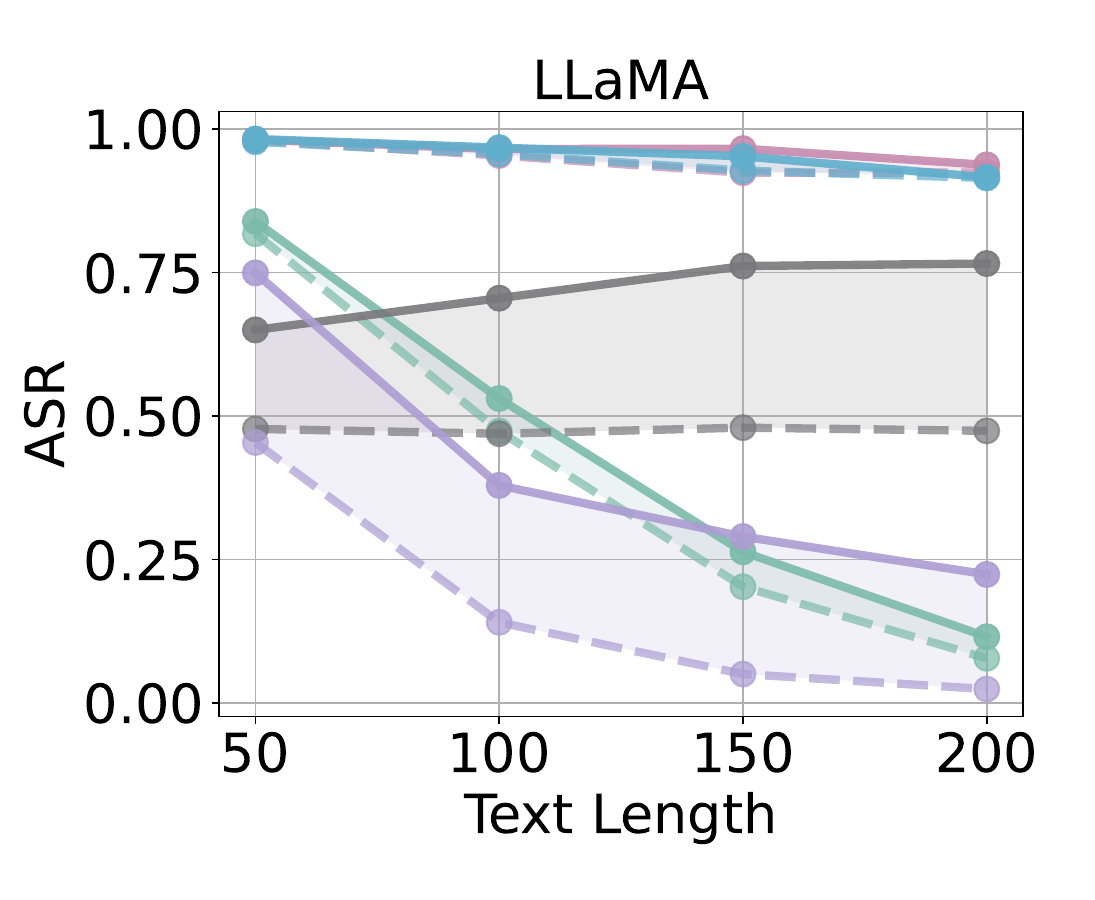}
    \caption{
    $\mathrm{ASR}$ of token-level and character-level watermark removal attacks under varying text lengths (from 50 to 200 tokens) with $\mathrm{ER} = 0.2$. 
    Solid lines represent character-level attacks, and dashed lines represent token-level attacks. 
    }
    \label{fig:text_len_asr}
\end{figure}

\subsection{Impact of Text Length for Random Watermark Removal}
\label{sec:text_len_asr}
Figure~\ref{fig:text_len_asr} shows how input length (ranging from 50 to 200 tokens) affects the ASR at a fixed editing rate of $\mathrm{ER} = 0.2$. For each watermarking scheme, character-level attack results are shown as solid lines, and token-level attacks as dashed lines.
Character-level attacks consistently outperform token-level attacks across all input lengths and watermark schemes. This performance gap is especially evident for KGW, Unigram, and SynthID. 
Notably, $\mathrm{ASR}$ decreases with increasing input length for KGW and Unigram, but remains relatively stable, or even slightly increases for SynthID. This behavior stems from differences in detection mechanisms. 
As described in Section~\ref{sec:detect_wm}, KGW and Unigram use a z-test formulation where the denominator depends on text length $m$, meaning longer texts require a greater proportion of green tokens to be detected as watermarked. In contrast, SynthID aggregates per-token watermark scores, making it less sensitive to text length under fixed $\mathrm{ER}$. 
For DIP and Unbias, $\mathrm{ASR}$ remains consistently high across all lengths, suggesting weaker robustness. 

\begin{table}[htbp]
  \centering
  \caption{
  \ndssmt{Watermark removal performance on French watermarked text under the \textbf{AC1} threat model. We evaluate both token-level and character-level attacks on LLaMA-generated text embedded with KGW and SynthID watermarks. Results are reported with input lengths of 100 and 200 tokens, and $\mathrm{ER}$ of 0.1 and 0.5.} }
  \resizebox{0.85\linewidth}{!}{
  {\ndsstt
    \begin{tabular}{l|c|cc|cc}
    \toprule
          &       & \multicolumn{2}{c|}{Token} & \multicolumn{2}{c}{Char} \\
          & $\mathrm{ER}$ & $\mathrm{WDR}(\uparrow)$ & $\mathrm{ASR}(\uparrow)$ & $\mathrm{WDR}(\uparrow)$ & \multicolumn{1}{c}{$\mathrm{ASR}(\uparrow)$} \\
    \midrule
    \multirow{2}[2]{*}{KGW (100)} & 0.1   & 0.0693 & 0.0736 & 0.0940 & 0.1207 \\
          & 0.5   & 0.2865 & 0.9264 & 0.3472 & 0.9816 \\
    \midrule
    \multirow{2}[2]{*}{KGW (200)} & 0.1   & 0.0674 & 0.0040 & 0.0907 & 0.0080 \\
          & 0.5   & 0.2740 & 0.7062 & 0.3304 & 0.9135 \\
    \midrule
    \multirow{2}[2]{*}{SynthID (100)} & 0.1   & 0.1342 & 0.0825 & 0.1719 & 0.1538 \\
          & 0.5   & 0.3802 & 0.9537 & 0.3953 & 1.0000 \\
    \midrule
    \multirow{2}[2]{*}{SynthID (200)} & 0.1   & 0.1596 & 0.0584 & 0.1889 & 0.0966 \\
          & 0.5   & 0.4359 & 0.9879 & 0.4547 & 1.0000 \\
    \bottomrule
    \end{tabular}%
    }}
  \label{tab:french}%
\end{table}%

\subsection{Generalization to Other Languages}
\label{sec:random-char-other-language}

\ndssm{Beyond English, we further evaluated whether our approach generalizes to other Latin-script languages. 
Table~\ref{tab:french} presents results on French text generated by LLaMA and watermarked by KGW and SynthID. 
We compare the effectiveness of token-level and character-level attacks under the \textbf{AC1} setting. 
Specifically, we evaluate text lengths of 100 and 200 tokens, and for each length, we test two editing rates: 0.1 and 0.5. 
The results show that character-level attacks consistently achieve higher ASR than token-level attacks under the same editing rate. This trend holds across both watermark schemes and text lengths. 
These findings are consistent with our observations on English data: character-level perturbations effectively remove watermarks with minimal edit distance by disrupting the tokenization process.}

\begin{figure}
    \centering
    \includegraphics[width=\linewidth]{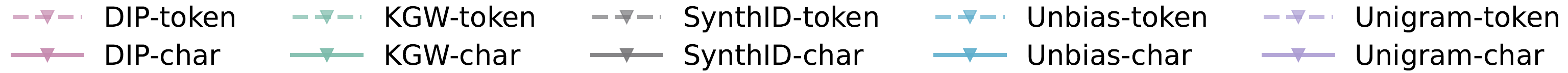}
    \includegraphics[height=0.4\linewidth]{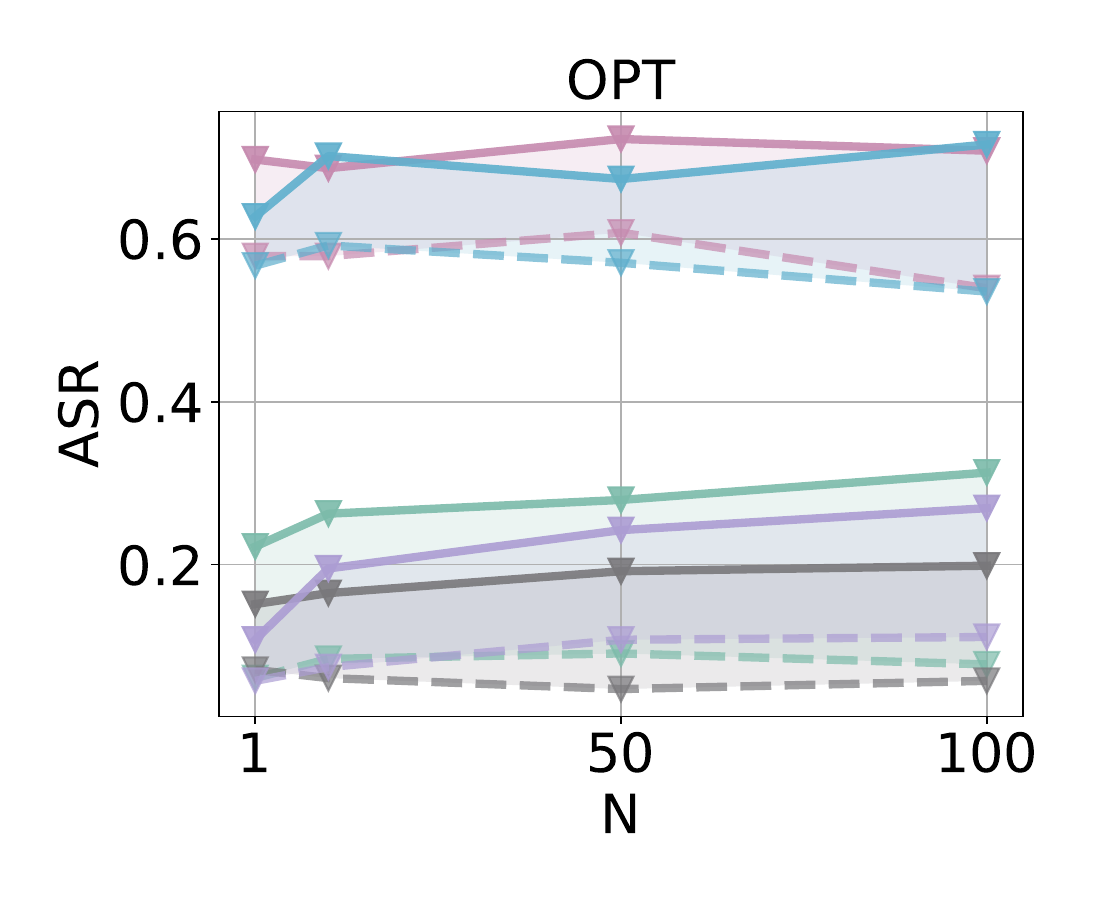}
    \includegraphics[height=0.4\linewidth]{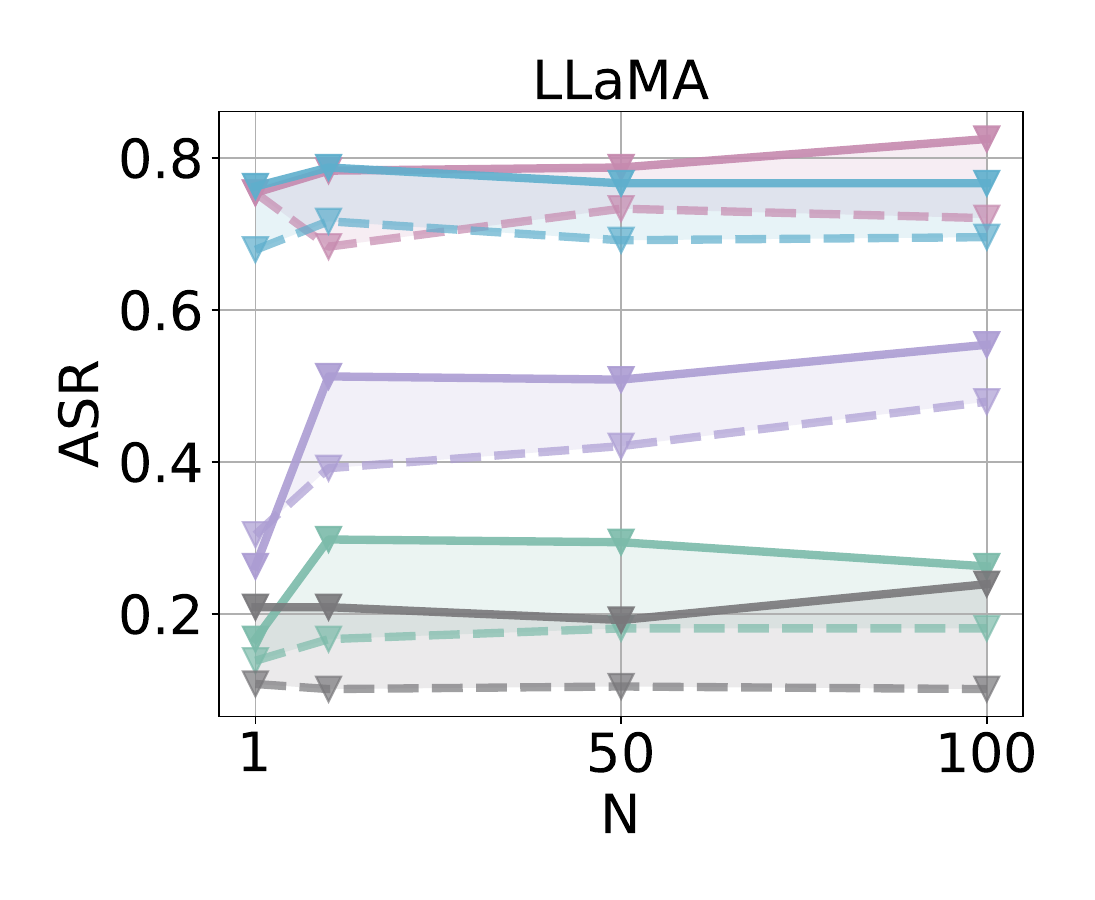}
    \caption{
    $\mathrm{ASR}$ comparison under the Best-of-$N$ attack with increasing $N \in \{1,10,50,100\}$ for five watermark schemes. Evaluated using Ref-9 as the reference detector and editing rate $\mathrm{ER}=0.1$. Solid lines represent character-level attacks; dashed lines represent token-level attacks. 
    }

    \label{fig:asr_n}
\end{figure}

\begin{figure*}
    \centering
    \includegraphics[width=0.18\linewidth]{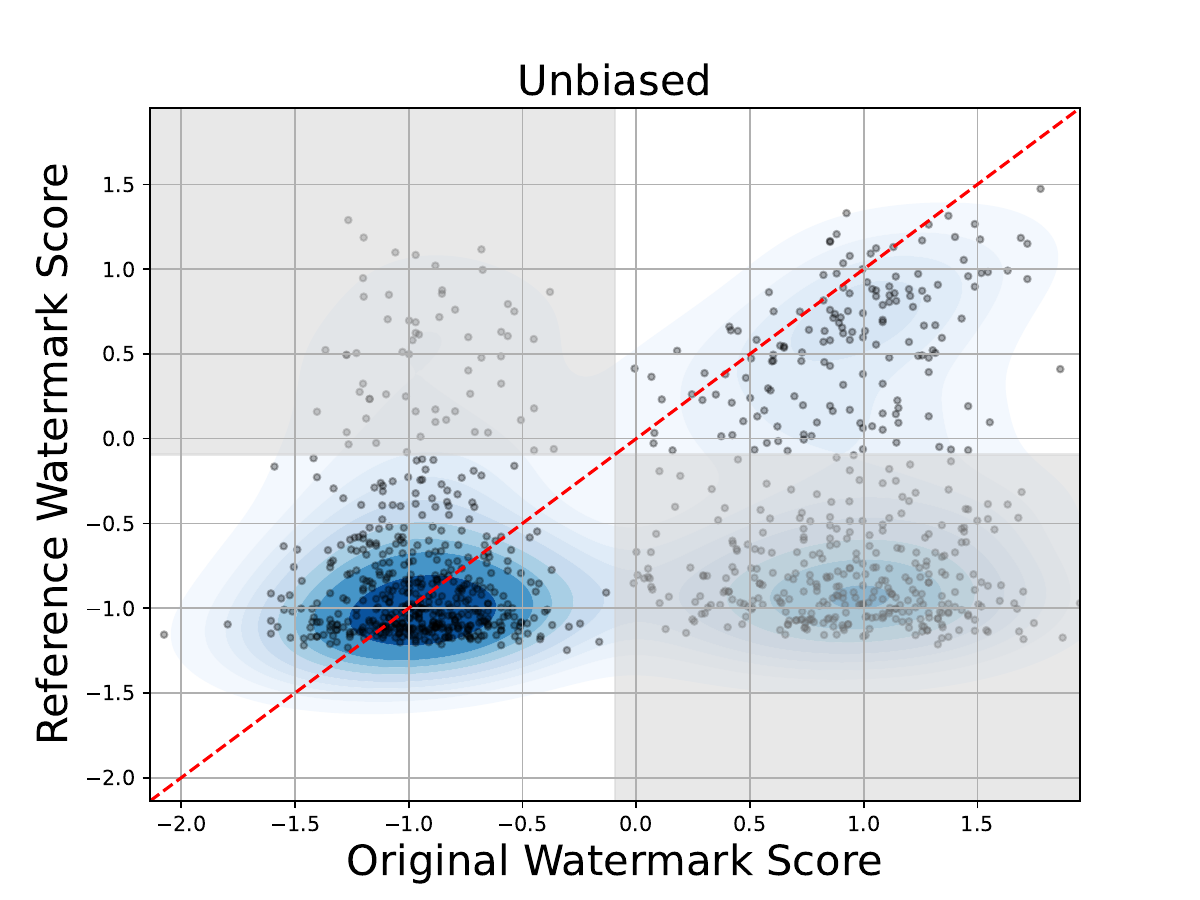}
    \includegraphics[width=0.18\linewidth]{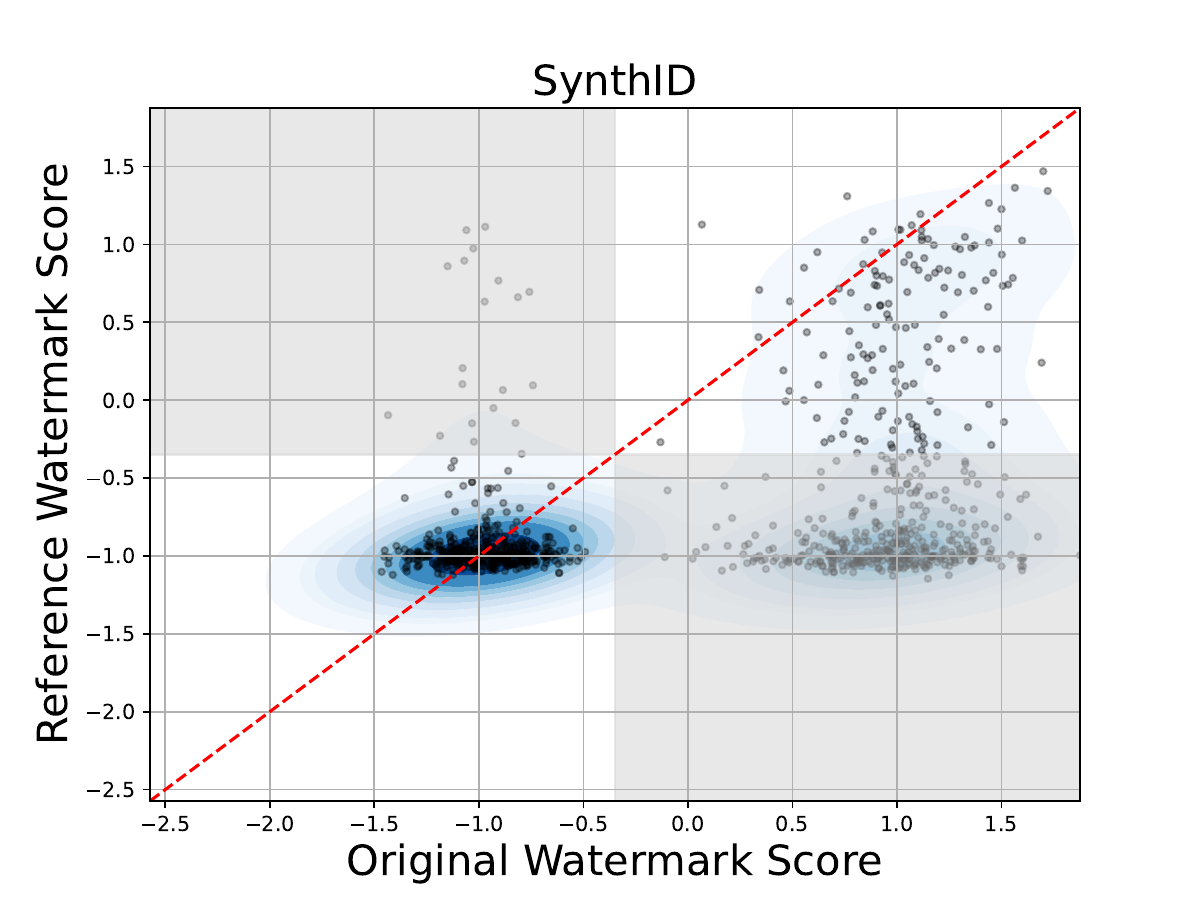}
    \includegraphics[width=0.18\linewidth]{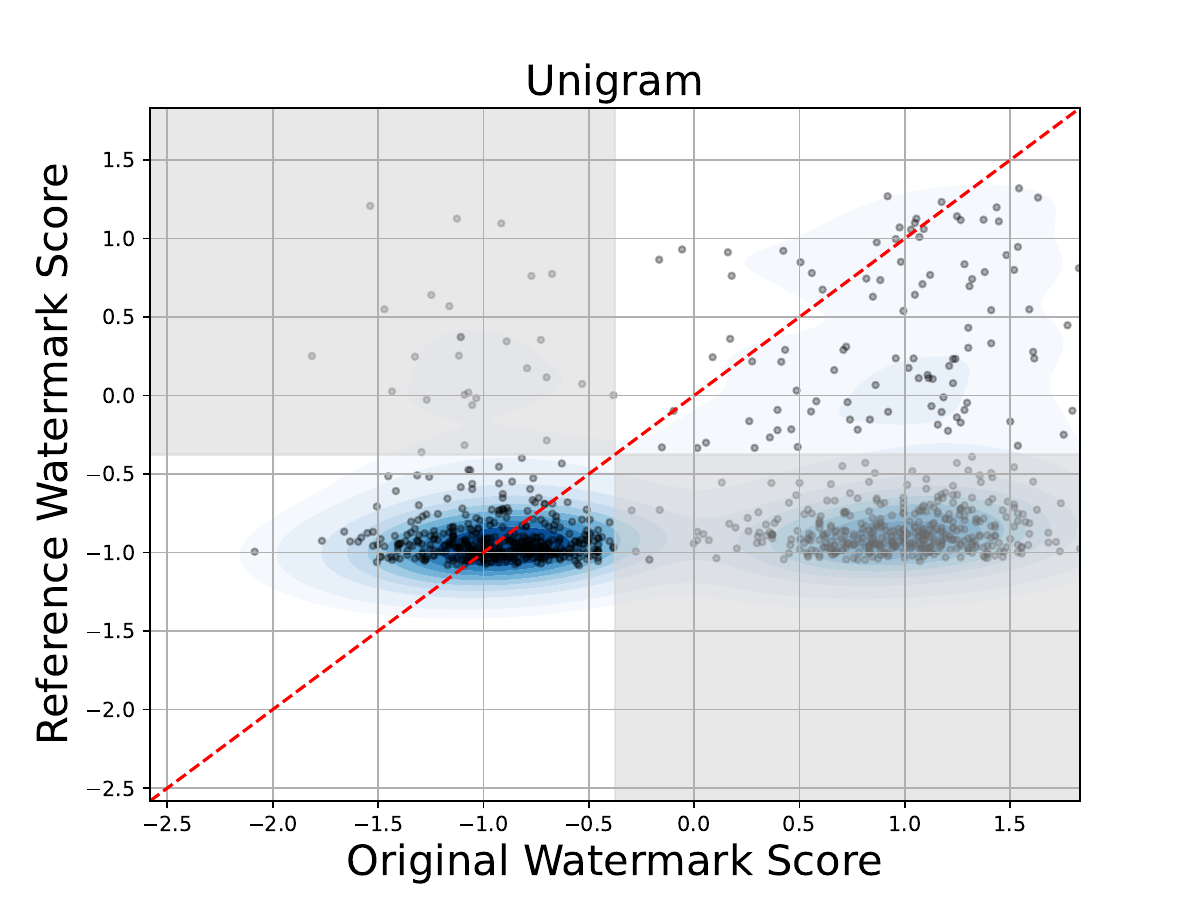}
    \includegraphics[width=0.18\linewidth]{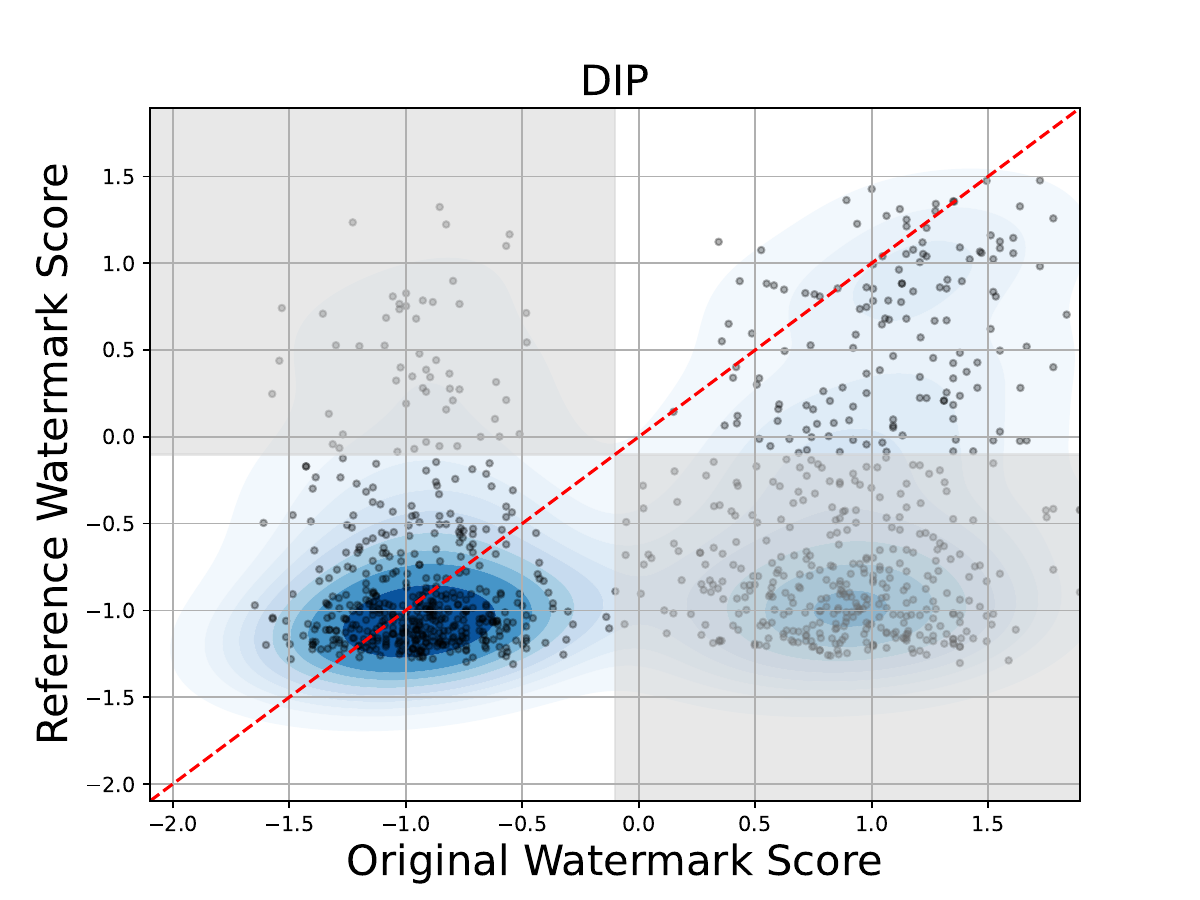}

    \includegraphics[width=0.18\linewidth]{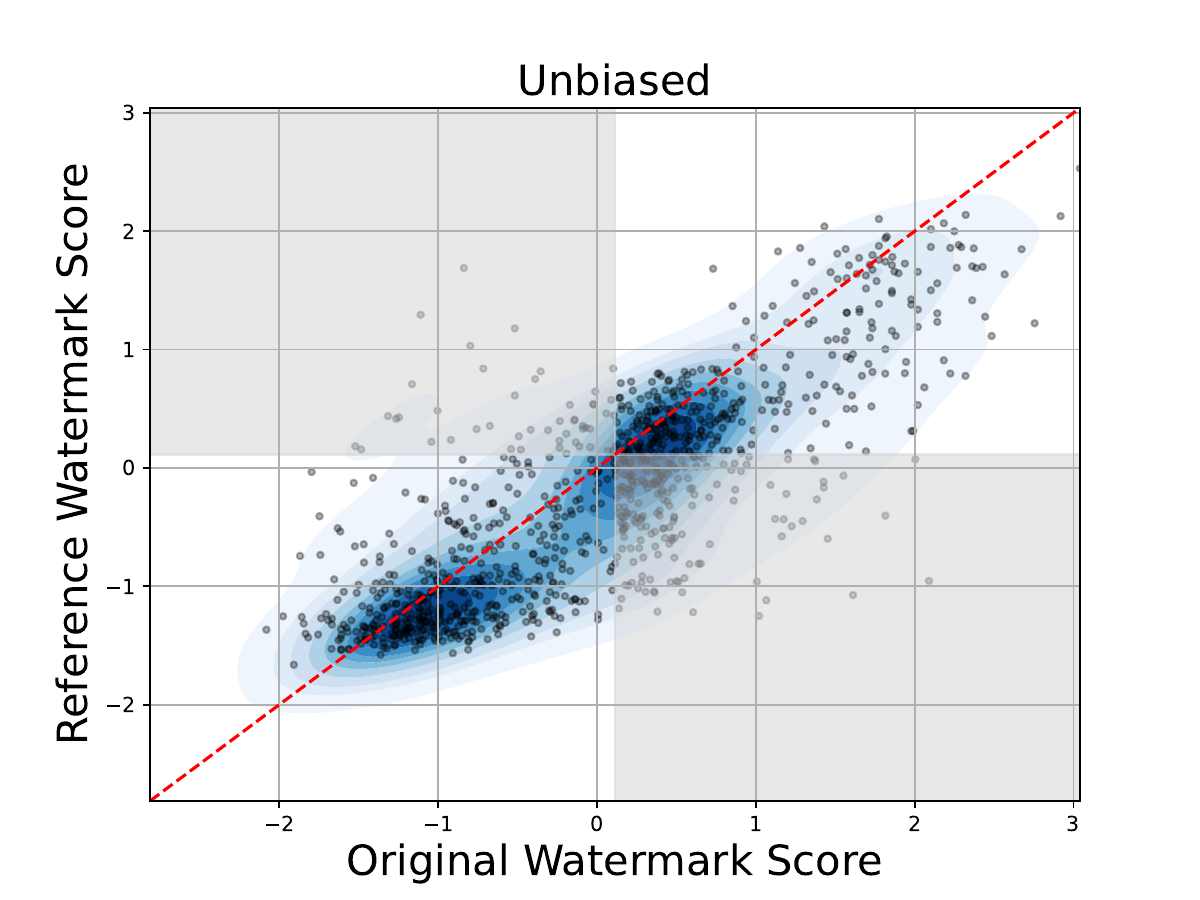}
    \includegraphics[width=0.18\linewidth]{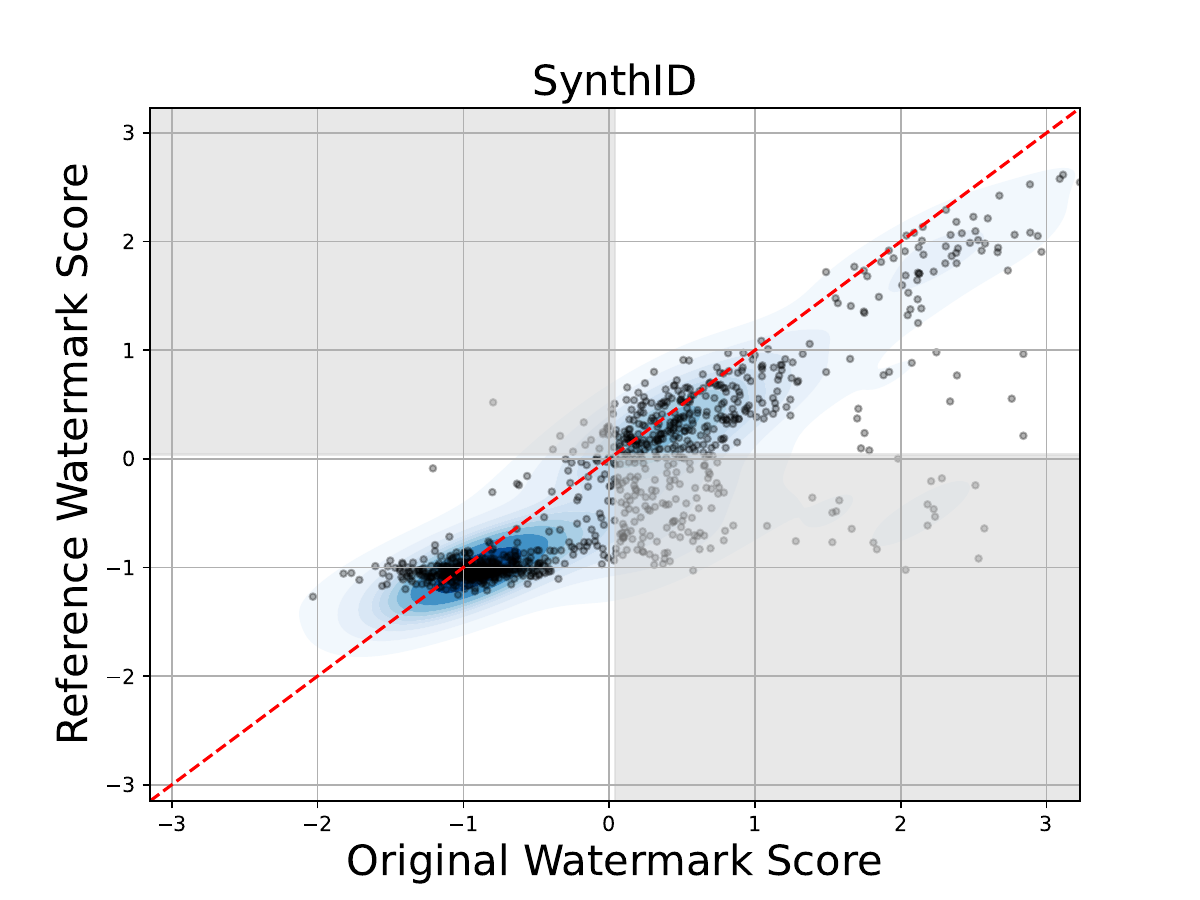}
    \includegraphics[width=0.18\linewidth]{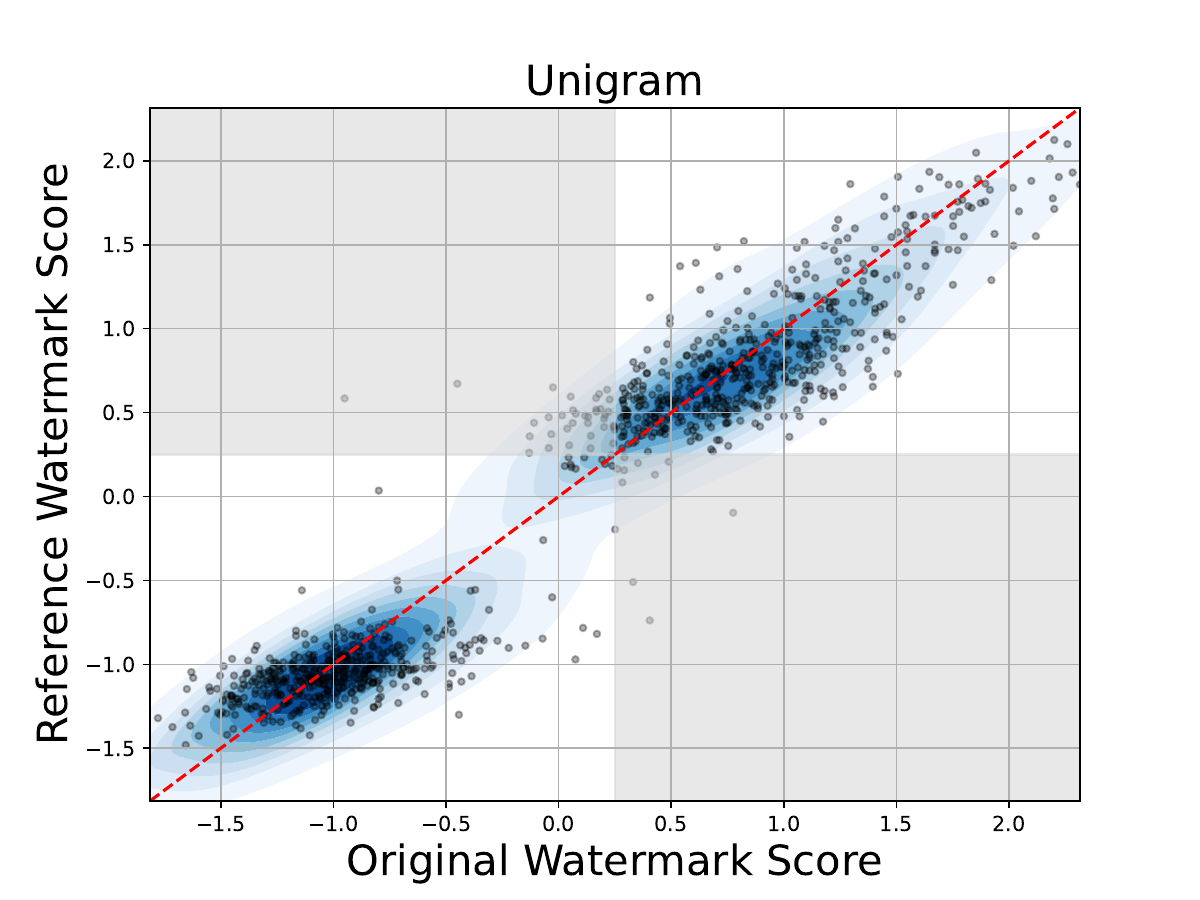}
    \includegraphics[width=0.18\linewidth]{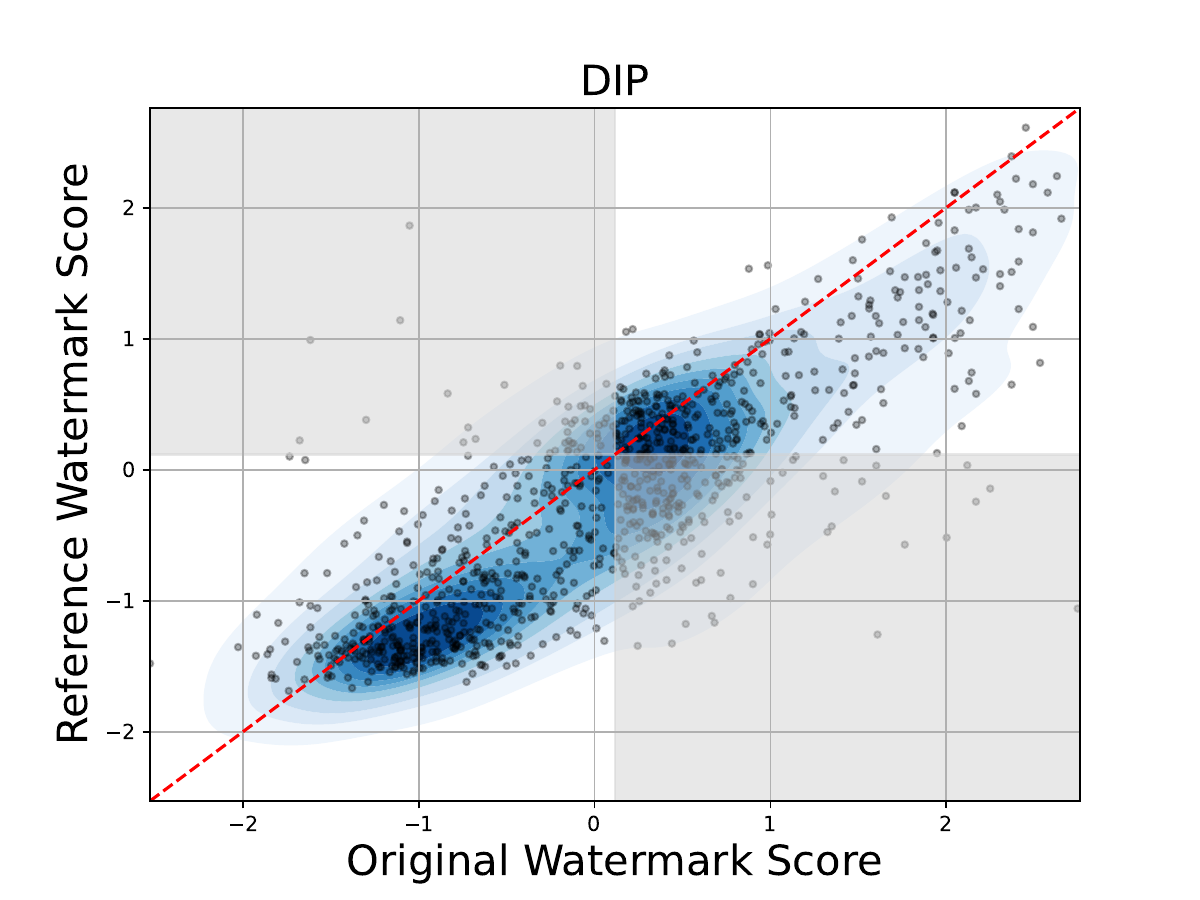}

    \includegraphics[width=0.18\linewidth]{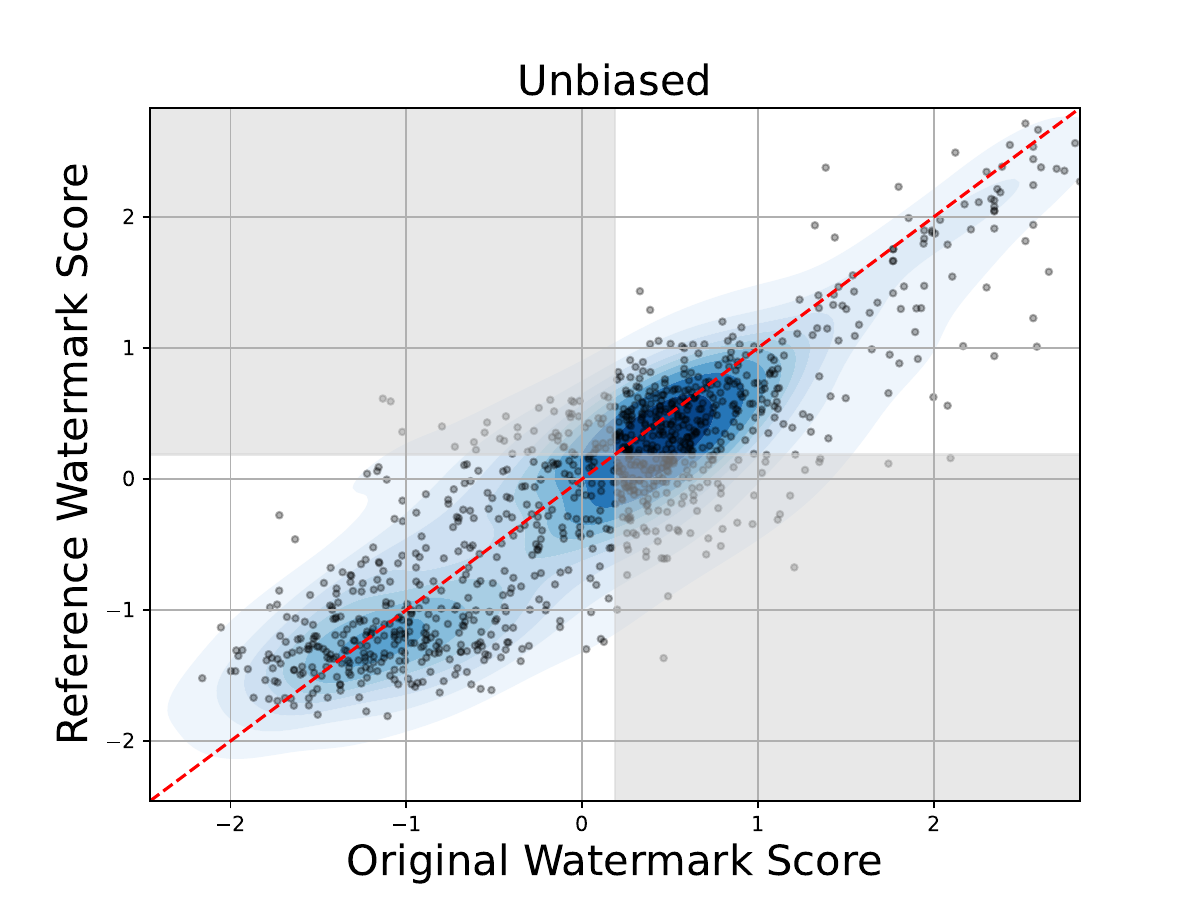}
    \includegraphics[width=0.18\linewidth]{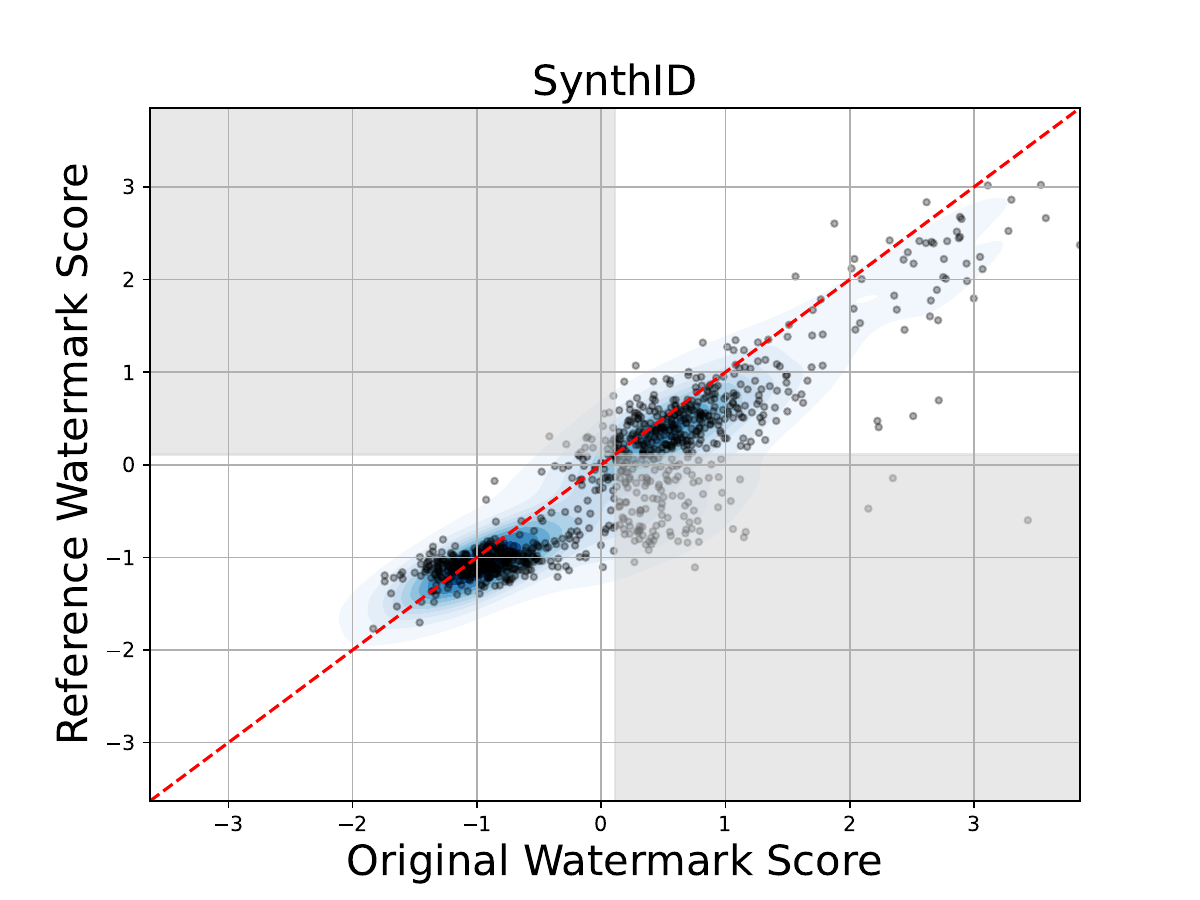}
    \includegraphics[width=0.18\linewidth]{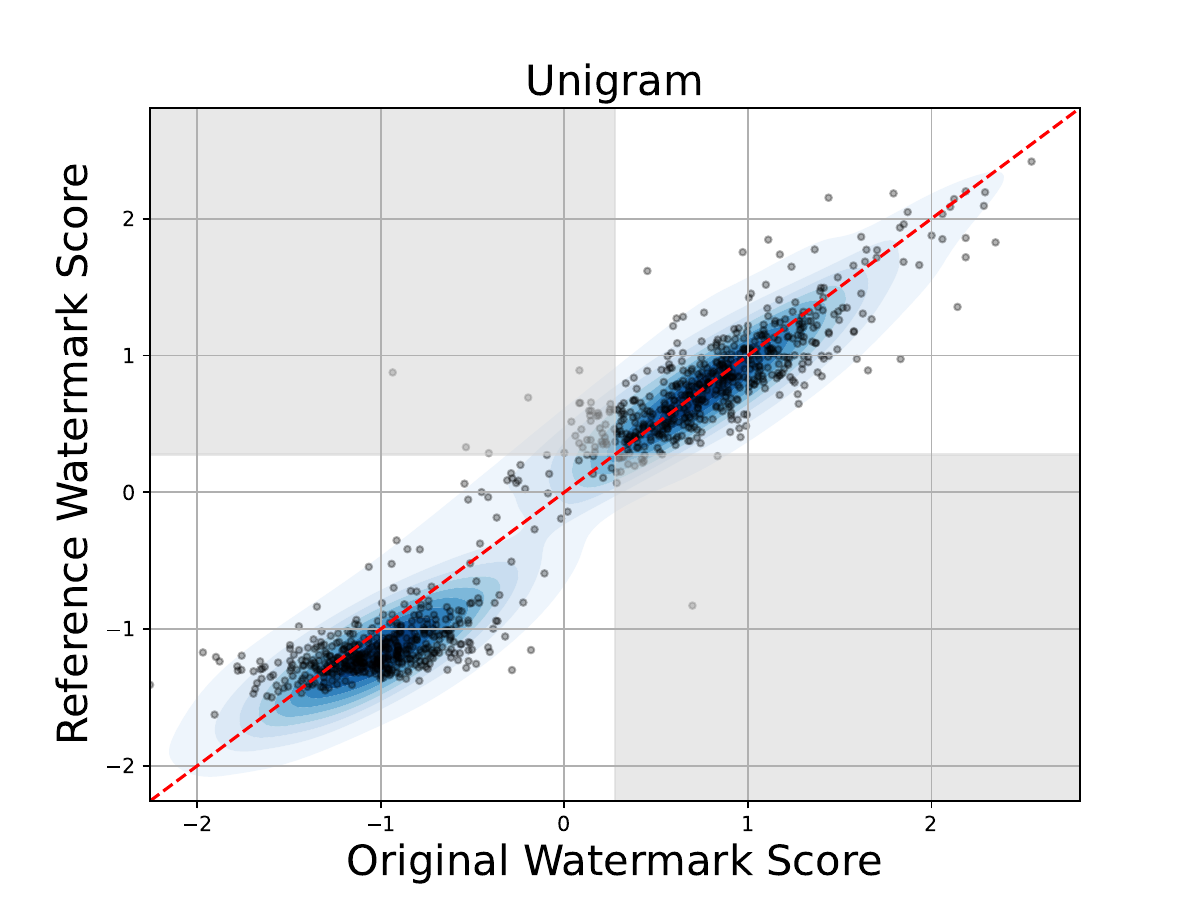}
    \includegraphics[width=0.18\linewidth]{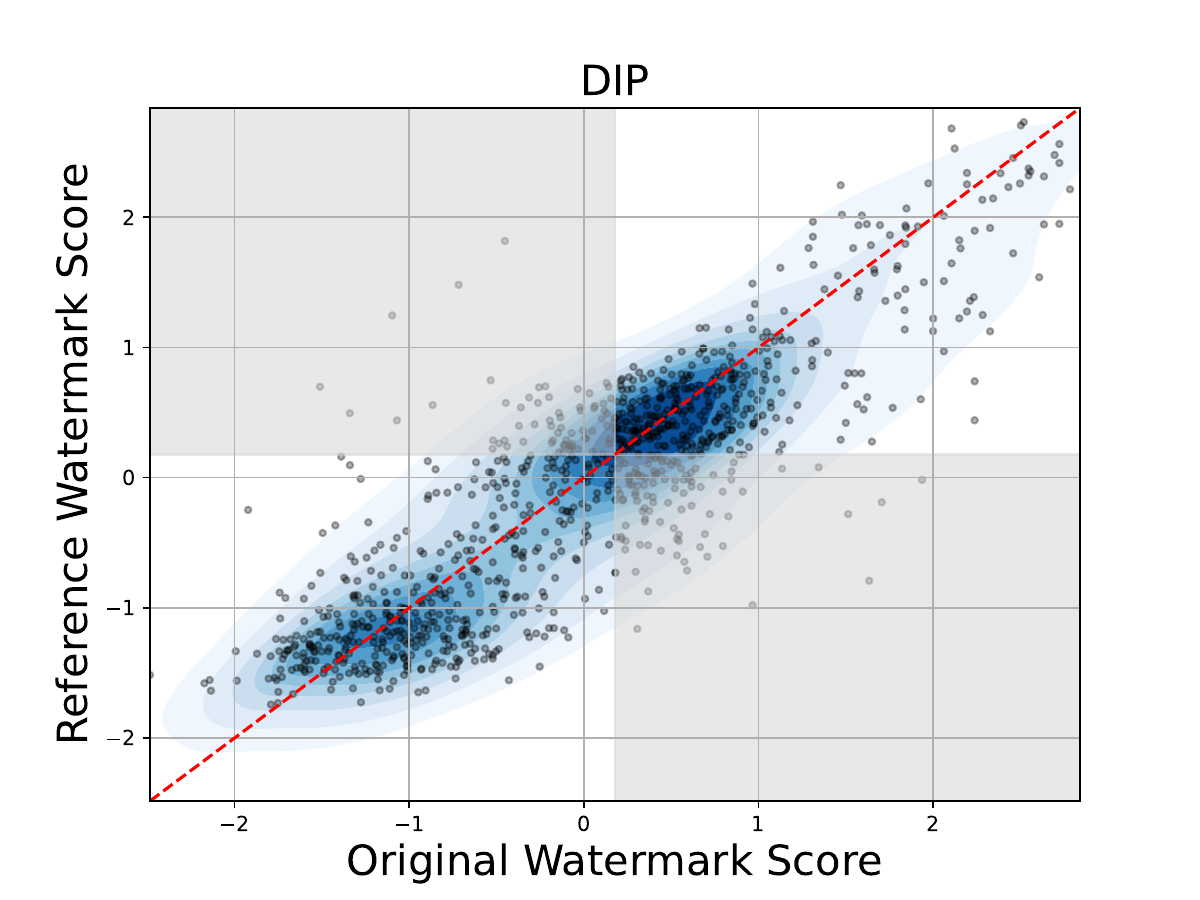}
    \caption{
    Scatter plots comparing predicted watermark scores from the reference detector (y-axis) against ground-truth scores from the original detector (x-axis) for the rest watermark schemes (columns). Each row corresponds to a reference model trained with increasing levels of data augmentation: Ref-0 (top), Ref-5 (middle), and Ref-9 (bottom). The red dashed line represents perfect prediction ($y = x$). White regions indicate correct classifications (top-right and bottom-left), while shaded regions show misclassified samples. 
    }

    \label{fig:ref_distri}
\end{figure*}

\subsection{Genetic Algorithm-Based Attack with Unlimited Access to the Original Watermark Detector}
\label{sec:ga_ori}

We begin by evaluating the GA-based attack in an excessive setting, where the adversary has unrestricted access to the original watermark detector $D_{\text{ori}}$. 
Due to the use of pseudorandom functions in its detection rules (refer to Section \ref{sec:detect_wm}), $D_{\text{ori}}$ is non-differentiable. 
So we adopt a Genetic Algorithm (GA) to optimize it. 
In this setting, the GA directly leverages the true detector scores to identify the most impactful perturbation positions for watermark removal. Specifically, it aims to find a minimal subset of token positions whose perturbation leads to a substantial reduction in the watermark score predicted by the original detector $D_{\text{ori}}$. The objective function is defined as: 
\begin{IEEEeqnarray}{lll}
\begin{aligned}
    \argmin_{\tilde{P} \subset \{1, \cdots, m\}} & D_{\text{ori}}(\mathcal{A}_{\tilde{P}}^C(X))+\lambda \cdot \frac{|\tilde{P}|}{|X|},
    \label{eq:ga_object_ori}
\end{aligned}
\end{IEEEeqnarray}
where $\lambda$ is the weight for editing rate,  $\tilde{P}\subset \{1, \cdots, m\}$ is an individual in the GA population that represents a set of token positions of the input text, and $m$ is the number of tokens in $X$. 
In each iteration of GA, it generates perturbed texts, evaluates their watermark scores using $D_{\text{ori}}$, and updates the best solution if a significant improvement is observed. The best individuals are selected as parents $Q$, which are then used to generate the next population through crossover and mutation. The process continues until a maximum number of iterations is reached. 
We summarize the full procedure in Algorithm~\ref{alg:genetic_base}. 

\begin{algorithm}
\small
    \caption{GA-based Removal with Original Detector}
    \begin{algorithmic}[1]
        \Require Watermarked text $X$, the token number of text $m$, original detector $D_{\text{ori}}$, iteration rounds $n$, population size $p$, parent size $p_s$, weight for editing rate $\lambda$
        \State Initialize population $\textbf{P}_1=\{\tilde{\mathcal{P}}^{(q)}\}_{q=1}^{p},\tilde{\mathcal{P}}^{(q)} \subset \{1,\cdots, m\}$
        \For {iteration $j = 1$ to $n$}
            \For {$q = 1$ to $p$}
                \State $\tilde{X}^{(q)} = \mathcal{A}_{\tilde{\mathcal{P}}^{(q)}}^C(X), \tilde{\mathcal{P}}^{(q)} \in \textbf{P}_j$
                \State Compute reference score $w^{(q)} = D_{\text{ori}}(\tilde{X}^{(q)})$
                \State Compute editing rate $e^{(q)}=\mathrm{ER}(\tilde{X}^{(q)}, X)$
                \State $\mathcal{L}^{(q)} = w^{(q)} + \lambda \cdot e^{(q)}$
            \EndFor
            \State Select parents $Q_j = \text{top-}{p_s}(\textbf{P}_j)$ in ascending $\mathcal{L}^{(q)}$
            \State Best perturbed text $\tilde{X}=\argmin_{\tilde{X}^{(q)}} \mathcal{L}^{(q)}, q \in [1,p]$
            \State Next population: $\textbf{P}_{j+1} \xleftarrow[\text{crossover}]{\text{mutation}} Q_j$
        \EndFor
        \State \Return Final perturbed text $\tilde{X}$
    \end{algorithmic}
    \label{alg:genetic_base}
\end{algorithm}

\paragraph{Best-of-$N$ Attack}
To highlight the advantages of the full Genetic Algorithm (GA), we introduce a simplified version as a baseline, referred to as the Best-of-$N$ Attack. This method removes the iterative refinement of GA and instead performs a one-shot random search. 
Specifically, Best-of-$N$ corresponds to a special case of the GA-based approach in Algorithm~\ref{alg:genetic_base}, where the number of optimization iterations is set to 1. In this setting, $N$ candidate texts are generated using the random perturbation strategy defined in Equation~(\ref{eq:bad_char_base}), and the one with the lowest watermark score $D_{\text{ori}}(\tilde{X})$ is selected. As $N$ increases, the likelihood of sampling removal-relevant tokens improves, leading to stronger candidates for watermark removal. The strategy is formally defined as follows:
\begin{IEEEeqnarray}{lll}
\begin{aligned}
    \tilde{X} &= \argmin_{\substack{\tilde{X}^{(q)} \in \mathcal{B}_{\epsilon}(X) \\ q \in [1, N]}} 
D_{\mathrm{ori}}\left( \tilde{X}^{(q)} \right),\\
    \tilde{X}^{(q)} &= \mathcal{A}_{\tilde{P}^{(q)}}^C(X), \quad
    \tilde{P}^{(q)} \xleftarrow{\text{rand}} \{1, \cdots, m\},
    \label{eq:bon_ori}
\end{aligned}
\end{IEEEeqnarray}
where $\mathcal{A}_{\tilde{P}^{(q)}}^C$ denotes the character-level perturbation method introduced in Section~\ref{sec:bad_char_rm}, which applies perturbations to the central characters of tokens at randomly sampled positions. The set $\mathcal{B}$ represents the collection of candidate texts constrained by a maximum editing rate, $\mathrm{ER} \leq \epsilon$.
Although simple, this approach lacks a guidance and relies entirely on random sampling. In contrast, the full GA performs guided selection and iterative refinement, allowing it to consistently identify more effective perturbation positions at a lower editing cost.

\begin{table*}[htbp]
  \centering
  \caption{
  Comparison of Best-of-$N$ and Genetic Algorithm attacks under unlimited access to the original watermark detector. The table reports $\mathrm{WDR}$ and $\mathrm{ASR}$ for each watermark scheme. Best-of-$N$ is evaluated with different values of $N=1,10,50,500$. \ndssmt{GA uses a population size of 100 and 5 iterations (query budget: $5*100$), matching the query budget of Best-of-$N$ (500).} GA consistently achieves higher performance, even when the total number of detector queries is matched.
  }
  \resizebox{0.75\linewidth}{!}{
  {
    \begin{tabular}{l|cc|cc|cc|cc|cc}
    \toprule
          & \multicolumn{2}{c|}{KGW} & \multicolumn{2}{c|}{DIP} & \multicolumn{2}{c|}{SynthID} & \multicolumn{2}{c|}{Unigram} & \multicolumn{2}{c}{Unbias} \\
          & \ndssmt{$\mathrm{WDR(\uparrow)}$} & $\mathrm{ASR}(\uparrow)$ & \ndssmt{$\mathrm{WDR(\uparrow)}$} & $\mathrm{ASR}(\uparrow)$ & \ndssmt{$\mathrm{WDR(\uparrow)}$} & $\mathrm{ASR}(\uparrow)$ & \ndssmt{$\mathrm{WDR(\uparrow)}$} & $\mathrm{ASR}(\uparrow)$ & \ndssmt{$\mathrm{WDR(\uparrow)}$} & $\mathrm{ASR}(\uparrow)$ \\
    \midrule
    Best-of-$N$ (1) & \ndssmt{0.1278} & 0.2238 & \ndssmt{0.2045} & 0.6883 & \ndssmt{0.2165} & 0.1318 & \ndssmt{0.1090} & 0.2053 & \ndssmt{0.1959} & 0.6223 \\
    Best-of-$N$ (10) & \ndssmt{0.1958} & 0.5503 & \ndssmt{0.3103} & 0.9805 & \ndssmt{0.2961} & 0.4645 & \ndssmt{0.1616} & 0.4797 & \ndssmt{0.3108} & 0.9828 \\
    Best-of-$N$ (50) &\ndssmt{ 0.2278} & 0.7372 & \ndssmt{0.3634} & 0.9978 & \ndssmt{0.3308} & 0.6815 & \ndssmt{0.1845} & 0.6280 & \ndssmt{0.3631} & 0.9979 \\
    Best-of-$N$ (500) & \ndssmt{0.2602} & 0.8706 & \ndssmt{0.4142} & 1.0000 & \ndssmt{0.3978} & 0.8451 &\ndssmt{0.2111} & 0.7534 & \ndssmt{0.4171} & 1.0000 \\
    GA ($5*100$)& \ndssmt{0.3232} & 0.9144 &\ndssmt{ 0.4253} & 1.0000 & \ndssmt{0.4083} & 0.8788 & \ndssmt{0.2245} & 0.7979 & \ndssmt{0.4256} & 1.0000 \\
    \bottomrule
    \end{tabular}%
    }
    }
  \label{tab:ori_N}%
\end{table*}%

\paragraph{Evaluation}
Table~\ref{tab:ori_N} compares the watermark removal effectiveness of Best-of-$N$ and the Genetic Algorithm under the setting where the original watermark detector is fully accessible for the adversary. We evaluate Best-of-$N$ under four different values of $N$ (1, 10, 50, 500), and set the GA with a population size of $100$ and $5$ iterations, which means it needs to query the original detector 500 times. Across all five watermark schemes, GA consistently achieves higher $\mathrm{WDR}$ and $\mathrm{ASR}$ scores than Best-of-$N$. 
While the performance of Best-of-$N$ improves with larger $N$, e.g., for KGW, $\mathrm{ASR}$ increases from $0.2238$ (at $N=1$) to $0.8706$ (at $N=500$). Note that GA outperforms Best-of-$N$ even when both approaches make an equal number of queries (500) to the original detector. So, GA’s advantage stems not only from more query budget but also from its ability to efficiently identify removal-relevant tokens through optimization.

\begin{table}[htbp]
  \centering
  \caption{
  Effect of GA iterations ($n$) on watermark removal. Watermarked texts are generated by OPT. We report $\mathrm{ASR}(\uparrow)$ for both token-level and character-level perturbations with $\epsilon = 0.13$ in GA. Increasing $n$ improves ASR, with character-level perturbations showing greater gains.
  }
  \resizebox{\linewidth}{!}{
    \begin{tabular}{c|cc|cc|cc|cc|cc}
    \toprule
          & \multicolumn{2}{c|}{KGW} & \multicolumn{2}{c|}{DIP} & \multicolumn{2}{c|}{SynthID} & \multicolumn{2}{c|}{Unigram} & \multicolumn{2}{c}{Unbias} \\
          $n$& Token & Char  & Token & Char  & Token & Char  & Token & Char  & Token & Char \\
    \midrule
    $5$ & 0.3567 & 0.4144 & 0.5578 & 0.8273 & 0.0610 & 0.3468 & 0.4106 & 0.5959 & 0.5385 & 0.7766 \\
    $10$ & 0.3511 & 0.4384 & 0.5624 & 0.8273 & 0.0633 & 0.3737 & 0.4279 & 0.6404 & 0.5650 & 0.7979 \\
    $15$ & 0.3744 & 0.4966 & 0.5622 & 0.8453 & 0.0634 & 0.4209 & 0.4343 & 0.6575 & 0.5795 & 0.8369 \\
    \bottomrule
    \end{tabular}%
    }
  \label{tab:ga_n_asr}%
\end{table}%

\subsection{Mismatch Between Reference and Original Detectors}
\label{sec:mismatch} 
An intuitive approach for leveraging a reference detector in watermark removal is to use its gradients to identify important tokens and apply perturbations. This idea follows the standard practice in adversarial NLP, where gradients are often used to rank token importance \cite{li2018textbugger, morris2020textattack, gao2018black}. 
However, this strategy fails in our setting due to a fundamental mismatch between the reference detector and the original detector. In original detectors, such as watermark during logits generation (e.g., KGW \cite{kirchenbauer2023watermark}), each token is categorized as either green or red. The global watermark score is computed using token counts: $S_w(X) = \frac{(1-\gamma)|X|_G - \gamma |X|_R}{\sqrt{\gamma(1-\gamma)|X|}}$, where $|X|_G + |X|_R = |X|$. Under the common setting $\gamma = 0.5$, all tokens contribute equally to the score, regardless of their position.

In contrast, reference detectors are typically neural networks trained to approximate the original detector’s output. Due to the nature of neural networks, their predictions are often dominated by a small number of high-gradient tokens. 
As a result, gradient-based strategies may focus on tokens important only to the reference detector but insignificant to the original, limiting their effectiveness.

\subsection{Experiment for Genetic Algorithm-Based Attack with Limited Access to the Original Watermark Detector}
\label{sec:app_ref}

\subsubsection{Impact of Iterations in GA}
\label{sec:ga_n_asr}
Table~\ref{tab:ga_n_asr} shows the impact of the number of GA iteration rounds ($n$) on watermark removal effectiveness, with maximum editing rate $\epsilon = 0.13$. Results are reported for $n = 5$, $10$, and $15$, using both token-level and character-level perturbations across five watermarking schemes. As $n$ increases, the $\mathrm{ASR}$ improves consistently. For instance, in the KGW scheme with character-level perturbation, $\mathrm{ASR}$ increases from $0.4144$ (at $n = 5$) to $0.4966$ (at $n = 15$), yielding a gain of $0.0822$. 
Moreover, character-level attacks not only achieve higher ASR, but also benefit more from increased iterations. On average, $\mathrm{ASR}$ of token-level perturbation improves by $0.0178$ from $n = 5$ to $n = 15$, while character-level improves by $0.0592$. These results further highlight the advantage of character-level perturbations in GA-based watermark removal.

\subsubsection{Impact of N in Best-of-N}
\label{sec:bon_ref}
Figure~\ref{fig:asr_n} illustrates the effect of increasing $N$ in the Best-of-$N$ strategy on attack success rate ($\mathrm{ASR}$), evaluated under five watermark schemes with $\mathrm{ER}$ set to 0.1 and Ref-9 used as the reference detector. We compare both token-level (dashed lines) and character-level (solid lines) perturbations. 
Across all watermark schemes, character-level attacks consistently outperform token-level attacks. Moreover, the $\mathrm{ASR}$ gap between the two methods widens as $N$ increases. 
For example, on OPT-generated watermarked text, at $N=1$, the average $\mathrm{ASR}$ across 5 watermark schemes is 0.2670 for token-level attacks and 0.3614 for character-level attacks, yielding a gap of 0.0944. At $N=100$, the token-level $\mathrm{ASR}$ remains similar ($0.2642$), while the character-level $\mathrm{ASR}$ increases to $0.4412$, expanding the gap to $0.1771$. This trend highlights that reference detector guidance is especially beneficial for character-level perturbations, and its advantage grows with larger search budgets.
These results demonstrate that reference detectors effectively enhance watermark removal performance for both perturbation types.

\begin{table}[t]
  \centering
  \caption{
   $\mathrm{ASR}(\uparrow)$ of the GA on OPT-generated watermarked texts under different filtering thresholds for high-gradient tokens. 
    Smaller $\alpha$ values result in more aggressive filtering. 
    \text{None} indicates no filtering is applied. The results show that appropriate filtering improves $\mathrm{ASR}$.
    }

  \resizebox{0.75\linewidth}{!}{
    \begin{tabular}{r|ccccc}
    \toprule
    $\alpha$ & KGW   & DIP   & SynthID & Unigram & Unbias \\
    \midrule
    1     & 0.4315 & 0.8129 & 0.4108 & 0.5788 & 0.7908 \\
    2     & 0.4555 & 0.8129 & \textbf{0.4209} & 0.6096 & 0.8121 \\
    3     & \textbf{0.4966} & \textbf{0.8453} & 0.4175 & 0.6267 & 0.8191 \\
    4     & 0.4555 & 0.8058 & 0.4074 & \textbf{0.6575} & \textbf{0.8369} \\
    None  & 0.4555 & 0.8237 & 0.3737 & 0.6370 & 0.7943 \\
    \bottomrule
    \end{tabular}%
    }
  \label{tab:high_grad_asr}%
\end{table}%

\subsubsection{Effect of Filtering High-Gradient Tokens}
\label{sec:high_grad}
Table~\ref{tab:high_grad_asr} shows the impact of filtering high-gradient tokens on the performance of the GA-based watermark removal. As introduced in Section~\ref{sec:ga_ref}, we identify high-gradient tokens based on their gradient magnitudes with respect to the reference detector, and exclude those whose gradients exceed $\mu + \alpha \cdot \sigma$, where $\mu$ and $\sigma$ are the mean and standard deviation of gradient values, respectively. A smaller $\alpha$ results in more aggressive filtering. 
The table reports $\mathrm{ASR}$ for various $\alpha$ values ranging from $4$ to $1$, as well as the case with no filtering (``None''). We observe that filtering high-gradient tokens leads to improved $\mathrm{ASR}$ across most watermark schemes. Without filtering, the average $\mathrm{ASR}$ across the five watermark schemes is $0.6169$. The best-performing filtered setting achieves an average $\mathrm{ASR}$ of $0.6514$, corresponding to an absolute improvement of $0.0346$. 
However, the optimal value of $\alpha$ varies across watermark schemes, suggesting that the effect of gradient-based token is scheme-dependent.

\begin{table}[t]
  \centering
  \caption{
  Performance of reference detectors with three levels of data augmentation across five watermark schemes. 
  Each model is evaluated by the Pearson correlation between predicted and ground-truth watermark scores, and detection accuracy ($\mathrm{ACC}$) when using the predicted scores for binary watermark classification. 
  }
  \resizebox{\linewidth}{!}{
    \begin{tabular}{cl|cc|cc|cc}
    \toprule
          &       & \multicolumn{2}{c|}{Ref-0} & \multicolumn{2}{c|}{Ref-5} & \multicolumn{2}{c}{Ref-9} \\
          &       & \multicolumn{1}{c}{Pearson$(\uparrow)$} & \multicolumn{1}{c|}{$\mathrm{ACC}(\uparrow)$} & \multicolumn{1}{c}{Pearson$(\uparrow)$} & \multicolumn{1}{c|}{$\mathrm{ACC}(\uparrow)$} & \multicolumn{1}{c}{Pearson$(\uparrow)$} & \multicolumn{1}{c}{$\mathrm{ACC}(\uparrow)$} \\
    \midrule
    \multirow{5}[1]{*}{OPT} & KGW   & 0.5941 & 0.7129 & 0.9471 & 0.9003 & 0.9677 & 0.9369 \\
          & DIP   & 0.2980 & 0.5351 & 0.7942 & 0.7002 & 0.8777 & 0.8188 \\
          & SynthID & 0.4832 & 0.4947 & 0.8891 & 0.8906 & 0.9308 & 0.9233 \\
          & Unigram & 0.9527 & 0.9739 & 0.9923 & 0.9970 & 0.9933 & 0.9977 \\
          & Unbias & 0.2743 & 0.5298 & 0.8201 & 0.7527 & 0.8850 & 0.8181 \\
    \midrule
    \multirow{5}[1]{*}{LLaMA} & KGW   & 0.6611 & 0.7607 & 0.9762 & 0.9792 & 0.9806 & 0.9850 \\
          & DIP   & 0.3240 & 0.5871 & 0.8106 & 0.7687 & 0.8721 & 0.8225 \\
          & SynthID & 0.3920 & 0.6261 & 0.8683 & 0.8241 & 0.9022 & 0.8312 \\
          & Unigram & 0.2422 & 0.5516 & 0.9606 & 0.9459 & 0.9673 & 0.9557 \\
          & Unbias & 0.2915 & 0.6057 & 0.8043 & 0.7582 & 0.8730 & 0.8069 \\
    \bottomrule
    \end{tabular}%
    }
  \label{tab:ref_performance}%
\end{table}%

\subsubsection{Performance of Reference Detector}
\label{sec:ref_preformance_app}
Figure~\ref{fig:ref_distri} compares the watermark scores predicted by the reference detectors with those from the original detectors for Unbiased, SynthID, Unigram, and DIP. For each row, as data augmentation increases, the reference detector's predictions become closer to the original detector's scores. However, samples near the classification threshold (around the boundaries of the shaded regions), are very sparse, which makes it difficult for the reference model to predict reliably in these areas. Even in denser regions, the reference detector often struggles to accurately capture small variations in the original watermark scores.

Table~\ref{tab:ref_performance} presents the performance of reference detectors trained to regress the watermark scores produced by the original detectors. 
This table reports the Pearson correlation between predicted and ground-truth watermark scores, as well as detection accuracy when using the reference model as a binary classifier. The results show that data augmentation significantly boosts performance. On OPT-generated text, Ref-5 and Ref-9 achieve average Pearson improvements of $0.3681$ and $0.4104$, and accuracy improvements of $0.1988$ and $0.2497$, respectively. On LLaMA-generated text, Ref-5 and Ref-9 improve average Pearson by $0.5018$ and $0.5862$, and accuracy by $0.2290$ and $0.2540$, respectively. 

\clearpage
\section{Artifact Appendix}

\subsection{Description \& Requirements}
In this artifact evaluation, we provide a scaled-down version of the experiments to validate the two primary claims presented in the paper: (1) character-level perturbations achieve superior watermark removal effectiveness compared to token-level perturbations under the same editing rate, and (2) Genetic Algorithm-based optimization can leverage the guidance of a reference detector to further improve removal attack performance. 
Our paper received a ``Major Revision'' decision, and in the revised version we are suggested to include more ablation studies and evaluation metrics to systematically investigate the effectiveness of character-level removal. These new components have already been integrated into the source code, and do not affect the artifact evaluation presented here.

\subsubsection{How to access}
The artifact can be accessed via GitHub (\url{https://github.com/plll4zzx/CharacterRemoval4WM}) or DOI (\url{https://doi.org/10.5281/zenodo.15872569}). 
We recommend downloading the code using the following command:
git clone https://github.com/plll4zzx/CharacterRemoval4WM. 
The datasets required for artifact experiment are available through the DOI and Google Drive ({\url{https://drive.google.com/file/d/1ZRPbyv8vHs_rh4fxIPEE3TzHNa-SU54E/view?usp=sharing}}).

\subsubsection{Hardware dependencies}
A commodity desktop machine with at least 8 CPU cores and 16GB RAM. GPU with at least 10GB of memory (e.g., NVIDIA GPU with CUDA) is strongly recommended for faster execution, especially for collecting watermarked text and reference detector training. 

\subsubsection{Software dependencies}
Ubuntu, Python 3.9, Conda, other dependencies are listed in \texttt{requirements.txt}

\subsubsection{Benchmarks}
We use the C4 dataset as the source of prompts to query the victim LLMs and generate watermarked text\footnote{The C4 dataset is publicly available at \url{https://huggingface.co/datasets/allenai/c4}. We recommend downloading it via git for convenience. The commands are:
GIT\_LFS\_SKIP\_SMUDGE=1 git clone https://huggingface.co/datasets/allenai/c4; cd c4; git lfs pull --include ``realnewslike/*''}. In this evaluation, we employ OPT-1.3B\footnote{\url{https://huggingface.co/facebook/opt-1.3b}} as the victim language model to produce watermarked outputs. 
Our reference detectors are finetuned from BERT\footnote{\url{https://huggingface.co/google-bert/bert-base-uncased}}. 
All model weights can be obtained from HuggingFace model repositories.

\subsection{Artifact Installation \& Configuration}
The high-level configuration steps consist of two main components: dependency installation and data preparation: 
\begin{enumerate}
    \item \textbf{Dependency Installation.} Install all required dependencies by running \texttt{sh install.sh}. After installation, using \texttt{conda activate test\_char} to activate the Conda environment.

    \item \textbf{Data Preparation.} Download the C4 dataset and generate watermarked text using the victim LLMs. For convenience, we provide sample data via Google Drive and the DOI link. After downloading, you can extract and place the \texttt{saved\_data}, \texttt{saved\_attk\_data}, and \texttt{saved\_model} directories into the project folder.
\end{enumerate}

\subsection{Experiment Workflow}
The artifact supports a 2-step experimental workflow to reproduce the main results presented in the paper:

\begin{enumerate}
    \item \textbf{Baseline Removal Evaluation.} Execute the scripts for random attacks to validate that character-level perturbations outperform token-level perturbations in removing watermark (refer to Table \ref{tab:compare_token_sentence}). 
    
    \item \textbf{Guided Removal Evaluation.} Train reference detectors on dataset of the watermarked data. Then, run the guided removal attacks, including Best-of-N and Genetic Algorithm (GA) -based optimization, to demonstrate their superior performance (refer to Table \ref{tab:guided_results}).
\end{enumerate}

\subsection{Major Claims}
The artifact is designed to reproduce the following major claims of the paper:
\begin{itemize}
    \item \textbf{C1:} Character-level perturbations have significantly higher attack success rates and watermark score reduction than token-level approaches under the same editing rate, as reported in Table~\ref{tab:compare_token_sentence} (see Appendix~\ref{sec:exp1}).
    \item \textbf{C2:} Guided removal attacks based on reference detector, including Best-of-N and Genetic Algorithm (GA) -based attacks, improve removal effectiveness under black-box conditions, as reported in Table~\ref{tab:guided_results} (see Appendix~\ref{sec:exp2}).
\end{itemize}

\subsection{Evaluation}

\subsubsection{Experiment (E1)}
\label{sec:exp1}
In E1, we aim to evaluate whether character-level removal attacks outperform token-level approaches under the AC1 setting, which assumes no access to the original watermark detector. Consistent with the paper, we test five representative watermarking schemes: KGW, DIP, SynthID, Unigram, and Unbias.

\par\noindent\textit{[Preparation]}
To run this experiment, the C4 dataset is needed. We recommend storing it in the path \texttt{"../../dataset/c4/realnewslike"}, where \texttt{".."} refers to the parent directory relative to the current working directory. Watermarked text is generated using the OPT-1.3B model. The model weights do not require a separate download, as they are automatically retrieved during script execution. 
This process can be performed with the script \texttt{collect\_wm\_text.py}, which requires specifying the watermark name, dataset path, model name, the number of text, and the GPU device. For example: 
\begin{verbatim}
python collect_wm_text.py --device 0 \
--wm_name "KGW" --dataset_name \
"../../dataset/c4/realnewslike" \
--model_name "facebook/opt-1.3b" \
--file_num 50 --file_data_num 100 
\end{verbatim} 
This command generates $5000$ (\texttt{file\_num} $*$ \texttt{file\_data\_num}) watermarked samples using the \texttt{facebook/opt-1.3b} model with the KGW watermark on GPU device 0. To produce watermarked text with other watermarking schemes, simply modify the \texttt{--wm\_name} parameter\footnote{the alias of ``Unbias'' is ``Unbiased'' in the code}. We recommend generating at least $5000$ examples to support reference detector training in E2.

Although using a GPU can accelerate text generation, the process remains time-consuming. For convenience, pre-generated watermarked data are provided in the \texttt{saved\_data} directory to facilitate artifact evaluation.

\par\noindent\textit{[Execution]}
The following command can be used to evaluate the effectiveness of token-level and character-level perturbations in the Random strategy removal attack:
\begin{verbatim}
python test_rand_sh.py \
--llm_name "facebook/opt-1.3b" \
--wm_name_list "['KGW']" \
--atk_style_list "['token','char']" \
--max_edit_rate_list "[0.1, 0.5]" \
--do_flag "True" --atk_times_list "[1]" \
--max_token_num_list "[100]"
\end{verbatim} 
In this example, the editing rates are set to 0.1 and 0.5, and the length of the watermarked text is fixed to 100 tokens, consistent with the settings reported in the paper. 
In this random strategy, the adversary performs only a single random attack per text sample, so \texttt{atk\_times\_list}$=1$. 
To evaluate additional watermarking schemes, simply add their names to the \texttt{wm\_name\_list}.

\par\noindent\textit{[Results]}
The results of this script are saved as log files in \texttt{attack\_log/Rand} and as JSON files in \texttt{saved\_attk\_data}. Alternatively, setting the \texttt{do\_flag} parameter to \texttt{"False"} prints the results directly to the terminal. The output demonstrates that, for the same editing rate, character-level perturbations consistently achieve higher watermark score dropping rate (WDR) and attack success rates (ASR) compared to token-level perturbations (refer to Table~\ref{tab:compare_token_sentence}  in the paper). 
Additionally, by adding more values into \texttt{max\_edit\_rate\_list} and \texttt{max\_token\_num\_list} reproduces the results in Figure \ref{fig:editing_dist_asr}\&\ref{fig:text_len_asr}, respectively.

\subsubsection{Experiment (E2)}
\label{sec:exp2}

In Experiment (E2), we aim to evaluate the effectiveness of watermark removal under the AC2 scenario, where the adversary has limited access to the original watermark detector. In this setting, a reference detector is first trained using a limited number of queries to the original detector. Subsequently, the trained reference detector is used to guide the watermark removal process.

\par\noindent\textit{[Preparation]}
Command for training the reference detector:
\begin{verbatim}
python train_ref_detector.py  --device 0 \
--wm_name "KGW" --num_epochs 15 \
--rand_char_rate 0.15 --rand_times 9 \
--llm_name "facebook/opt-1.3b" --ths 4
\end{verbatim}
In this example, the watermarking scheme is set to KGW. Each sample is augmented 9 times, with an editing rate of 0.15 (\texttt{rand\_char\_rate}). The \texttt{ths} is detection threshold used to evaluate the performance of the reference detector. The number of training epochs is set to 15. Due to both data augmentation and model training are computationally intensive and time-consuming, preprocessed datasets and pre-trained reference detector are provided in the \texttt{saved\_data} and \texttt{saved\_model} to facilitate artifact evaluation.

\par\noindent\textit{[Execution]}
We evaluate the two proposed guided removal attacks (Best-of-N and GA) along with a baseline method (Sand). All of these attacks leverage the trained reference detector to provide guidance.  
\begin{verbatim}
Best-of-N: python test_rand_sh.py \
--llm_name "facebook/opt-1.3b" \ 
--wm_name_list "['KGW']" \
--atk_style_list "['token','char']" \
--do_flag "True" --atk_times_list "[10]" \
--max_token_num_list "[100]" \
--max_edit_rate_list "[0.1]" \
--data_aug_list "[9]"
\end{verbatim}
In this command, the \texttt{atk\_times\_list} parameter specifies the value of $N$ in the Best-of-$N$ strategy, determining how many perturbation candidates are sampled for each input. The \texttt{data\_aug\_list} is used to choose reference detector. Due to we set \texttt{rand\_times} to $9$ when training the reference detector, \texttt{data\_aug\_list} is also set to $9$.

\begin{verbatim}
Sand: python test_rand_sh.py \
--llm_name "facebook/opt-1.3b" \
--wm_name_list "['KGW']" \
--atk_style_list \
 "['sand_token', 'sand_char']" \
--data_aug_list "[9]"  --do_flag "True" \
--max_edit_rate_list "[0.1]" \
--atk_times_list "[1]" \
--max_token_num_list "[100]"
\end{verbatim}

\begin{verbatim}
GA: python test_ga_sh.py \
--llm_name "facebook/opt-1.3b" \
--wm_name "KGW" --atk_style "char" \
--num_generations 15 --do_flag "True" \
--max_edit_rate 0.13 --data_aug 9 \
--max_token_num_list "[100]"
\end{verbatim}

\par\noindent\textit{[Results]}
As in E1, the results of these three methods are saved by default as log files in the \texttt{attack\_log/Rand} and \texttt{attack\_log/GA} directories and as json files in the \texttt{saved\_attk\_data} folder. Setting the \texttt{do\_flag} parameter to \texttt{False} will print the results directly to the terminal. 

The output shows that, under comparable editing rates, the GA consistently achieves higher attack success rates (ASR) than both the Best-of-N and sand attacks (refer to Table~\ref{tab:guided_results}). Additionally, for all three methods, character-level perturbations yield higher ASR compared to token-level perturbations. 

Because this evaluation process can be time-consuming, precomputed json results are provided in the \texttt{saved\_attk\_data} directory to facilitate artifact verification.
\end{document}